\documentclass[usenatbib, useAMS]{mn2e}
\usepackage{graphicx,amssymb,amsmath,epsfig, epstopdf, xfrac, subfigure}

\title[P-MaNGA: Emission Line Properties] {P-MaNGA: Emission Lines Properties - Gas Ionisation and Chemical Abundances from Prototype Observations}
\author[F. Belfiore et al.] {F.~Belfiore,$^{1,2}$\thanks{fb338@cam.ac.uk, Cavendish Laboratory, Cambridge, CB3 0HE, UK} R.~Maiolino,$^{1,2}$ K.~Bundy,$^3$ D.~Thomas,$^{4}$ C.~Maraston,$^{4}$ \newauthor 
 D.~Wilkinson,$^{4}$ S. F.~S\'anchez,$^{5}$  M.~Bershady,$^6$ G. A.~Blanc,$^{7}$ M.~Bothwell,$^{1,2}$  \newauthor 
 S. L. Cales,$^{8}$ L.~Coccato,$^9$  N.~Drory,$^{10}$ E.~Emsellem,$^{9, 11}$ H.~Fu,$^{12}$ J.~Gelfand,$^{13,14}$\newauthor
D.~Law,$^{15}$ K.~Masters,$^{4,16}$ J.~Parejko,$^{8}$ C.~Tremonti,$^6$ D.~Wake,$^{6, 17}$  A.~Weijmans,$^{18}$ \newauthor
R.~Yan,$^{19}$ T.~Xiao,$^{20}$ K.~Zhang,$^{21}$ T.~ Zheng,$^{12}$ D.~Bizyaev,$^{22}$ K.~Kinemuchi,$^{22}$ \newauthor D.~Oravetz$^{22}$ and A.~Simmons$^{22}$
\\
\\
(Affiliations can be found after the references)
}
\begin{document}

\maketitle

  \begin{abstract}
MaNGA (Mapping Nearby Galaxies at Apache Point Observatory) is a 6-year SDSS-IV survey that will obtain spatially resolved spectroscopy from 3600 \AA\ to 10300 \AA\ for a representative sample of over 10000 nearby galaxies. In this paper we present the analysis of nebular emission line properties using observations of 14 galaxies obtained with P-MaNGA, a prototype of the MaNGA instrument. 
By using spatially resolved diagnostic diagrams we find extended star formation in galaxies that are centrally dominated by Seyfert/LINER-like emission, which illustrates that galaxy characterisations based on single fibre spectra are necessarily incomplete.
We observe extended LINER-like emission (up to $\rm 1 R_{e}$)  in the central regions of three galaxies. We make use of the $\rm EW(H \alpha)$ to argue that the observed emission is consistent with ionisation from hot evolved stars.
We derive stellar population indices and demonstrate a clear correlation between $\rm D_n(4000)$ and $\rm EW(H \delta_A)$ and the position in the ionisation diagnostic diagram: resolved galactic regions which are ionised by a Seyfert/LINER-like radiation field are also devoid of recent star formation and host older and/or more metal rich stellar populations. We also detect extraplanar LINER-like emission in two highly inclined galaxies, and identify it with diffuse ionised gas.
We investigate spatially resolved metallicities and find a positive correlation between metallicity and star formation rate (SFR) surface density. We further study the relation between N/O vs O/H on resolved scales. We find that, at given N/O, regions within individual galaxies are spread towards lower metallicities, deviating from the sequence defined by galactic central regions as traced by Sloan $3''$ fibre spectra. We suggest that the observed dispersion can be a tracer for gas flows in galaxies: infalls of pristine gas and/or the effect of a galactic fountain.

 \end{abstract}

\section{Introduction}
The Sloan Digital Sky Survey (SDSS, \citealt{York2000}) has proved to be a transformational tool for the study of galaxy evolution. The statistical power of the SDSS galaxy sample has shed light on the complex interplay of phenomena that shape the formation and evolution of galaxies -  from their history of star formation and chemical enrichment, the impact of inflows and outflows, to the role of active galactic nuclei (AGN), the effect of environment on galaxy properties and cessation of star formation \citep{Kauffmann2003, Kauffmann2003a, Brinchmann2004, Tremonti2004, Kewley2006, Asari2007, Mannucci2010, Peng2010, Thomas2010, Johansson2012}. SDSS currently provides the most comprehensive picture of galaxies in the low-redshift Universe, against which theories of galaxy evolution are tested and properties of high redshift objects are compared.

Despite its tremendous success, Sloan spectroscopy suffers from a major drawback: the single $3''$-diameter Sloan fibre can only trace circum-nuclear galactic properties. For nearby galaxies the derived properties might not be representative of the galaxy as a whole, since the fibre might just be targeting the galactic centre or other areas of high surface brightness. By investigating the full 2D spectroscopic structure of galaxies, integral field spectroscopy (IFS) provides the ideal observational tool to circumvent this problem, while adding a whole new dimension to the study of galaxy evolution by providing spatially resolved spectral information in the optical and near-IR.

In the last decade the internal structure and kinematics of galaxies in the local universe have been successfully studied using IFS by the SAURON \citep{TimdeZeeuw2002}, ATLAS$^\mathrm{3D}$\citep{Cappellari2011}, DiskMass \citep{Bershady2010} and VENGA \citep{Blanc2013a} surveys, just to mention a few. The ongoing CALIFA \citep{Sanchez2012a} and SAMI \citep{Croom2012} surveys, planning to observe respectively 600 and 3400 galaxies, are currently extending IFS data to much more representative samples of nearby galaxies of all morphologies. Moreover new instruments with IFS capabilities are now available on large telescopes (for example KMOS (\citealt{Sharples2006}) and MUSE (\citealt{Bacon2010})) and future large ground-based and space telescopes (e.g., E-ELT, JWST) are planning to provide IFS capability for the study of the high-redshift Universe. In this framework it is clear that a large IFS survey of the local Universe is needed to provide a suitable benchmark for future high-z observations.

Intending to provide the next step forward towards a statistically representative IFS survey of the nearby Universe, one of the three major programs of the fourth generation SDSS (SDSS-IV) is dedicated to a large IFS survey of local galaxies called MaNGA (Mapping Nearby Galaxies at APO). MaNGA plans to observe 10000 galaxies with average redshift $<z>=0.03$, systematically covering galaxies out to $1.5$ and $2.5$ effective optical radii ($R_e$) in the wavelength range $\mathrm{3600 \AA <  \lambda < 10300 \AA }$. A detailed survey overview, describing MaNGA's key science questions, is presented in \cite{Bundy2015}. The MaNGA survey started on 1st July 2014, and the program will run for 6 years, utilising half of the dark time available in SDSS-IV.  Details on the instrumentation including the design, testing, and assembly of the IFUs are given in \cite{Drory2015}. Wake et al.~(in preparation) presents the sample design, optimisation, and final selection of the survey.  The software and data framework as well as the reduction pipeline is described in Law et al.~(in preparation).  A description of the commissioning, the quality of survey observations and further operational details will be given in Yan et al.~(in preparation).

In this work we exploit observations of 14 galaxies obtained with a prototype of the MaNGA instrument (P-MaNGA), which were secured through a generous donation of observing time by the SDSS-III Collaboration \citep{Eisenstein2011}. This data and the P-MaNGA setup is described in more detail in \cite{Bundy2015}. In this work we focus on two broad science areas: the study of the spatially resolved properties of the ionised gas and the study of spatially resolved gas chemical abundances.
 
The study of gas ionisation and chemical abundances using IFS has benefited from the efforts of previous surveys \citep{Sarzi2006, Sarzi2010, Sharp2010, Alonso-Herrero2010, Arribas2014}, which were, however, limited by small sample size and/or narrow wavelength coverage. Recent publications from the CALIFA and SAMI collaborations \citep{Sanchez2012b, Kehrig2012, Fogarty2012, Sanchez2013,Papaderos2013, Singh2013, Sanchez2014} have considerably extended on previous work and demonstrated the power of IFS in disentangling different sources of gas ionisation and probing chemical abundances of star-forming regions at different galactocentric radii. The MaNGA survey, with its extended and continuous wavelength coverage and large galaxy sample, will provide the necessary step forward to understand how the trends studied in previous work behave statistically in the overall galaxy population and vary as a function of environment and other galactic properties. The work presented here is intended as a demonstration of the insights that the MaNGA survey might deliver in this area.

More specifically, in this work we present resolved classical Baldwin-Phillips-Terlevich (BPT, \citealt{Baldwin1981, Veilleux1987}) diagrams for the 14 emission-line galaxies in our sample. Four of these galaxies are classified as Seyfert/LINER based on Sloan single-fibre spectra, and 10 as star formation dominated. We further investigate the relation between the gas excitation conditions and properties of the underlying stellar population. Finally, we investigate the relation between spatially resolved metallicity and star formation rate as well as the spatially resolved distribution of the nitrogen-to-oxygen abundance ratio. These tracers, we argue, might provide insights into the star formation history and gas flows in galaxies

This paper is structured as follows. In Section \ref{obs} we describe the P-MaNGA observations of our galaxy sample and the reduction steps leading to the science-level datacubes. In Section \ref{dat_an} we describe the subsequent analysis performed on the datacubes to derive emission line fluxes. In Sections \ref{results} and \ref{results2} we describe our results on ionised gas properties and chemical abundances. Finally in Section \ref{conc} we present the overall conclusions.

\section{Observations and Data Reduction}
\label{obs}

\begin{table}
\caption{Summary of galaxy fields observed during the MaNGA prototype observations (P-MaNGA, January 2013).}
\centering
\begin{tabular}{ c c c c c c }

\hline 
Field& RA & Dec & $\mathrm{t_{exp}}$ & Seeing & Airmass \\ %& Date\\
    & deg & deg & hr & arcsec& \\
\hline 

9	& 143.74001	& 21.788594   & 3.0 & 1.7 & 1.0 \\ %& 20-21/01/12\\
11	& 207.87862	& 14.175544   & 1.0 & 2.0 & 1.4-1.9 \\ %& 18/01/12\\
4	& 163.98026  	& 36.944852   & 2.0  & 1.3  & 1.0 \\ %& 12-17/01/12\\
\hline

\end{tabular}
\label{table_data}
\end{table}

Observations were carried out using the P-MaNGA instrument in January 2013. P-MaNGA was designed to explore a variety of instrument design options and observing strategies, hence P-MaNGA data differ substantially from the MaNGA survey data in several ways. The P-MaNGA instrument consisted of a set of 8 hexagonal fibre bundles, deployed across a field of view of 3 degrees in diameter on the Sloan 2.5 m telescope at Apache Point Observatory. Individual science Integral Field Units (IFUs) consisted of 19 fibres (12\farcs5 diameter, 5 per field), 61 fibres (22\farcs5 diameter, 1 per field) and 127 fibres (32\farcs5 diameter, 2 per field), feeding one of the two BOSS spectrographs \citep{Smee2013}. This IFU size distribution differs from that adopted by the MaNGA survey, and hence the masses and sizes of the P-MaNGA galaxies are not representative of the final MaNGA sample. While some of the P-MaNGA targets were drawn from an earlier version of the MaNGA sample selection (see target column in Table \ref{table4}), some targets were chosen for specific testing tasks. In particular three galaxies were chosen to provide overlap with the CALIFA DR1 galaxy sample \citep{Husemann2013}.

P-MaNGA observations were obtained of three galaxy fields (labelled 9, 4 and 11, see summary in Table \ref{table_data}). In each case, observations were obtained in sets of three 20-minute exposures, dithered by roughly a fibre radius along the vertices of an equilateral triangle to provide uniform coverage across each IFU. These three fields were observed to varying depths, and in varying conditions, as required by the P-MaNGA engineering tasks that they were designed for. Field 9 was observed to a depth comparable to what will be regularly achieved during MaNGA operations, but both Field 11 and Field 4 are significantly shallower than MaNGA survey data. Moreover Field 11 was intentionally observed at high airmass, resulting in particularly poor image quality. For each field one of the 19-fibre IFU was targeted at a region of blank sky and a second one at a standard star, leading to a final sample of 18 galaxies. Of these, 4 galaxies will not be discussed in this work, since their spectra do not show significant emission lines. Table \ref{table4} provides an overview of the 14 galaxies observed by P-MaNGA which will be discussed in this work.

The raw data was reduced using a prototype of the MaNGA Data Reduction Pipeline (DRP), which is described in detail by Law et al. (in prep). In brief, individual fibre spectra were extracted using a row-by-row algorithm, wavelength calibrated using a series of Neon-Mercury-Cadmium arc lines, and flatfielded using internal quartz calibration lamps.  Sky-subtraction of the IFU fibre spectra was performed by constructing a cubic basis spline model of the sky background flux as seen by the 41 individual fibres placed on blank regions of sky, and subtracting off the resulting composite spectrum shifted to the native wavelength solution of each IFU fibre.  Sky subtraction in the P-MaNGA data redward of $\sim$8000 \AA\ suffered from a coma aberration in the red channel of the BOSS spectrograph used. This issue has now been fixed and will not affect MaNGA survey data, which is expected to reach Poisson-limited sky subtraction at all wavelengths (see \citealt{Drory2015}, Law et al. in prep).

Flux calibration of the P-MaNGA data was performed by fitting Kurucz model stellar spectra to the spectra of calibration standard stars covered with single fibres at each of the three dither positions. The flux calibration vectors derived from these single-fibre spectra were found to vary by $\sim$10\% from exposure to exposure, depending on the amount of light lost from the fibre due to atmospheric seeing and astrometric misalignments.  While this uncertainty is acceptable for the present science purposes, the flux calibration uncertainty of the single fibres ultimately drove the design decision of the MaNGA survey to instead use 7-fibre IFU `mini-bundles' for each calibration standard star, which result in a photometric accuracy of $2.5 \%$  between $\mathrm{H \alpha}$ and  $\mathrm{H \beta}$ and of $7 \%$ between [NII] $\mathrm{ \lambda 6584}$ and [OII] $\mathrm{ \lambda \lambda 3726, 29 }$ for MaNGA survey data (see Yan et al., in prep.).

Flux calibrated spectra from the blue and red cameras were combined together across the dichroic break using an inverse-variance weighted basis spline function.  Astrometric solutions were derived for each individual fibre spectrum that incorporate information about the fibre location within an IFU, dithering, and atmospheric chromatic differential refraction, among other effects. Fibre spectra from all exposures for a given galaxy were then combined together into a single datacube (and corresponding error vectors) using these astrometric solutions and a nearest-neighbour sampling algorithm similar to that used by the CALIFA survey.  For the P-MaNGA datacubes, a spaxel size of 0\farcs5 was chosen. The typical effective spatial resolution in the reconstructed datacubes can be described by a Gaussian with ${\rm FWHM}\approx2$\farcs5.

As a check on our spectrophotometry, a spectrum was extracted from the P-MaNGA datacubes in a $3"$ aperture at the position of the SDSS-I Legacy fibre spectrum. We find that P-MaNGA spectrophotometry agrees with that of Sloan spectra within better than 10 \% over the whole wavelength range of the Sloan spectra for Plates 4 and 9. In the case of Plate 11, taken at high airmass, a distortion in the spectral shape of the P-MaNGA data is observed blueward of 4500 \AA\ with respect to the Sloan single fibre spectra. The amplitude and shape of the distortion is approximately constant over all the galaxies on this plate. Given the poorer spectrophotometric accuracy of P-MaNGA Plate 11 observations with respect to the legacy Sloan spectra, the P-MaNGA spectral shape was anchored to that of the Sloan spectrum blueward of 4500 \AA. This procedure leads to an overall spectrophotometric accuracy of 15\% for Plate 11 in the blue with respect to the legacy Sloan spectra.
Absolute flux calibration of the P-MaNGA data was anchored to that of Sloan photometry in the r-band.

For more details on the P-MaNGA hardware and observations, see \cite{Bundy2015}, sections 4.7 and 7.1.

\begin{table*}
\caption{Galaxy sample. Data from SDSS DR7 \protect\cite{Abazajian2009}. Stellar masses are from the MPA-JHU catalogue, based on \protect\cite{Kauffmann2003}. LTG=late type galaxy, IRR=irregular.}

\begin{tabular}{ l  c c c r c r c c c c c}

\hline 
\hline 
Galaxy ID	 & RA 			& 	Dec 		 &  z	& $\mathrm{\log (M_\star/M_{\odot})}$	& g-i 		& $\mathrm{R_{e}}$ 		& target 			& $\mathrm{R_{IFU} }$  	& Comments\\
	       &J2000 deg	&J2000 deg  	&  	& 							& mag  	&arcsec 				& type$^{\dagger}$ 	& $\mathrm{R_{e}}$  	& 		\\
\hline

\hline
\multicolumn{10}{c}{Field 9: 3.0 hr, seeing 1\farcs7} \\
\hline

P9-127A 	 & 143.74000 		& +21.705262 	& 0.013 	& 10.7  	& 0.70 	& 23.7 	& m 		& 0.7   	& NCG 2916; LTG, face-on  \\		%spiral; CALIFA \\
P9-127B &  	 143.77632 		& +21.627660 	& 0.013 	& 9.1 	&  0.51 	& 6.8 	& 1 			& 2.4  	& CGCG122-022; LTG, edge-on \\		%HI (HR89)\\
P9-61A & 		 144.29931 		& +21.669221 	& 0.019 	& 10.1 	&  1.3 	& 9.3 	& m		& 1.2   	&  UGC 5124; LTG, edge-on\\  		%bright radio source (AGN?)\\ 
%p9-19B & 		ma005 & 142.790		& +22.747 	& 0.051 	& 10.6 	& 0.86 	& 4.0 	& 1 			& 1.6   	& ETG  \\
%p9-19D &	ma007 & 142.788 		& +20.917 	& 0.034 	& 10.3 	& 0.79 	& 3.2 	& 1 			& 2.0   	& ETG\\
P9-19E & 145.12595 		& +21.253809 	& 0.024 	& 9.7 	&  0.73 	& 2.9 	& 1 			& 2.2  	& IRR \\

\hline
\multicolumn{10}{c}{Field 11: 1.0 hr, seeing 2\farcs0, (airmass 1\farcs5)} \\
\hline

P11-127A &		207.87863 		& +14.092194		& 0.024 	& 10.9 	&  0.85 	& 16.95 	& m 		& 1.0    	& IC 0944; LTG, edge-on \\		%CALIFA; dust \\
P11-127B &	 	209.23090 		& +14.142252 	& 0.016 	& 9.4 	&  0.42 	& 9.61 	& m 		&1.7  	& KUG 1354+143; LTG, face-on\\		%ALFALFA HI\\
P11-61A &			208.04816		& +13.999926 	& 0.024 	& 10.1 	& 0.77 	& 7.32 	& 1 			& 1.5   	& LTG \\
%p11-19A &		ma004& 207.581 & +14.141 & 0.024 & 9.5  & 0.85 & 1.12 & $\star$ & 2.8    & Edge-on disc (too small) \\
P11-19B &			207.29175	& +13.347008 	& 0.024 	& 9.2  	& 0.42 	& 2.37 	& 1 			& 2.6    	& LTG, face-on \\
P11-19C &		206.70605 	& +14.400476 	& 0.021 	& 10.0  	& 0.54 	& 2.82 	& m 		& 2.2    	& LTG, strong bar \\

\hline
\multicolumn{10}{c}{Field 4: 2.0 hr, seeing 1\farcs3} \\
\hline

P4-127A		&  163.98025 	& +36.861503 	& 0.022 	& 10.7  	& 0.84 	& 10.3 	& m  		&1.6    & UGC 6036; LTG, edge-on,  \\ 
%& & & & & & & & & &  & & CALIFA; dust \\
P4-127B &	163.24608 	& +37.613401 	& 0.042 	& 11.0 	&  0.77 	& 13.9 	&m 		& 1.2  	& LTG, face-on\\
P4-61A &		164.44418 	& +36.282683 	& 0.030 	& 9.7  	& 0.71 	& 3.4 	& m 		& 3.4   	& IRR, red core \\
P4-19A &		165.05043 	& +36.387252 	& 0.027 	& 9.4 	& 0.49 	& 2.5 	&  1 		& 2.5   	& LTG, edge-on\\
P4-19B &		162.49460 	& +36.415032 	& 0.023 	& 9.5 	& 0.53 	& 4.6 	& 1 		& 1.4    	& LTG, edge-on\\
%p4-19C &		ma006	& 164.024 	& +36.960 	& 0.022 	& 9.5 	& 0.72 	& 4.8 	&1 			& 1.3    	& ETG\\

\hline
\end{tabular}

\begin{flushleft}
\small
\textbf{Notes:} 
$^{\dagger}$ Target type 1 indicates the galaxy would be selected in the MaNGA Survey's Primary sample. ``m'' indicates a galaxy that was chosen manually for the prototype run. $\mathrm{R_{IFU}}$ indicates the radial coverage of the allocated bundle in units of $\mathrm{R_{e}}$. $\rm R_e$ values are taken from an extended version of the NASA-Sloan Atlas (NSA, http://www.nsatlas.org, \citealt{Blanton2011}).
\end{flushleft}

\label{table4}
%%\end{minipage}
\end{table*}

\section{Data Analysis}
\label{dat_an}

The final products of the Data Reduction Pipeline are sky subtracted, wavelength calibrated and interpolated data and error cubes. In this work we extract emission line fluxes by first fitting simple stellar population (SSP) models to the spectra and then fitting the emission lines with Gaussian functions. We expand and adapt a suite of routines from the preliminary MaNGA analysis pipeline (see Wilkinson et al., submitted, for further details on the MaNGA analysis pipeline).

The detailed sequence of steps implemented to extract the emission line fluxes in each P-MaNGA galaxy is described in the rest of this section.
\begin{enumerate}
\item{For each spaxel in the datacube, the signal and noise per pixel are computed on the r-band continuum. Spaxels with S/N lower than a specified threshold (here $\mathrm{S/N = 4}$) are discarded.}

\item{The remaining spaxels are binned using a centroidal Voronoi binning algorithm \citep{Cappellari2003} to reach a target r-band S/N of 30\footnote{From now on we call `Voronoi bins' all the bins that have been created by this procedure, even when they consist of single spaxels, as is often the case in centres of galaxies or other regions of high S/N.}. The noise from different spaxels is coadded in quadrature, neglecting the fact that errors in different spaxels are correlated because of the interpolation procedure used to create the datacube.}

\begin{figure*}
\includegraphics[width=0.8\textwidth, trim=100 100 120 100, clip]{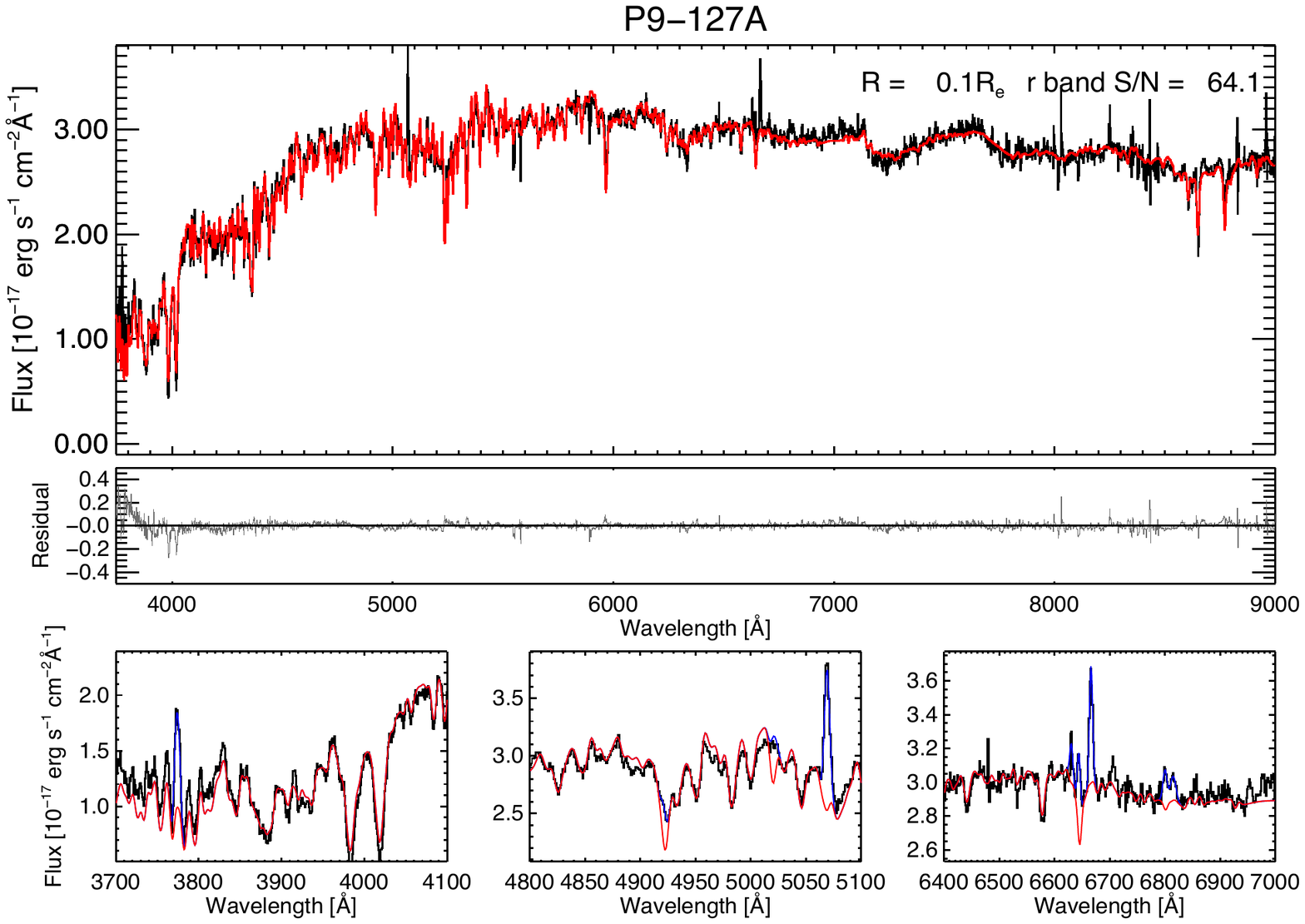} 
\includegraphics[width=0.8\textwidth, trim=100 100 120 100, clip]{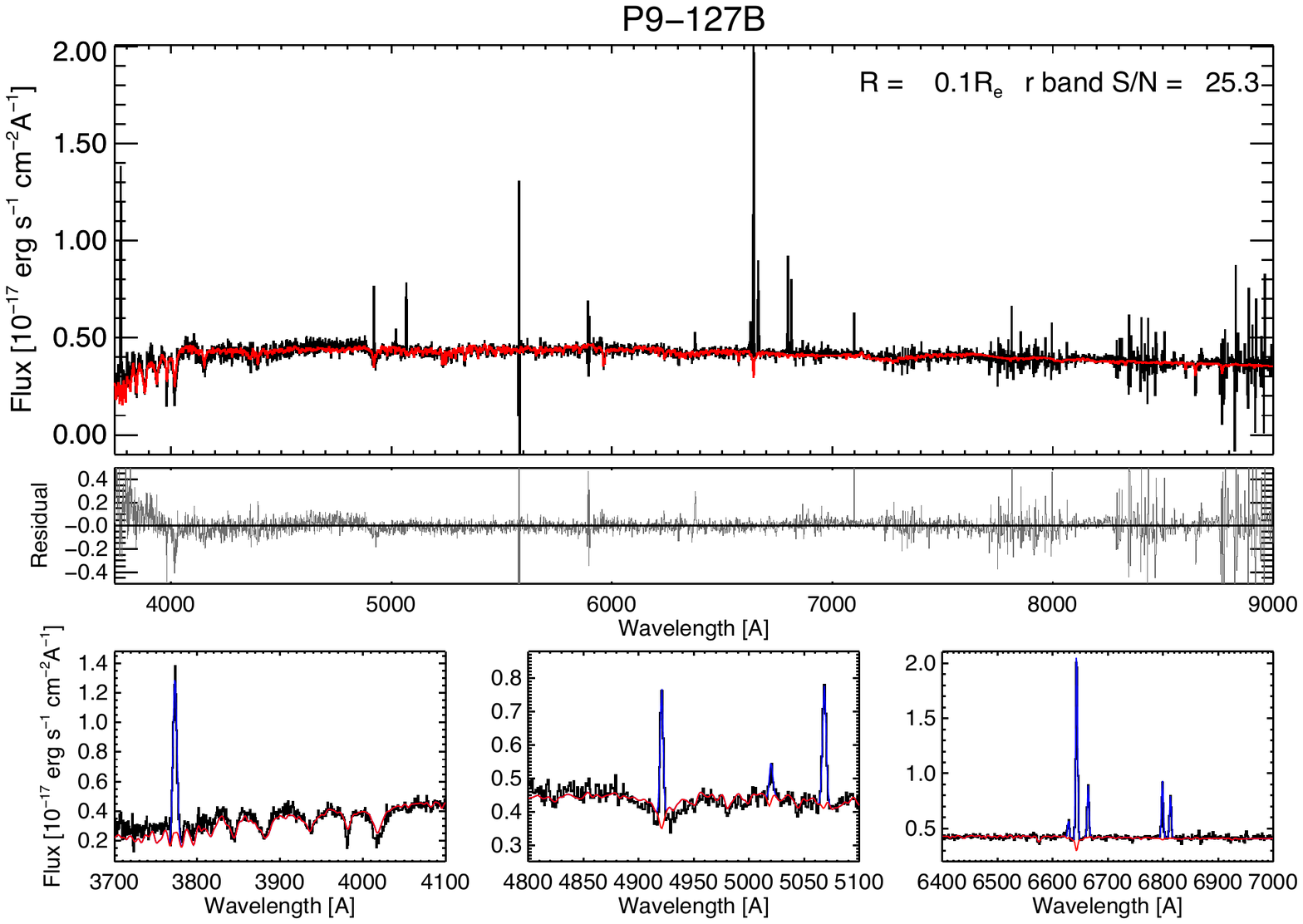}

\caption{Spectral fitting of two spectra, corresponding to Voronoi bins in the central regions of two galaxies: P9-127B and P9-127A. For each galaxy, the top panel shows the spectrum (black) with the SSP fitting (red). Fit residuals are shown in the middle panel. The smaller panels at the bottom show the details of the fit in three narrow wavelength ranges: around 4000 \AA \ (note the prominent [OII] $\rm \lambda 3727$ unresolved doublet),  5000 \AA \ (with prominent $\rm H \beta$ and $\rm [OIII] \lambda \lambda 4945, 5007 $ lines)  \ and 6700 \AA \ (with prominent $\rm H \alpha$, $\rm [NII] \lambda \lambda 6548, 6584$ and [SII]$\lambda\lambda$6717,6731 lines). The observed spectrum is in black, the SSP fit in red and the emission line fit (summed to the SSP fit) in blue. The galactocentric distance in units of $\rm R_e$ and the S/N in the r-band (from fit residuals) is shown in the top right corner.}
\label{spectra}
\end{figure*}

\item{In each bin, a linear combination of SSP templates is fitted to the spectrum, using Penalised Pixel Fitting \citep{Cappellari2004}, after correcting for systemic velocity (as estimated from the redshift) and instrumental dispersion (taken as $\rm 2.73 \ \AA$ FWHM). For the purposes of this work we use a set of 29 models generated using \cite{Maraston2011} templates based on the STELIB empirical stellar library \citep{LeBorgne2003}, chosen to represent a wide range in age and stellar metallicity (ages from 0.2 to 15 Gyr and three metallicities: $\rm [Z/H] = -2.25, 0.00, 0.35$).} A region of $600 \ \mathrm{ km \ s^{-1}} $ is masked around each fitted emission line and strong sky lines. Additive and multiplicative polynomials are also used to take into account residual imperfections in the spectrophotometry, stellar exctinction and possible contribution to the continuum emission from non-stellar (e.g. AGN) sources. In this work we allowed the fit to use polynomials of up to fourth degree.

\item{In each bin, the stellar population fit is subtracted from the observed spectrum to obtain a pure emission line spectrum. This is then fitted with a set of Gaussian functions, one per emission line. The initial guesses for the emission lines velocities and velocity dispersion are set to be equal to the ones derived from the SSP fitting of the stellar continuum. Multiplets are set to have the same velocity and velocity dispersion and the ratio of intensities is fixed to the theoretical one (here only in the case of $\mathrm{[OIII] \lambda \lambda 4945, 5007}$ and $\mathrm{[NII] \lambda \lambda 6548, 6584}$), while for the other lines no constraints are set. Figure \ref{spectra} shows spectra from the nuclear regions of two different galaxies (P9-127B and P9-127A), fitted using this procedure. We calculate emission line fluxes by integrating the flux under the fitted Gaussian. Errors are calculated for each emission line flux.}

\item{The reddening (for the gas component) is calculated from the Balmer decrement, for those spaxels where we detect both $\rm H \alpha$ and $\rm H \beta$ with S/N $>$ 3, and a \cite{Calzetti2001a} reddening curve with $\rm R_V = 4.05$. The theoretical value for the Balmer line ratio is taken from \cite{Osterbrock2006}, assuming case B recombination ($ \mathrm{H \alpha  / H \beta=2.87 }$). We note that the use of extinction curve of \cite{Cardelli1989} (or the modification by \citealt{O'Donnell1994}) with $\rm R_V=3.1$ yields very similar results for the 3000 $\rm \AA$ to 7000 $\rm \AA$ wavelength range considered in this work.}

\end{enumerate}

\begin{figure}
\includegraphics[width=0.5\textwidth, trim=90 100 120 100, clip]{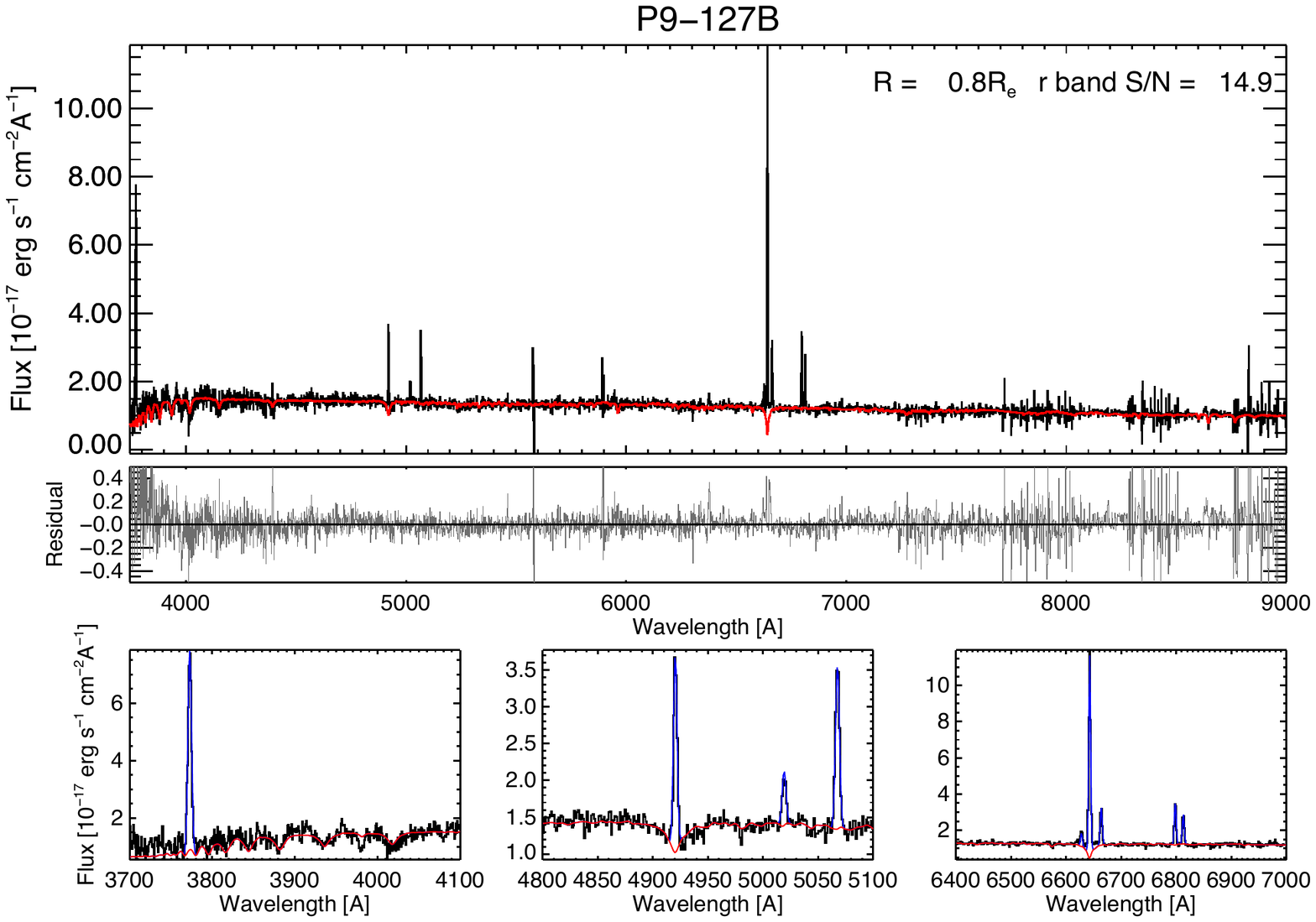} 
\includegraphics[width=0.5\textwidth, trim=100 100 120 100, clip]{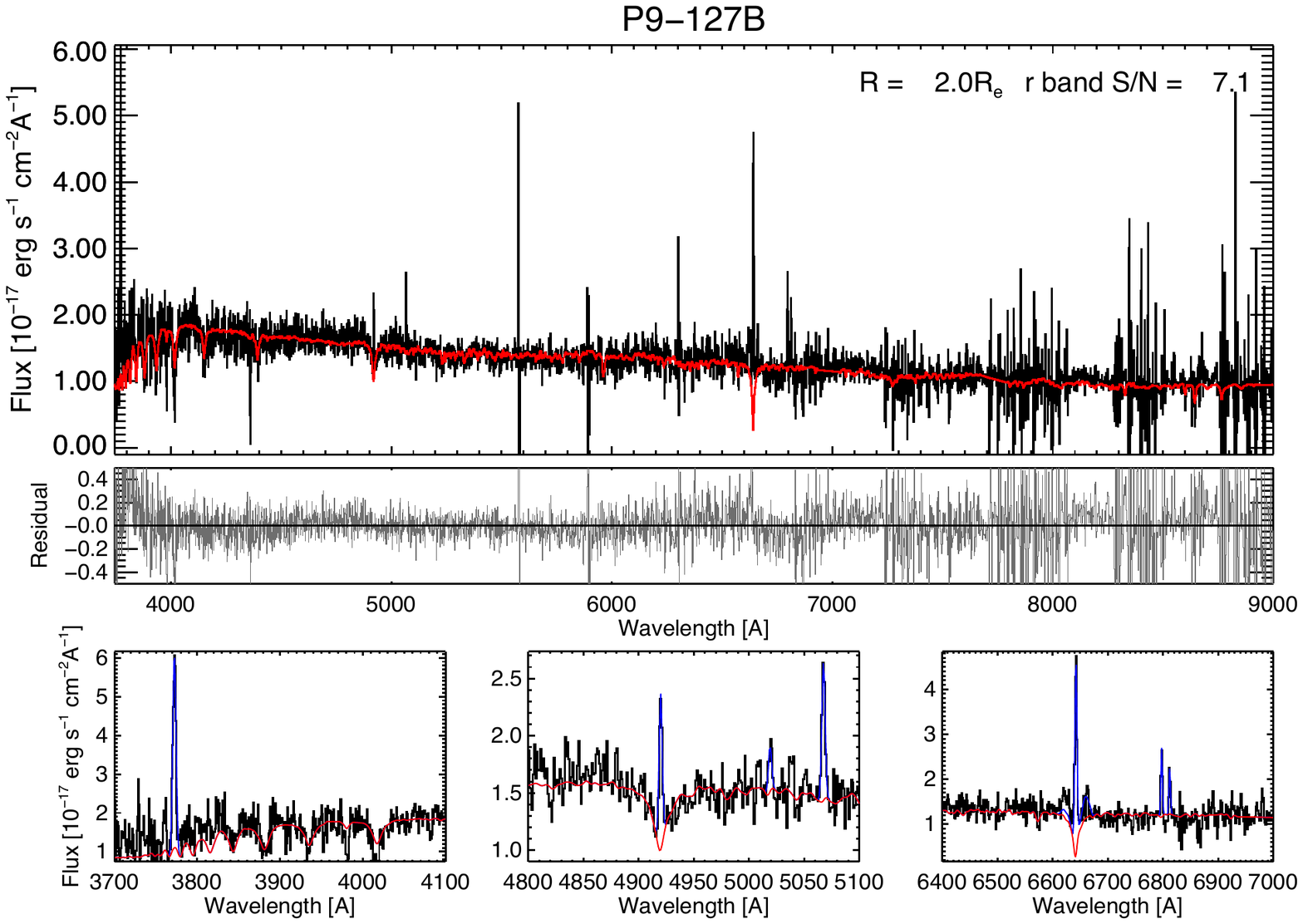} 

\caption{Spectral fitting of two off-nuclear spectra, corresponding to Voronoi bins in the outskirts of P9-127B. Galactocentric distance and S/N in the r-band are shown for each spectrum in the top right corner (see caption in Fig. \ref{spectra} for full details).}
\label{spectra2}

\end{figure}

To test the reliability of the emission line fluxes, the fitting procedure detailed above was repeated using the \cite{Bruzual2003} templates (BC03, used in previous work with SDSS data, for example: \citealt{Kauffmann2003, Tremonti2004, Brinchmann2004}). Both the fit residuals and the emission line fluxes obtained with the M11 STELIB and BC03 templates were found to be in good statistical agreement. The reduced $\chi^2$ for the continuum fit favours the M11 STELIB models. We will therefore present results obtained using the M11 STELIB templates, although the use of BC03 templates does not significantly affect any of our conclusions.

To evaluate the effect of neglecting noise correlation between different spaxels when building the Voronoi bins, we estimate of the noise `a posteriori' using the standard deviation of the fit residuals in the r-band. While this noise estimate will be affected by template mismatch, it should represent an upper limit on the true noise vector. We observe that the S/N calculated using the fit residuals is up to a factor of 2-3 lower than that estimated from the error vectors associated with each bin, and follows a functional form similar to that found in CALIFA data \citep{Husemann2013, Garcia-Benito2014}. The error vectors were hence rescaled to match the fit residuals (S/N shown in Fig. \ref{spectra} and \ref{spectra2} refers to the fit residuals). This approach differs from the one recently implemented by the SAMI collaboration \citep{Sharp2014}, who deliver an estimate of the covariance matrix generated by their cube-reconstruction algorithm. To date, no work has attempted to compare these different approaches and estimate the effect that the choice of method has on the final error vectors. This comparison, although important to validate the soundness of our error analysis and timely in the era of large IFS surveys, goes beyond the scope of this prototype data paper and will be presented in future publications from the MaNGA collaboration.

In Fig. \ref{spectra2} we show two off-nuclear spectra from P9-127B, the galaxy in our sample that most closely resembles MaNGA survey quality data in terms of exposure time, airmass and sample selection. Two spectra are shown, belonging to bins at deprojected galactocentric radii of $\rm 0.8\ R_e$ and $\rm 2 \ R_e$. In both of these spectra all strong emission lines ($\rm [OII]  \lambda 3727, [OIII] \lambda \lambda 4945, 5007$, $\rm H \beta$, $\rm H \alpha$, $\rm [NII] \lambda \lambda 6548, 6584$ and $\rm [SII]\lambda\lambda 6717,6731$) are detected above the noise level even at such large galactocentric distances.

Maps of  $\mathrm{H \alpha}$ flux, velocity and velocity dispersion for all the galaxies are presented in the Appendix.

\section{Gas Ionisation Diagnostics}
\label{results}

\subsection{The BPT Diagram}
\label{intro_BPT}

The BPT diagram is a powerful tool to identify the dominant mechanism of gas ionisation using ratios of strong optical lines \citep{Baldwin1981, Veilleux1987}. The most popular version of this diagnostic diagram makes use of the $\mathrm{ [OIII] \lambda 5007/H \beta}$ and $\mathrm{ [NII] \lambda 6584/H \alpha} $ line ratios ([NII] BPT). Other commonly used ionisation diagnostic diagrams substitute the $\mathrm{ [NII] \lambda 6584/H \alpha}$ with the $\mathrm{ [SII] \lambda \lambda 6717, 31/H \alpha}$ ([SII] BPT) or $\mathrm{ [OI] \lambda 6300/H \alpha}$ line ratios. 

\cite{Kewley2001} (K01) used population synthesis models together with photoionisation models to construct a division line, arguing that points above the line cannot be reproduced using star forming models alone. With the advent of the SDSS survey it became apparent that galaxies tend to lie in two well-defined sequences in the BPT diagram, leading to the characteristic `sea gull' shape. The left-wing sequence is associated with star-forming galaxies, while the right-wing sequence is associated with other ionisation mechanisms, including photoionisation by AGN (Type 2 Seyfert nuclei), Low Ionisation Nuclear Emission-line Regions (LINER) and shocks. 

\cite{Kauffmann2003a} (K03) proposed a modification of the K01 line to better match the observational data derived from more than $10^5$ galaxies from SDSS. More recently \cite{Stasinska2006} studied the demarcation lines of K03 and K01 using composite AGN+SF photoionisation models and concluded that the K03 demarcation line allows for contamination from an AGN to the $\rm H \beta$ flux of $< 3 \%$ while the K01 line allows for a contamination up to roughly $20 \%$. Qualitatively similar conclusions are obtained by \cite{Schawinski2010}, who argue that for an AGN sample with L[OIII] $ \rm = 10^{40} \ erg \ s^{-1}$, about $13 \%$ of the objects would appear as left-wing sources if observed through the $3''$ Sloan fibre, based on the K03 delimiter.

The right-wing of the BPT diagram is known to show a bimodality, and is generally divided into an upper branch (we will refer to this as the Seyfert branch, although we only wish to state that these objects have Seyfert-like line ratios) and a lower branch (the LINER branch, populated by object with LINER-like line ratios). \cite{Kewley2006} studied this bimodality from an empirical perspective and concluded that the division between the two classes is most evident by using the $\mathrm{ [OI]/H \alpha}$ and $\mathrm{ [SII]/H \alpha}$ line ratios. Hence in this work we will be using the demarcation in the $\mathrm{ [SII]/H \alpha}$ diagram to distinguish between the Seyfert-like and LINER-like branches.
 
The LINER-like branch, accounting for up to a third of all galaxies \citep{Ho1997, CidFernandes2010}, is generally characterised by an ionisation mechanism significantly different from that of known Seyfert nuclei. Nuclear LINERs are thought to be associated with low-luminosity accreting black holes with a very hard ionisation spectrum, and some of them are known to host an X-ray detected AGN \citep{Gonzalez-Martin2009}. However, other possible explanations have been put forward, especially to account for LINER-like emission on extended scales, including ionisation by post-asymptotic giant branch (post-AGB) stars \citep{Binette1994, Goudfrooij1999}, shocks \citep{Heckman1980, Lipari2004, Rich2011, Rich2014} and hot (possibly Wolf-Rayet) stars \citep{Terlevich1985, Barth2000}. 

Post-AGB stars represent a particularly promising explanation, since they are the natural product of stellar evolution and have been shown to produce the required LINER-like line ratios, at least within the uncertainties of modern stellar population modelling \citep{CidFernandes2010, CidFernandes2011}. Moreover it has been found that extended LINER-like sources \citep{Sarzi2010, Yan2012, Singh2013} have radial profiles of line emission surface brightness shallower than $r^{-2}$. Some authors have argued that this represents evidence against ionisation from a central point source. However, \cite{Yan2012} correctly point out that line emission surface brightness gradients are chiefly dependant on several unknown parameters (including the gas filling factor, the spatial distribution of gas clouds, and the gas density), and hence do not provide sufficient evidence for ruling out ionisation from a central point source.

LINER-like line ratios can also be produced by hardening of the ionising spectra due to photoelectric absorption and the low density of the gas above and below the disc of star-forming galaxies. This phase of the ISM is generally referred to as extraplanar Diffuse Ionised Gas (DIG). DIG is warm ($10^4$K), low density ($\mathrm{10^{-1} cm^{-3}}$) ionised gas with low ionisation parameter, which is known to surround the plane of the Milky Way and other spiral galaxies. Its ionisation state is generally compatible with the ionising flux from stellar sources (O and B stars, but possibly also hot evolved stars), hardened by photoelectric absorption \citep{Hoopes2003, Rossa2003, Oey2007, Haffner2009, Blanc2009, Flores-Fajardo2011}. Extensive work on edge-on galaxies (with NGC 891 as a prototypical case, see for example \citealt{Rand1990, Rossa2004, Bregman2013}) has shown that DIG emission is detectable up to several kpc from the disc and that DIG contains also dust and metal enriched material.

An extra source of confusion, specific to the [NII] BPT (but absent in the [SII] BPT), is the role of what is often called the `Composite' region, lying between the K03 and K01 lines. There is no physical reason to associate this region with ionisation from a mixture of AGN and SF, since both purely SF and bone-fide Seyfert ionised regions can be found there. It has also been suggested that an increase in the nitrogen abundance due to population ageing can shift H\textsc{ii} regions into this intermediate region \citep{Kennicutt1989, Ho1997c, Sanchez2014b}. However, it is found observationally that the distance from the star-forming sequence along the right wing of the BPT diagram is an empirically well-motivated parameter, which correlates with other galactic properties \citep{Stasinska2006, Kewley2008}. We therefore decided to keep the so-called `Composite' class as a rough proxy for sources in the right-wing characterised by small distance from the SF sequence. We will, however, not call objects in this class `Composite' but rather `Intermediate', since we will be using the term `Composite' for sources where we resolve both SF and Seyfert/LINER-like emission in different regions within the same galaxy.

\subsection{Resolved Ionisation Diagnostics: the P-MaNGA Perspective}
\label{BPT_Pmanga}

\begin{figure}
\centering
\includegraphics[width=0.5\textwidth, trim=0 80 0 130, clip]{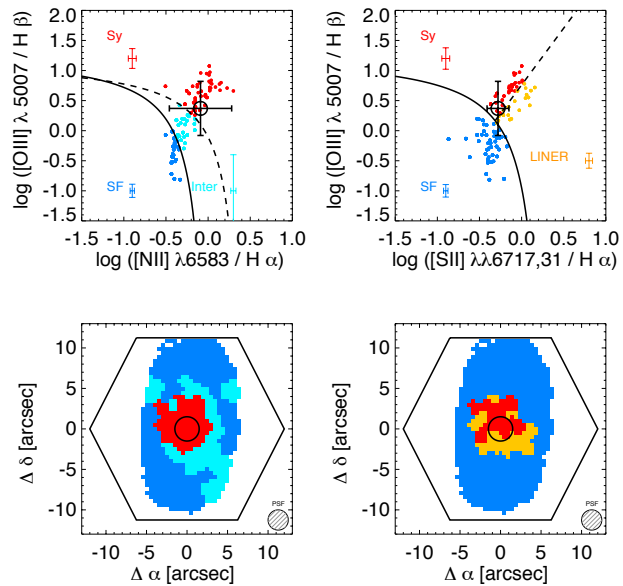} 
\caption{The ionised gas properties in P11-61A. The top panels show the BPT diagnostic diagrams using canonical strong line ratios and the demarcation lines of Kauffmann 2003 (solid line in the [NII] BPT diagram) and Kewley 2006 (dashed line in the [NII] BPT diagram), as described in the text. Points in the BPT diagrams correspond to individual Voronoi bins within the galaxy (97 in total in this galaxy) and are colour-coded depending on their position in the diagram. The open circle shows the line ratios extracted from the Sloan $3''$ fibre (using the MPA-JHU emission line fluxes). The coloured error bars show the median error in the different areas of the BPT diagram (the colour reflects the area it corresponds to). The bottom panels show, with the same colour-coding, the position of these regions on the sky. The black hexagon shows the coverage of the P-MaNGA fibre bundle, while the black circle shows the position of the Sloan $3''$ fibre. The P-MaNGA PSF is represented in the bottom right corner.}
\label{example_BPT}
\end{figure}

In this work we present results based on the classical BPT diagrams using the line ratios $\mathrm{ [OIII]/H \beta}$ vs $\mathrm{ [NII]/H \alpha} $ and $\mathrm{ [SII]/H \alpha} $. We do not use the [OI]$\mathrm{\lambda6300}$ line since it is weaker than the other lines and it would require a coarser binning to be detected above the noise level.  We discard regions where either $\mathrm{H \beta}$ or $\mathrm{[OIII]\lambda5007}$ are not detected to $\mathrm{3 \sigma }$. We do not correct the line fluxes for extinction when calculating line ratios in the BPT diagram, since the line ratios considered are nearly unaffected by the extinction correction due to their proximity in wavelength space.

Overall, our sample spans a wide variety of ionisation conditions. Seven galaxies are entirely dominated by star formation. P11-61A is the only galaxy in which we confirm the SDSS Seyfert classification. In three galaxies we detect extended LINER-like ionisation, traced by both diagnostic diagrams, extending out to 0.5-1 $\mathrm{R_{e}}$. Finally, three galaxies have a peculiar ionisation structure and we classify them as `Composite', since they present both LINER-like line ratios and SF-dominated regions, but the spatial distribution of the ionisation is more varied. We summarise the differences between the Sloan single fibre classification and the resolved P-MaNGA analysis in Table \ref{table2}. 

In the Appendix we show BPT diagrams and maps of the ionisation conditions as derived from the BPT diagrams for all the galaxies, together with line ratio maps ($\mathrm{ [NII]/H \alpha} $ and $\mathrm{[OIII]\lambda5007}$).

\subsubsection{Resolved Seyfert-like Ionisation in a Star Forming Disc}

In Fig. \ref{example_BPT} we show the [NII] and [SII] BPT diagrams for galaxy P11-61A. Points in the BPT diagram correspond to individual Voronoi bins (and are colour-coded according to the demarcation lines shown). The open circle corresponds to the line ratios obtained from the Sloan $3''$ diameter fibre (emission line fluxes from the MPA-JHU DR7 catalogue, based on \citealt{Brinchmann2004, Kauffmann2003}\footnote{The MPA-JHU catalogue is accessible at http://www.mpa-garching.mpg.de/SDSS/DR7/}). We note that in both the [NII] and [SII] BPT diagrams the galaxy is classified as a Seyfert using the Sloan spectra. 

Immediately below each BPT diagram we show how regions with different ionisation conditions, according to the BPT diagrams, are spatially distributed in the galaxy, by using the same colour-coding. The black hexagon corresponds to the size on the sky of the P-MaNGA bundle. 

In the [NII] BPT we observe Seyfert-like ionisation (red) in the central regions, with a gradual transition to star-forming environments (blue) as we move further out into the disc. There is also evidence for LINER-like line ratios (orange), mostly in the transition region between Seyfert-dominated and SF-dominated regions. If we assume that ionisation is due mainly to a point source in the centre, then the observed Seyfert $\rightarrow$ LINER transition could be explained in terms of a decrease in the ionisation parameter because of geometrical dilution of the ionising radiation ($\rm f_{ion} \propto r^{-2}$) and/or hardening of the photoionising radiation caused by photo-electric absorption. We also recall that, despite the fact that the ionising radiation of a nuclear source decreases as $\rm f_{ion} \propto r^{-2}$, the line emissivity of the resulting nebular line can easily depart from the $\rm r^{-2}$ profile because of the clumpy gas density distribution, changes in the covering factor, self shading and projection effects. We therefore wish to stress that within individual objects there is no reason to interpret the Seyfert/LINER demarcation as a clear-cut distinction between different ionisation mechanisms. \textit{The Seyfert/LINER demarcation line is an intrinsically porous demarcation}.

Overall, this observation shows the potential of MaNGA in spatially resolving star formation from the AGN-dominated nuclear regions in active galaxies, opening up new possibilities to study AGN host galaxies and their star formation properties.

\begin{figure}
\centering
\includegraphics[width=0.43\textwidth, trim=100 230 150 90, clip]{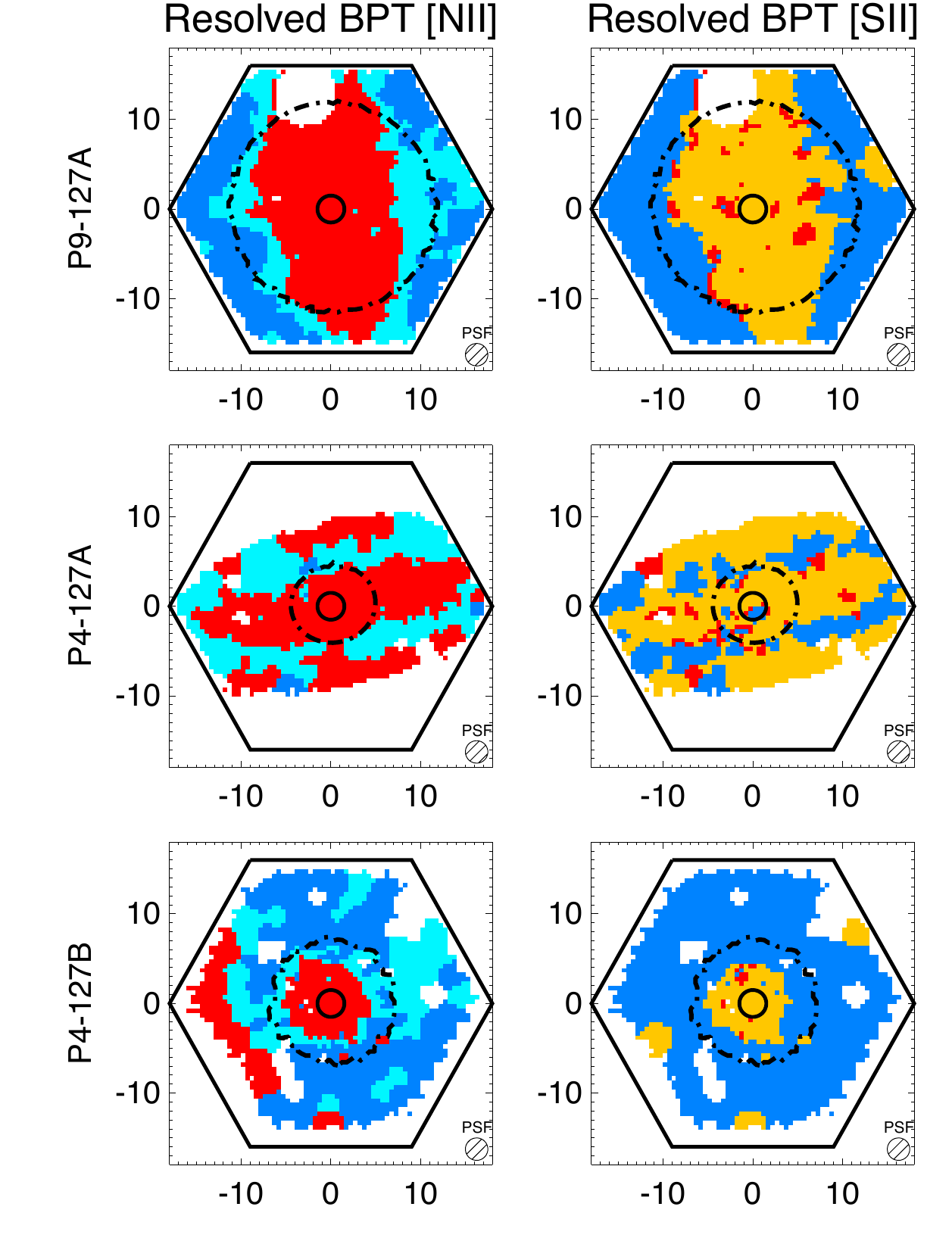}
\includegraphics[width=0.43\textwidth, trim=100 230 150 105, clip]{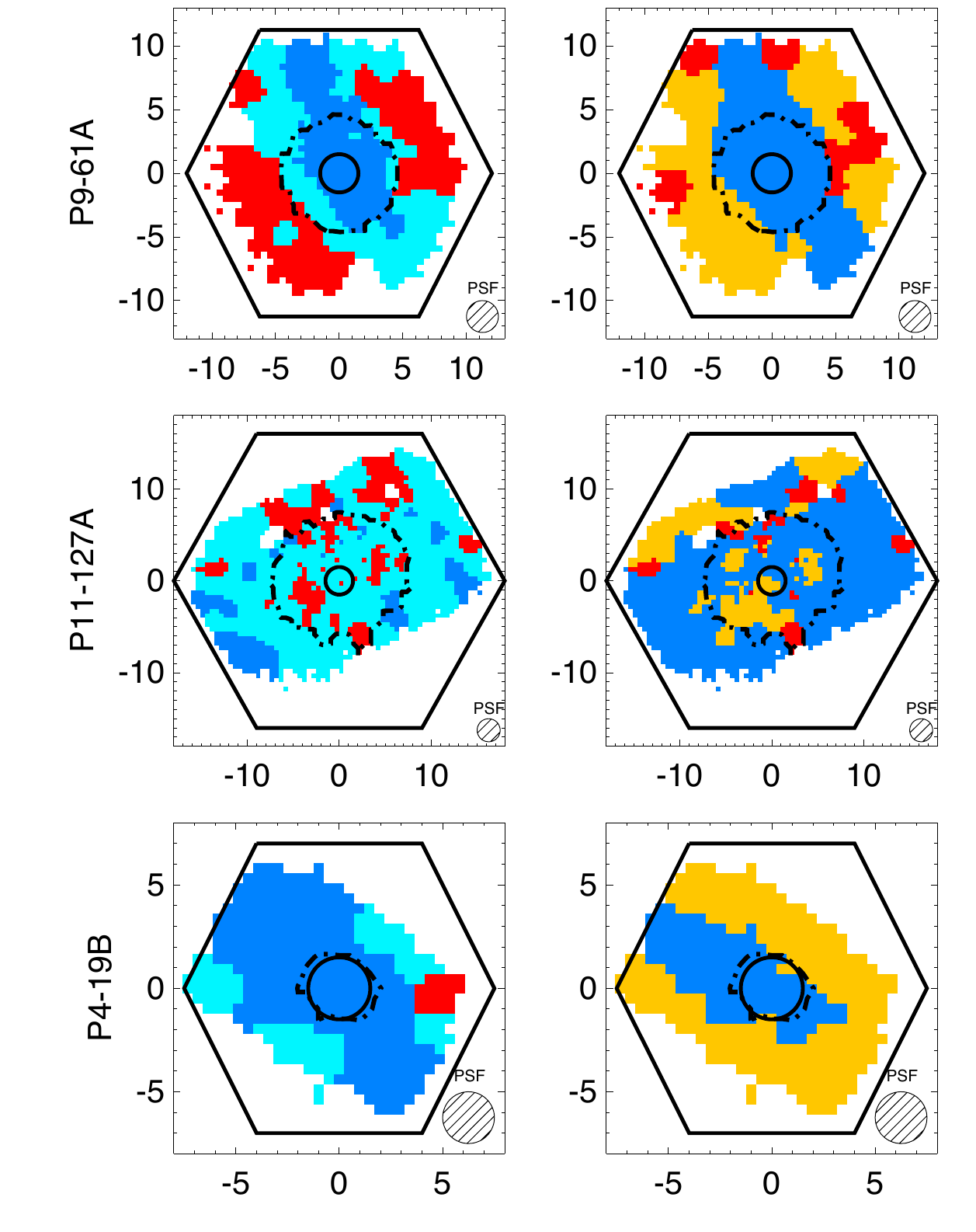}
\caption{BPT classification of P-MaNGA galaxies (purely SF galaxies are not shown). The colour-coding reflects the position of each region in the [NII] (left) and [SII] (right) BPT diagrams. The colour-coding is the same as in Fig. \ref{example_BPT}. In the [NII] BPT blue corresponds to SF regions, cyan to intermediate and red to Seyfert/LINER-like. In the [SII] BPT blue corresponds to SF, orange to LINER-like and red to Seyfert-like. The dot-dashed black contour encloses all the bins with a galactocentric distance smaller than $\rm 0.5 R_e$. The black hexagon corresponds to the P-MaNGA IFU, while the black open circle corresponds to the position and size of the Sloan $3''$ fibre. The PSF of the P-MaNGA data is represented in the bottom right corner. The spatial scale on the maps is in arcsec.}
\label{example_BPT_seyfert}
\end{figure}

\begin{table}
\caption{Properties of the ionised gas in our galaxy sample, using Sloan spectroscopy and P-MaNGA data.}

\begin{tabular}{ l c c c }

\hline 
\hline 
Galaxy   & Sloan class & P-MaNGA& P-MaNGA \\
		&    		& (nucleus) & (disc/outskirts) \\
\hline

\hline 
\multicolumn{4}{c}{SF-dominated galaxies} \\
\hline

P9-127B				&SF		&SF		&SF 		\\
P9-19E				& SF		& SF    	&SF 		\\
P11-127B				& SF		& SF    	&SF 		\\
P11-19B				&SF 		& SF 	& SF		\\
P11-19C				&SF/Int	& SF	/Int	& SF/Int		\\
P4-61A 				& SF		& SF    	&SF (Int?) 		\\
P4-19A				& SF		& SF    	&SF 		\\

\hline 
\multicolumn{4}{c}{Seyfert galaxy} \\
\hline

P11-61A				&Sy		& Sy/LINER   	&SF 					\\

\hline 
\multicolumn{4}{c}{LINER-like galaxies} \\
\hline

P9-127A				& Sy 	& LINER 		&SF 					\\
P4-127A		 		&	 	& LINER		&LINER (some SF) 		\\
P4-127B				&Sy 		&LINER		&SF					\\

\hline
\multicolumn{4}{c}{Composites} \\
\hline

P9-61A				& SF		& SF 		& LINER  	\\
P11-127A				& SF		& SF/LINER	& SF			\\
P4-19B				& SF		& SF 		&LINER		\\

\hline

\end{tabular}

\small
\textbf{Notes:} 
The classification is performed using the [SII] and [NII] BPT diagrams. SF=star forming, Sy=Seyfert-like, Int=Intermediate (between the K01 and K03 lines in the [NII] BPT). Where the Sloan classification from both diagram is ambiguous (e.g. SF in [NII] BPT and Sy in [SII] BPT) the field is left blank.

\label{table2}
\end{table}

\subsubsection{Extended Central LINER-like Ionisation}
\label{Ext-LINER}

Three galaxies in our sample (P9-127A, P4-127A, P4-127B) show extended central LINER-like ionisation. Sloan single fibre fluxes (from the MPA-JHU catalogue) classify two of these galaxies as Seyfert-like, while P4-127A has an ambiguous classification. However, considering the errors associated with the emission line ratios (and the intrinsic porosity of the LINER-Seyfert demarcation line), the Sloan line ratios are also consistent with a LINER-like classification.

We show the BPT-derived maps of these galaxies using P-MaNGA data in Fig. \ref{example_BPT_seyfert} (top three galaxies, with the same colour-coding of regions as in Fig. \ref{example_BPT}). Both in P9-127A and P4-127B the LINER-like ionisation dominates within $\rm 0.5 \ R_e$, while in the case of P4-127A, the LINER-like ionisation extends from the central regions out to galactocentric distances larger than $\rm 1.0 \ R_e$. Interestingly, all three LINER-like galaxies are massive disc galaxies, ($\mathrm{log M_\star/M_\odot > 10}$, masses from MPA-JHU DR7 catalogue), and host active star formation in regions at large galactocentric radii. We will argue in subsequent Sections that, although contribution from an AGN to the ionisation budget cannot be ruled out, the most likely source of the observed LINER-like ionisation is hot evolved stars.

These conclusions confirm recent work from the CALIFA collaboration \citep{Singh2013, Papaderos2013} on extended LINER-like emission. However, only full MaNGA sample will be able to study extended LINER emission in statistical sense, and observe correlations between the extended LINER-like emission and other galactic properties. This study may potentially have a large impact on the local census of AGNs.

\begin{figure*} 
\includegraphics[width=0.7\textwidth, trim=60 90 80 60, clip]{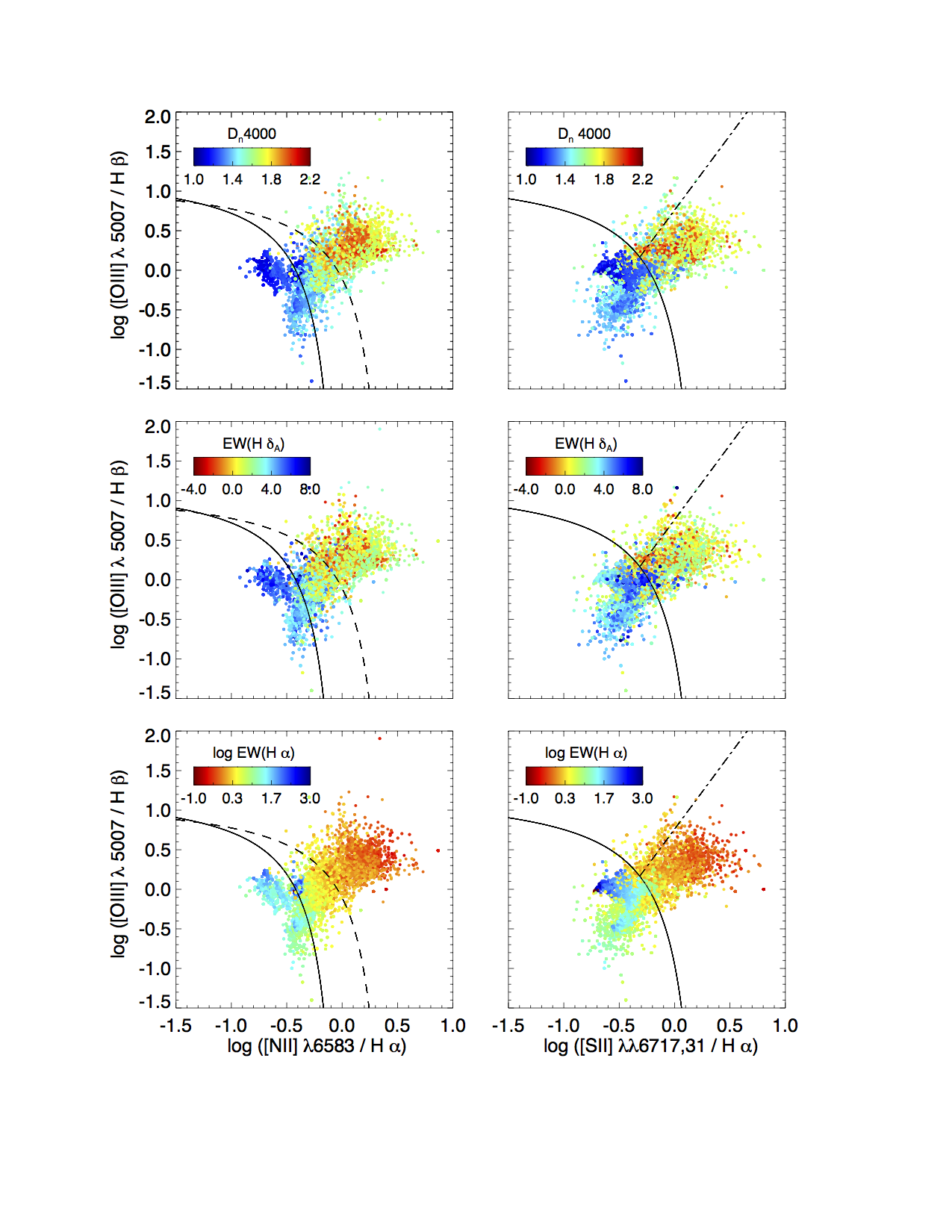} 
\caption{[NII] (left) and [SII] (right) BPT diagnostic diagrams colour-coded by $\rm D_n (4000)$ (top), $\rm EW(H \delta_A)$ (middle) and $\rm \log EW(H \alpha)$ (bottom). Each points corresponds to a Voronoi bin in one of the 14 galaxies considered. Note that, since single spaxels oversample the PSF, not all data points are statistically independent.} 
\label{BPT_tot}
\end{figure*}

\subsubsection{Extraplanar LINER-like Ionisation}

Galaxies P9-61A and  P4-19B (see Fig. \ref{example_BPT_seyfert}) present peculiar ionisation conditions, as traced by the BPT diagrams, displaying a star formation dominated (highly inclined) disc and LINER-like ionisation in the extraplanar direction. We suggest these line ratios can be interpreted as signature of diffuse ionised gas (DIG, i.e. gas ionised by stellar radiation, likely from O and B stars in the disc), which dominates the ionised gas budget in extraplanar regions of low continuum emission. This observation shows that using MaNGA data for highly inclined/edge on galaxies we might be able to shed new light on the disc-halo interaction. 

We see no signature of kinematic distortion or increase in velocity dispersion in these regions in either P9-61A or  P4-19B, at least within the relatively large errors associated with emission line fitting in these regions of low S/N. This argues against the identification of the LINER-like emission in the outer regions with shocked outflowing (or inflowing) gas. However a more detailed analysis, exploiting the emission lines detailed profile shapes (as in \citealt{Ho2014}), might be needed to detect outflow signatures and will be presented in future work.

\begin{figure}
\centering
\includegraphics[width=0.5\textwidth, trim=120 10 110 20, clip]{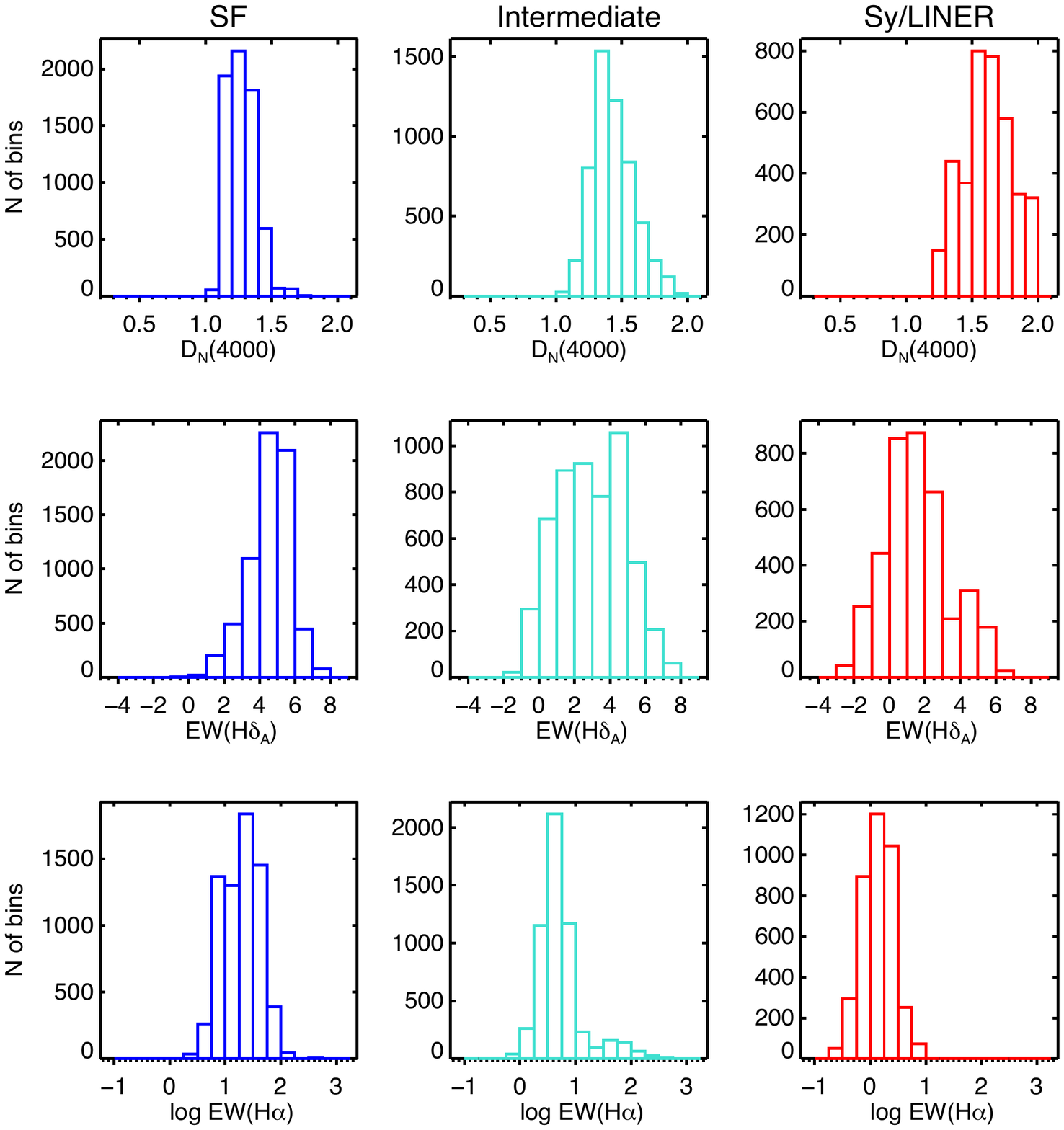}
\caption{Histograms of the distribution of  $\rm D_n(4000)$, $\rm EW(H \delta_A)$  and $\rm Log \ EW(H \alpha)$ for spaxels in different areas of the [NII] BPT diagram. }
\label{histo_BPT}
\end{figure}

\subsection{Comparison of the Ionised Gas Properties with Stellar Population Indices}
\label{st_indices}

In this section we will focus on the study of $\mathrm{D_n (4000)}$, $\rm EW(H \delta_A)$ and the $\rm EW(H \alpha)$ to gain insight into the nature of the underlying stellar population and the interplay between the gas ionisation and the continuum emission. We refer to Li et al. (submitted) for a more detailed analysis of these indices and a study of their radial gradients in the P-MaNGA galaxies. 

$\mathrm{D_n (4000)}$  (the so called `$\rm 4000 \ \AA$' break), calculated here using the definition given by \cite{Balogh1999}, is a proxy of recent star formation. For an instantaneous burst of star formation the $\mathrm{D_n (4000)}$ index increases monotonically with the age of the stellar population, and is roughly independent of metallicity until about 1-2 Gyr after the burst.  The equivalent width of $\mathrm{H \delta_A}$ (defined by \citealt{Worthey1997}), on the other hand, increases until about 300-400 Myr after the burst (post-starburst population) and then decreases again at later times \citep{Kauffmann2003, Delgado2005}. At intermediate ages, both indices are known to depend both on the age and the metallicity of the stellar population \citep{Thomas2004, Korn2005, Thomas2006}. Both $\mathrm{D_n (4000)}$  and $\mathrm{H \delta_A}$ are computed on the spectra after subtracting the emission lines. The equivalent width (EW) of $\rm H  \alpha$ is computed by dividing the $\rm H  \alpha$ flux by an interpolation of the underlying continuum, which is calculated as the average of the median fluxes in two bandpasses $\rm 200 \ km \ s^{-1}$ wide, at a distance of $\rm 200 \ km \ s^{-1}$ from the centre of the line (the model continuum is used). Here we choose a convention where the EW of the emission line is defined as positive when the line is in emission. 

The $\rm EW ( H\alpha)$ is a measure of the amount of ionising photons absorbed by the gas relative to the stellar mass. \cite{CidFernandes2011} used stellar population models combined with data from SDSS to argue that $\rm EW(H \alpha)$ can be used to distinguish ionisation from hot evolved stars (characterised by $\rm EW(H \alpha) < 3 \AA$) from other ionisation mechanisms (namely ionisation due to star formation or AGN activity, generally characterised by $\rm EW(H \alpha) > 3 \AA$). DIG ionised by escaping UV photons from the disc is also expected to show low $\rm EW ( H\alpha)$, as ionising photons can be coming from as far as one kpc away and DIG is only present in regions of continuum emission.

In Fig. \ref{BPT_tot} we present the [NII] and the [SII] BPT diagrams for all the sub-galactic regions in the galaxies in our sample (that meet the S/N cuts applied above) colour-coded by $\mathrm{D_n (4000)}$, $\mathrm{ EW(H \delta_A)}$ and $\rm EW ( H\alpha)$. Note that, since single spaxels oversample the PSF, not all data points shown are statistically independent.

We observe that within our sample there exists a strong correlation between the position in the BPT diagram and the indices considered. Regions classified as star-forming generally have low $\mathrm{D_n (4000)}$ ($\rm 1.2 < D_n (4000) <1.4$)  and large $\mathrm{EW(H \delta_A)}$ (peaking between 4 and 5): this is the expected signature of young stellar populations and it confirms that the BPT diagram correctly isolates regions of ongoing star formation, although at the typical distance of a MaNGA galaxy we are not able to resolve individual H\textsc{ii} regions. The scatter observed in these indices is most likely due to the effect of varying metallicity. We refer to Wilkinson et al. (submitted) for a more in-depth discussion of the relation between age, metallicity and stellar population indices for the P-MaNGA galaxies.

As regions move away from the star-forming sequence towards the AGN locus in the BPT diagram, they are characterised by a gradual increase in $\mathrm{D_n (4000)}$ and a parallel reduction in $\mathrm{EW(H \delta_A)}$. This trend is tracing an increase in the age (and possibly metallicity) of the underlying stellar population. The same effect is already well documented using SDSS data \citep{Kauffmann2003, Kewley2006}. This result is consistent with the analysis performed in \cite{Sanchez2014} using the largest sample of H\textsc{ii} regions to date extracted from the CALIFA survey. However a statistical study of the resolved ionised gas properties of LINER and Seyfert galaxies is still missing, and would be particularly interesting in the context of understanding how and where galactic regions stop forming stars. We also note that in regions ionised by a Seyfert/LINER-like ionisation field we consistently observe both a high $\mathrm{D_n (4000)}$ and a small $\mathrm{EW(H \delta_A)}$, indicating that these regions have not been `quenched' recently (within the last few 100 Myr): we are not observing the result of rapid cessation of star formation.

The analysis of the $\rm EW(H \alpha)$ reveals that, as expected, regions classified as star-forming present large EWs. In particular $99 \%$ of the regions classified as SF in the [SII] BPT have $\rm EW(H \alpha) > 3 \AA$, which is the delimiter proposed by \cite{CidFernandes2011} to classify regions as dominated by ionisation from young stars. This further confirms that, within the limitation of the considered diagnostics and the uncertainties in the extraction of line fluxes and EWs, we do not expect any significant contamination from other ionisation sources in regions that we define as SF. Of course the reverse statement is not necessarily true: contamination from SF regions could be present in regions classified as right wing (Seyfert/LINER-like) sources based on the [SII] BPT diagram.

Figure \ref{histo_BPT} summarises the above discussion by showing histograms of the distribution of $\rm D_n(4000)$, $\rm EW(H \delta_A)$  and $\rm Log \ EW(H \alpha)$ for spaxels in different areas of the [NII] BPT diagram. 

\subsection{The WHAN Diagram: a Complementary Diagnostic}
\label{WHAN}

\begin{figure} 
\centering
\includegraphics[width=0.5\textwidth, trim=40 120 80 150, clip]{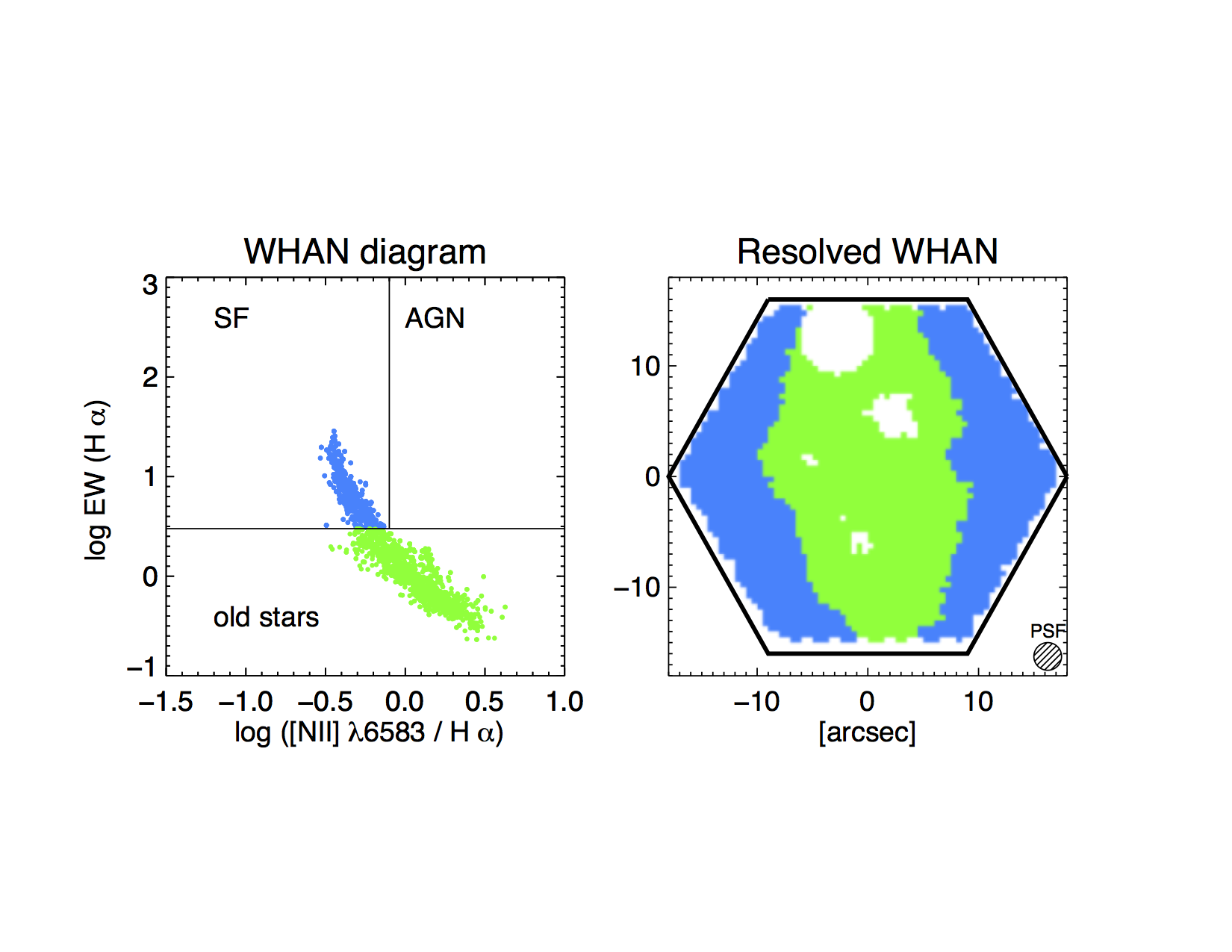} 
\caption{Example of a resolved WHAN diagram (as defined by \protect\citealt{CidFernandes2011}) for the galaxy P9-127A and the distribution of the sky of the various regions with the same colour-coding. This diagram classifies the nuclear regions of this galaxy as dominated by hot evolved stars.} 
\label{WHAN_ex}
%\end{figure}

%\begin{figure}
\includegraphics[width=0.5\textwidth, trim=130 80 130 80, clip]{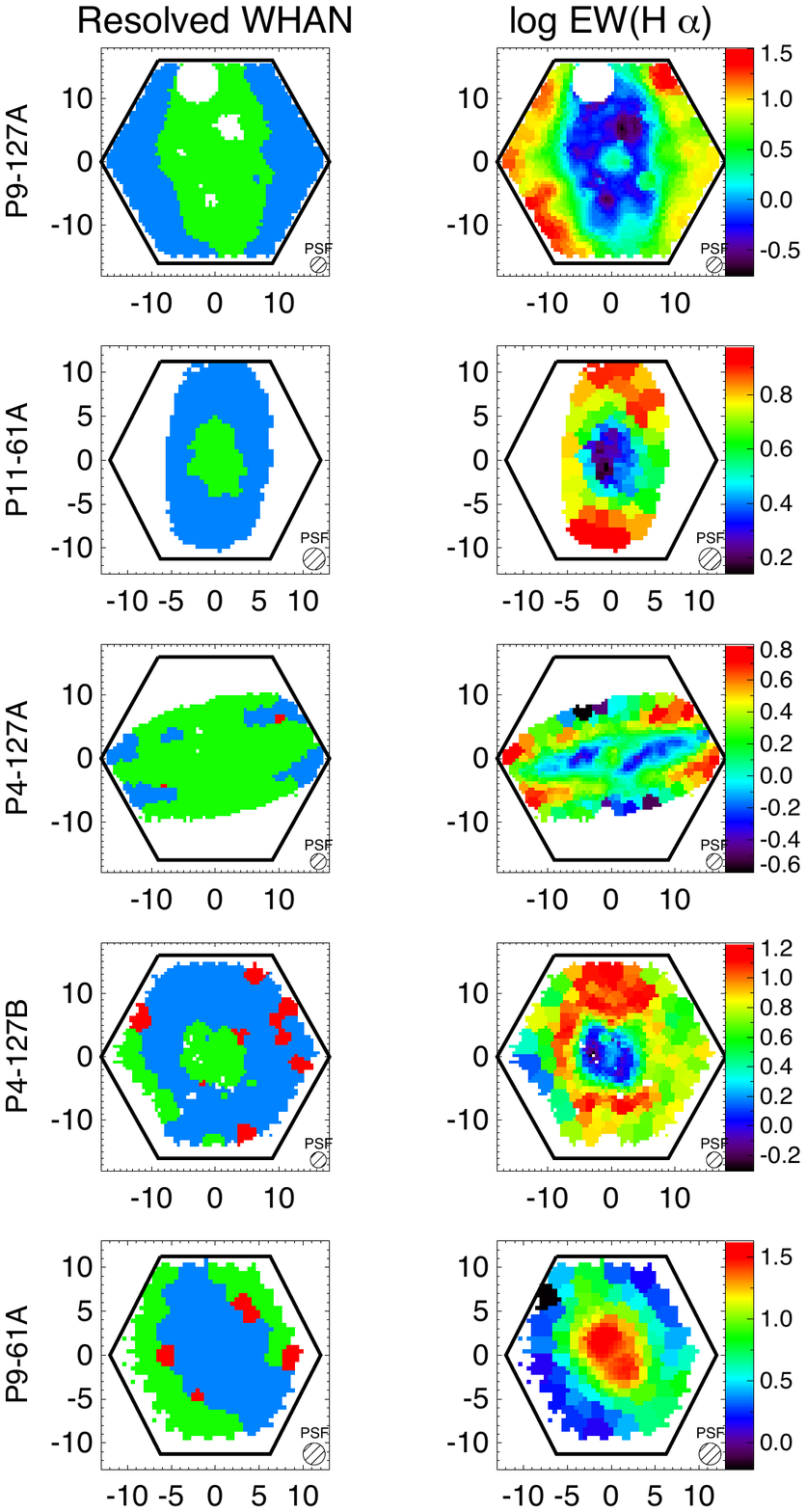} 
\caption{Left panels: Classification scheme for the P-MaNGA galaxies that show evidence for old hot stars/DIG ionisation using the `WHAN' diagram \protect\citep{CidFernandes2010}. Green regions are associated with hot evolved stars (or DIG), blue regions with SF and red regions with AGN. See text for details. Right Panels: Maps of $\rm log \ EW(H \alpha)$ for the same sources.}
\label{new_class}
\end{figure}

The comparison of the $\rm EW(H \alpha)$ with the $\rm [NII]/H \alpha$ can help us gain insight into the nature of the observed extended LINER ionisation. 
To this aim we tested the `WHAN'  diagram \citep{CidFernandes2010, CidFernandes2011} on the P-MaNGA galaxies. For consistency with our previous classification, which makes use of the Kewley 2001 demarcation lines, we used the best transposition of the Kewley 2001, as presented in \cite{CidFernandes2010}. We show an example of this classification for P9-127A in Fig. \ref{WHAN_ex}. In detail, we classify regions as follows:

\begin{enumerate}
\item{ionisation due to old hot stars or DIG, if $\rm EW(H \alpha) < 3 \AA$ (green in Fig. \ref{new_class}).}
\item{AGN if $\rm EW(H \alpha) > 3 \AA$  and $\rm [NII]/H \alpha > -0.1$ (red in Fig. \ref{new_class}).}
\item{SF, if $\rm EW(H \alpha) > 6 \AA$ and $\rm [NII]/H \alpha > -0.1$ (blue in Fig. \ref{new_class}).}
\end{enumerate}

We show the results of this classification for the galaxies classified as Seyfert and LINER-like in Fig. \ref{new_class}, together with the maps of $\rm \log EW(H \alpha)$ for the same galaxies. We note that using this revised classification the nuclear regions of P9-127A, P11-61A, P4-127A and P4-127B are all classified as dominated by hot evolved stars, with no AGN signature. Interestingly the external regions of P9-61A, which are classified as LINER-like in the classical BPT diagram, present low $\rm EW(H \alpha)$ and are consistent with our preferred interpretation of extraplanar DIG.

Echoing the remark in \cite{CidFernandes2011} we note that AGN might still be present in the nuclei of other galaxies (especially in P9-127A, which shows an increase in $\rm EW(H \alpha)$ in the nuclear regions, and P11-61A, which is classified as AGN using the classical BPT diagram), but, according to this classification scheme, the AGN does not constitute the major contribution to the ionising photon flux. It is clear, however, that more detailed photoionisation models, taking into account the information provided by stellar population modelling, is required in order to derive a self-consistent test of the photon budget. This analysis will be presented in future work from the MaNGA collaboration.

\subsection{Electron Density and Ionisation Parameter}

The flux ratio of the [SII]$\lambda\lambda$6717,6731 doublet is sensitive to the gas electron density. More specifically the ratio [SII]$\lambda$6731/[SII]$\lambda$6717 is expected to vary from $\sim$0.65, for $\rm n_e \sim 10^2~cm^{-3}$ to $\sim 2.5$, for $\rm n_e \sim 10^4~cm^{-3}$.

In the Appendix we show the maps of the [SII] doublet flux ratio for each galaxy. Although the doublet flux ratio can be converted into electron density, we prefer to show the measured flux ratio, since the conversion to electron density is non-linear and it saturates close to the boundaries, an effect which is difficult to illustrate graphically in the maps.

\begin{figure}
\centering
\includegraphics[width=0.5\textwidth, trim=20 280 20 300, clip]{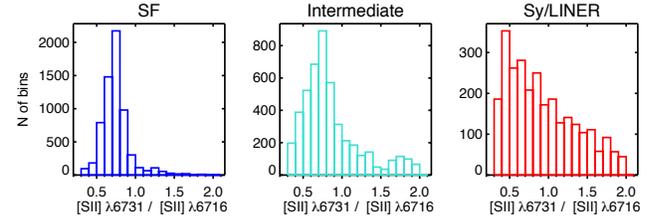} \quad
\caption{Histograms of the distribution of the [SII] doublet flux ratio $\rm [SII] \lambda 6731/ [SII] \lambda 6717$ for spaxels in different areas of the [NII] BPT diagram.}
\label{BPT_SII}
\end{figure}

Fig. \ref{BPT_SII} shows how the gas densities, as traced by the [SII] doublet, change in the various galactic regions, depending on the location on the BPT diagram. Quite interestingly, galactic regions located in the SF locus in the BPT diagrams are characterised by a relatively uniform and narrow range of [SII] ratio - between $\sim$0.60 (which is about the lower limit, the value at which the ratio saturates) and $\sim$0.85, implying densities below $\rm \sim 300 ~cm^{-3}$. The intermediate region is characterised by a larger scatter, while the Seyfert/LINER-like regions are found to have a very broad distribution of densities, ranging from low values of $\rm  <100 ~cm^{-3}$ out to $\rm >10^4 ~cm^{-3}$. 

We also investigate the spatial distribution of the extinction-corrected line ratio $\rm [OIII]\lambda5007 / [OII] \lambda \lambda 3726, 29$. This ratio is a good proxy for the ionisation parameter (U), defined as $\rm U=\frac{F_i}{n_e~c}$
(where $\rm F_i$ is the flux of ionising photons directed onto the gas clouds with electron density $\rm n_e$) in classical H\textsc{ii} regions. In regions dominated by other sources of ionisation the [OIII]/[OII] line ratio is also influenced by the hardness of the radiation field. For SF-regions we can relate the ionisation parameter to the [OIII]/[OII] ratio using the formula provided by \cite{Diaz2000}, $ \mathrm{\log U = -0.80 \log \frac{[OII]}{[OIII]} -3.02} $.

\begin{figure} 
\includegraphics[width=0.45\textwidth, trim=60 70 30 320, clip]{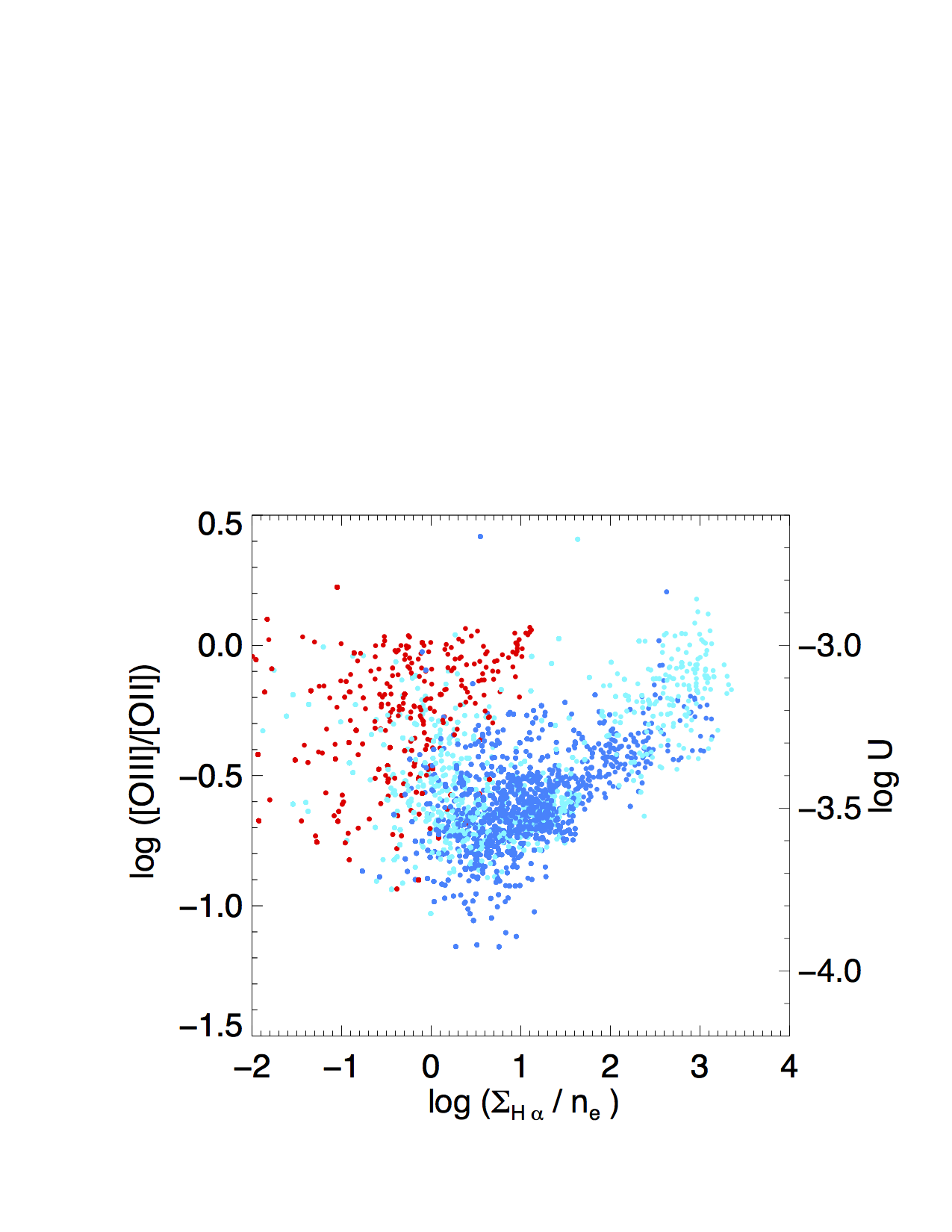}
\caption{Logarithmic plot of the  extinction-corrected line ratio $\rm [OIII]\lambda5007 / [OII] \lambda \lambda 3726, 29$ versus the ratio of extinction-corrected $\rm H \alpha$ flux divided by the electron density, as derived from the sulphur doublet flux ratio ($\rm [SII] \lambda 6731/ [SII] \lambda 6717$). Each point corresponds to a region (Voronoi bin) in one of the galaxies in the P-MaNGA sample and is colour-coded depending on its classification in the [NII] BPT diagram, with blue corresponding to SF-dominated regions, cyan to intermediate regions and red to Seyfert/LINER-like regions. The alternative y-axis on the right shows the value of log U corresponding to each [OIII]/[OII] ratio for H\textsc{ii} regions (valid only for the SF regions).} 
\label{b_d}
\end{figure}

For some of the galaxies that present nuclear Seyfert/LINER-like emission the [OIII]/[OII] ratio peaks in the nuclear region (P9-127A, P4-127A, P11-61A). For the star forming galaxies, the ionisation parameter has a broader range of distributions, often not peaking in the central region and offset with respect to the H$\alpha$ emission, suggesting that gas density variation might play a significant role in the distribution of the ionisation parameter. 

We have tried to quantify these effects by investigating in more detail the relation between ionisation parameter, extinction-corrected H$\alpha$ surface density ($\rm \Sigma _{H\alpha}$) and gas density. Assuming the geometry of H\textsc{ii} regions (and in particular the distribution of gas around young stars) to be similar on average, then the H$\alpha$ luminosity ($\rm L _{H\alpha}$) should be proportional to the local luminosity in ionising photons (Q): $\rm L_{H \alpha} \propto c_f Q$, where $\rm c_f$ is the photoionised clouds covering factor (this factor would take into account the possible effect of dust inside the H\textsc{ii} region absorbing ionising photons). If we assume that H\textsc{ii} regions are generally unresolved at the MaNGA pixel scale (the size of a pixel corresponds to around 250 pc at the median redshift), then, denoting the pixel size in physical units as p and the H\textsc{ii} region size as r we have

\[
\rm U\propto \frac{\Sigma _{H\alpha}}{c_f n_e}\frac{p^2}{r^2}
\]

We therefore expect a correlation between U and $\Sigma _{H\alpha}$, with scatter induced by changes in covering factor. The relation will also depend on distance (i.e. redshift) as, at higher redshift, as the size of one pixel in kpc (p) increases with redshift. The redshift dependance is lost in the limit where H\textsc{ii} regions are resolved out by the MaNGA pixel scale.

Fig. \ref{b_d} shows, for the various regions in the different galaxies, the value of [OIII]/[OII] as a function of the ratio $\rm \Sigma _{H\alpha}/n_e$. The various points are colour-coded according to their location on the [NII] BPT diagram as in Fig. \ref{example_BPT}. The star forming (blue) and intermediate regions (cyan) show a large scatter, but they tend to correlate with $\rm \Sigma _{H\alpha}/n_e$. The observed correlation is strongly sublinear, pointing towards systematic changes in covering factor with ionisation parameter. We have also tested for the presence of a redshift dependance, but have found no evidence for it within our current sample. The issue will be revisited with a sample covering a larger redshift range.

Interestingly a large fraction of Seyfert/LINER-like regions are located well above the SF sequence, inhabiting predominately the top left part of the plot.
We suggest two possible reasons responsible for this effect: 
\begin{enumerate}
\item{ The covering factor of the ionised clouds is generally lower than in typical H\textsc{ii} regions, thus moving the Seyfert nuclei towards the left in Fig. \ref{b_d}.}
\item{The ionising radiation of the AGN/hot stars is harder than the radiation ionising H\textsc{ii} regions. This produces a boost in the fraction of O$^{3+}$ relative to O$^{2+}$, boosting the [OIII]/[OII] ratio.}
\end{enumerate}

This result highlights that the [OIII]/[OII] line ratio can be successfully used to trace different ionisation conditions in the hot gas. In accordance with previous studies (e.g. \citealt{CidFernandes2011}) we suggest that this ratio might be a powerful probe of the dominant source of ionisation, complementary to the BPT diagram.

\section{Metallicity and Chemical Abundances}
\label{results2}

In this Section we study the relation between the spatially resolved metallicity and star formation rate surface density, as well as the distribution of the nitrogen-to-oxygen abundance ratio.

Gas phase metallicity (oxygen abundance) can be estimated from collisionally excited lines making use of ratios of nebular to auroral lines to estimate the electron temperature. This method (referred to as the $\mathrm{T_e}$ or `direct' method) for calculating metallicity is generally considered the most reliable in the low metallicity regime \citep{Pagel1992, Garnett1992, Izotov2006}. At higher metallicities, however, the electron temperature decreases and auroral lines are generally undetected. 

To overcome this limitation over the years a number of metallicity calibrators have been developed to estimate the oxygen abundance from ratios of strong nebular lines (including amongst others $\rm [OIII]\lambda\lambda4959,5007$, $\rm [OII] \lambda \lambda 3726, 29$, $\rm [NII]\lambda6584$, $\rm [SII]\lambda\lambda6717,31$), suitable to be used in the high metallicity regime. These metallicity diagnostics are generally calibrated against `direct' metallicity measurements \citep{Pettini2004, Pilyugin2005, Pilyugin2010}, photoionisation models \citep{Denicolo2002, Kobulnicky2004, Tremonti2004}, or a mixture of both \citep{Nagao2006, Maiolino2008, PerezMontero2014}. 

It is well known that the $T_e$-based calibrations generally yield a lower value of metallicity compared to the calibrations based on photoionisation models. The reasons for these discrepancies are still debated and thoroughly discussed in the literature, where they are attributed to uncertainties associated with the models, temperature fluctuations within H\textsc{ii} regions, deviations from the thermal Maxwellian distribution in electron plasma, different assumptions regarding oxygen depletion onto dust grains and other effects (e.g. \citealt{Stasinska2005, Kewley2008, Pilyugin2010, Rosales-Ortega2011, Binnette2012, Lopez-Sanchez2012, Dopita2013, Nicholls2013, PerezMontero2014}). 

In this work we are mainly interested in the scaling relations between metallicity and other galactic properties, hence the absolute value of the metallicity scale is not a major concern. However, given the possible systematics introduced by the choice of a particular metallicity calibrator, we will explore the effect that the choice of calibrator has on the observed trends. In the following section all line fluxes are considered to be extinction corrected using the procedure outlined in Section \ref{dat_an}.

To explore the parameter space of available calibrations in this work we will focus on three of them:
\begin{enumerate}
\item{The calibration presented in \cite{Maiolino2008} (M08), based on a combination of the \cite{Kewley2002} photoionisation models at high metallicities and a match to the direct metallicity measurements (using $T_e$) at low metallicities. Of the various strong line diagnostics presented in that work we use the one based on R23, defined as 
$ 
\mathrm{R23 =  ([OII]  \lambda\lambda 3726, 29+ [OIII] \lambda 4959 + [OIII] \lambda 5007) / H \beta}.
$
This calibration gives results very similar to the widely used fit to R23 presented in \cite{Tremonti2004}.}
\item{The calibration of $\rm O3N2 = [OIII]/[NII]$ presented in \cite{Marino2013}, based on a fit to $T_e$-derived metallicities. This calibration yields similar results to the one proposed by \cite{Pettini2004}.}
\item{The so-called `ONS' calibration of \cite{Pilyugin2010} (P10), which makes use of the [OII], [OIII], [NII] and [SII] lines to calculate the oxygen abundance based on best fit to a set of calibrating H\textsc{ii} regions with `direct' metallicity measurements.}

\end{enumerate}

All the metallicity calibrators above have been derived for classical H\textsc{ii} regions. We compute metallicity only on the spaxels that are classified as star forming regions using the K01 line in the [SII] BPT diagram. We choose to use this diagram rather than the [NII] BPT to avoid introducing any prior into our study of the N/O ratio in the subsequent section. In Section \ref{st_indices} we found that only $\rm 1 \%$ of the regions classified as SF in the [SII] BPT diagram have $\rm EW(H\alpha) < 3 \AA$, which is consistent with no significant contamination from ionisation due to hot evolved stars/diffuse ionised gas within the limits of the errors on the BPT line ratios. 

For the same set of spaxels, the star formation rate surface density ($\rm \Sigma_{SFR}$) is calculated from the extinction-corrected $\rm H \alpha$ surface brightness using the conversion presented in \cite{Kennicutt1998}.
\[
\mathrm{ \Sigma_{SFR} [M_{\odot} \ yr^{-1} \  kpc^{-2} ] = 7.9 \cdot 10 ^{-42} \Sigma_{H \alpha}  [erg \ s^{-1} \ kpc^{-2} ] }
\]

We note that $\rm \Sigma_{SFR}$ might be influenced by the imperfect flux calibration of the P-MaNGA data. We consider $\rm \Sigma_{SFR}$ to be uncertain to the 0.1 dex level.

\subsection{The Relation between Metallicity and Star Formation Rate Surface Density}
\label{results_2_1}

\begin{figure*}
\centering
\centering
\includegraphics[width=1.0\textwidth, trim=30 50 30 80, clip]{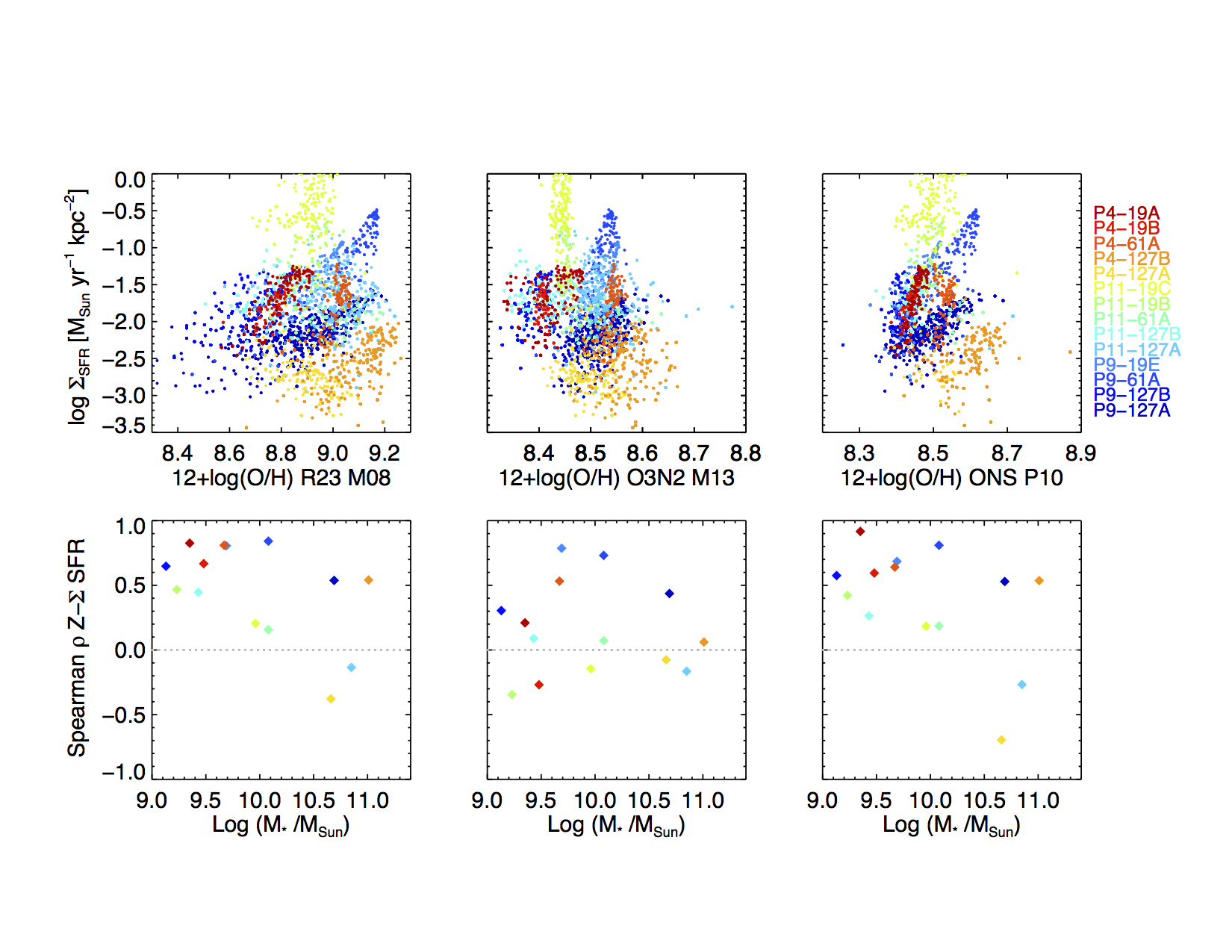} 
\caption{Top Panels: Star formation rate surface density ($\rm \Sigma _{SFR}$) plotted against gas phase metallicity ($\rm 12+log(O/H)$) using different calibrators: M08 is the \protect\cite{Maiolino2008} R23-based calibrator, O3N2 from \protect\cite{Marino2013} and P10 ONS from \protect\cite{Pilyugin2010}. Each point is colour-coded by the galaxy it belongs to. Regions that are not compatible with photoionisation by SF according to the [SII] BPT diagram are discarded. Bottom Panels: The Spearman $\rho$ rank correlation coefficients between metallicity and $\rm \Sigma _{SFR}$ as a function of the total stellar mass of the P-MaNGA galaxies.}
\label{met_SFR}
\end{figure*}

In this Section we investigate the relation between spatially resolved metallicity and star formation rate surface density ($\rm \Sigma_{SFR}$). Previous studies of integrated galactic properties have revealed, in addition to the well known mass-metallicity relation \citep{Lequeux1979, Tremonti2004}, a secondary relationship between gas metallicity and star formation rate \citep{Mannucci2010, Lara-Lopez2010, Yates2012,  Andrews2013}. This three-dimensional relation has been dubbed  the `Fundamental Metallicity Relation' (FMR). The secondary relation of metallicity on SFR may be resulting from a more fundamental relation between metallicity and gas content \citep{Moran2012, Bothwell2013, Hughes2013} and may be partially understood in terms of an equilibrium between gas inflows, outflows and star formation \citep{Mannucci2010, Dave2012, Dayal2013}. The extension of the relation to high redshift has been recently subject to some debate \citep{ Zahid2014, Wuyts2014, Steidel2014, Williams2014b, Troncoso2014, Cullen2014}.

The extension to spatially resolved scales is even more unclear. Spatially resolved observations of some star forming galaxies at high redshift \citep{Cresci2010, Troncoso2014} and blue compact dwarfs in the local universe \citep{Werk2010, SanchezAlmeida2014} have found an anti-correlation between SFR surface density and metallicity. This finding has been interpreted as tracing the effect of localised metal poor inflows in diluting the metallicity and boosting the SFR. In the larger galaxy population, on the other hand, it is well known that both SFR and metallicity have negative gradients \citep{Vila-Costas1992, Bigiel2008, Leroy2008, Sanchez2013, Pilyugin2014, Sanchez2013}, thus generating a global positive correlation between these two quantities on resolved scales. 
A systematic study of the relation between resolved metallicity and $\rm \Sigma_{SFR}$ would help shed more light on the plausibility of the inflow scenario for the wider galaxy population. Current and future IFS surveys are ideally placed for carrying out this work.

Using P-MaNGA data we make a preliminary study of the relation between metallicity and $\rm \Sigma_{SFR}$. For each galaxy we plot $\rm \Sigma_{SFR}$ against the gas phase metallicity, using the three metallicity calibrations introduced in Sec. \ref{results2}, and calculate the Spearman correlation coefficients between the two quantities. We show the results in Fig. \ref{met_SFR}. 

Using the M08 and P10 metallicity calibrators, 12 of the 14 galaxies in our sample show a positive correlation between metallicity and $\rm \Sigma _{SFR}$. This can be understood as a consequence of the fact that both metallicity and $\rm \Sigma _{SFR}$  generally present a negative radial gradient. Within our sample the only exception is P9-127A, where both the metallicity and $\rm \Sigma _{SFR}$ gradient are positive between 0.4 and 0.7 $\rm R_e$ (the largest radius sampled by the P-MaNGA observations). This is not completely surprising if we consider that for this galaxy we have limited radial sampling (in terms of $ R_e$), but better resolution, in terms of kpc, than the MaNGA survey data. Moreover we have checked that, taking advantage of the CALIFA IFS data, a negative metallicity gradient is observed in P9-127A at larger radii.

In Fig. \ref{met_SFR}, bottom panel, we show the Spearman correlation coefficients between metallicity and $\rm \Sigma _{SFR}$ as a function of the total stellar mass of the galaxy. We observe that the positive correlation between the two quantities is stronger at lower masses ($\rm log(M_\star/ M_\odot) < 10.5$), while at high stellar masses our data does not allow to pose any constrains on the presence of a correlation.

The fact that the trends discussed above are less evident or even absent using the O3N2 calibrator might be due to that fact that this calibrator is sensitive to changes in the N/O ratio, which are expected to occur on longer timescales than oxygen enrichment. However, the well known indirect correlation between metallicity and ionisation parameter \citep{PerezMontero2014}, which is assumed in the adopted strong-line calibrations, might also be playing a role in shaping the observed relation.

It is worth commenting here on the galaxy with the highest SFR in our sample: P11-19C (light green in Fig. \ref{met_SFR}, $\rm SFR \sim 12 \ M_{\odot} \ yr^{-1}$). Most of the bins in this galaxy are classified as intermediate using the [NII] BPT, while in the [SII] BPT it lies safely inside the SF sequence. Its median $\rm EW(H \alpha)$ is $\sim 10^2$ (in fact these points are clearly visible in the [NII] BPT colour coded by $\rm EW(H \alpha)$ in Fig. \ref{BPT_tot} as the blue points just across the K03 line) corroborating our conjecture that this system is experiencing a starburst. We suggest that the reason why this galaxy lies so much above the SF sequence in the [NII] BPT is related to its nitrogen enhancement, as will be discussed in the next section.

\subsection{Metallicity and Nitrogen Abundance}
\label{results_3}

Nitrogen and oxygen have different nucleosynthetic pathways in galaxies. Elements like oxygen, carbon and sulphur have a primary nucleosynthetic origin, i.e. their production yield is approximately independent from the amount of heavy elements already present in the galaxy. Therefore, to first order, the sulphur abundance will be proportional to the oxygen abundance, making sulphur lines emission lines precious tracers for abundance studies.

The nitrogen enrichment, on the other hand, is very sensitive to the details of the history of chemical evolution of a galaxy. It is predicted that in the low-metallicity regime most of the nitrogen has a primary origin (i.e. yield independent of metallicity), being generated in massive stars. At higher metallicities nitrogen becomes a secondary nucleosynthetic product (i.e. the yield is proportional to the metallicity), since production of nitrogen depends on the previous amount of oxygen (and carbon) synthesised in stars, via the CNO cycle. Moreover, since nitrogen is primarily produced by intermediate mass stars, it probes longer enrichment timescales relative to oxygen (which is promptly produced by short-lived massive stars). 

Overall modelling nitrogen enrichment remains a difficult task because of the uncertain stellar yields and the time delay for oxygen enriched with respect to nitrogen. The effect of inflows and outflows further complicate the task. \cite{VanZee2006}, for example, argue that because of the time delay in the production of nitrogen, supernova-driven winds in galaxies experiencing a starburst would primarily deplete the ISM of oxygen, shifting these galaxies at higher N/O ratios than expected. In summary, careful study of the N/O ratio as a function of O/H in principle have the potential of informing us about all the above processes, provided that we succeed in disentangling the different effects.

Observations of single H\textsc{ii} regions \citep{Garnett1990, Nava2006, Berg2012} and data from large spectroscopic surveys of galaxies like SDSS agree with the above models and show a constant N/O ratio (log(N/O) $\sim$ -1.5) as a function of O/H ratio at low metallicities and a steeper slope for higher metallicities where nitrogen production becomes secondary \citep{Edmunds1978, Vila-Costas1993, Henry2000, Perez-Montero2009, Andrews2013, Wu2013}. The relation between N/O and O/H also depends on stellar mass: more massive galaxies are found at higher metallicities and higher N/O ratio. This finding is consistent with massive galaxies being more evolved and hence dominated by secondary nitrogen production and by older intermediate mass stars. 

\subsubsection{The N/O Ratio as a Function of Metallicity in P-MaNGA}

In this section we compute the N/O ratio in conjunction with O/H using the `ONS' calibration presented in \cite{Pilyugin2010}. The advantage of this calibration framework is that it computes the oxygen and nitrogen abundance self-consistently. To ease comparison with previous work we compared the nitrogen abundances obtained this way with the nitrogen abundances computed using the calibration presented in \cite{Vila-Costas1993}, which makes use primarily of the $\mathrm{ [NII] \lambda 6584}$ to $\mathrm{[OII]} \lambda 3727$ flux ratio, concluding that the two calibrations are equivalent (scatter from the one to one relation of 0.02 dex).

\begin{figure} 
\includegraphics[width=0.5\textwidth,  trim=0 150 150 250, clip]{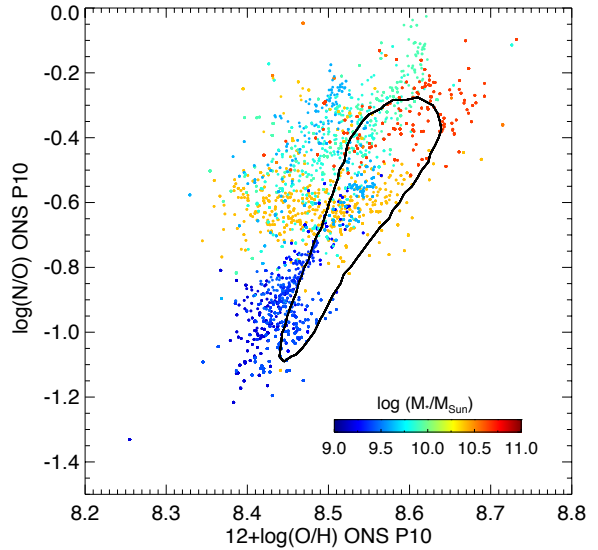} 
\caption{A plot of gas phase oxygen abundance ($\mathrm{12+log(O/H)}$) vs N/O ratio, colour-coded by the galaxy total stellar mass (as obtained from the MPA-JHU catalogue).}
\label{Nitrogen}
\end{figure}

In Fig. \ref{Nitrogen} we show N/O versus O/H for all the spaxels in the P-MaNGA galaxies that meet the selection cuts detailed above and in the previous section. The points are colour-coded by the total stellar mass of the galaxy they belong to (stellar masses from the MPA-JHU catalogue). For comparison, the black contour encircles 50 \% of the galaxies from Sloan DR7.

Three features are immediately evident:
\begin{enumerate}
\item{There is a sharp lower envelope in N/O vs O/H space, with very few points lying in the bottom-right corner of the diagram. The lower envelope qualitatively follows the theoretical expectation (N/O constant at low metallicity and increasing at high metallicity, though the observed slope is somehow steeper than prediction from simple models, see \citealt{Henry2000}), and is consistent with the data from Sloan DR7 (black contour in Fig. \ref{Nitrogen}).}
\item{A large scatter is present above the lower envelope. Regions from individual galaxies show large deviations from the main SDSS DR7 sample in the O/H direction at given N/O.}
\item{The position on the N/O versus O/H sequence of each particular galaxy is associated with its total stellar mass, with higher mass galaxies lying at higher metallicities but also higher N/O ratios.}
\end{enumerate}

\subsubsection{Modelling the N/O vs O/H Diagram: Signature of Gas Flows?}
\label{gas_flows}

Overall both i) and iii) above are consistent with theory and previous observations. The observed N/O scatter above the lower envelope (point ii above) is particularly interesting. Previous work has already reported the presence of this effect on resolved scales \citep{Vila-Costas1993, Diaz2007, PerezMontero2014}, but with much smaller statistics. Here we consider three possible explanations (and later discuss what are the most likely scenarios for the various galaxies):

\begin{figure*} 
\includegraphics[width=\textwidth, trim=0 160 0 180, clip]{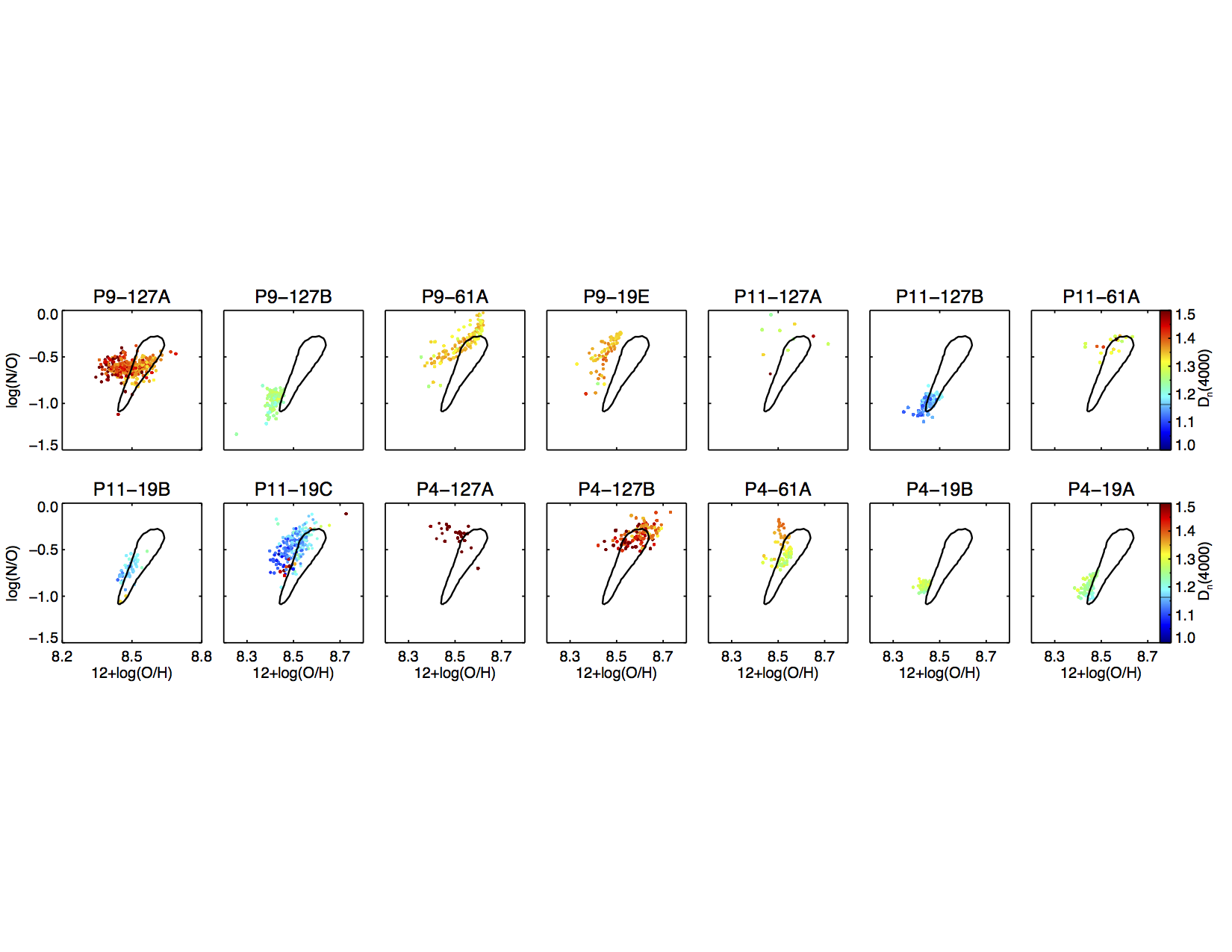}
\caption{N/O vs O/H for all the P-MaNGA galaxies. Individual points are colour-coded by $\rm D_n(4000)$.  Metallicities and N/O are calculated using the ONS calibration \protect\citep{Pilyugin2010}. }

\label{all_nitro}
\end{figure*}

\begin{enumerate}

\item{\emph{Bursts of SF}. Models have been proposed in which a sequence of discrete star formation bursts are responsible for the observed scatter \citep{Matteucci1986, Garnett1990, Coziol1999, Torres-Papaqui2012}: at early times after a burst mostly oxygen is produced (hence decreasing N/O and increasing O/H), while at later times intermediate mass stars produce mostly nitrogen, hence increasing N/O. We note that in this scenario such `bursts' must be associated with an increase in star formation efficiency, and not from fuelling star formation with additional, infalling low metallicity gas (which would instead also dilute the metallicity). This might be the favoured scenario in cases of interactions and mergers.}

\item{\emph{Infall of low metallicity gas}. Another possible explanation is that the observed scatter results from inflow of low-metallicity/pristine gas (e.g. high velocity H\textsc{i} clouds), which primarily dilutes the metallicity, hence lowering O/H, but does not affect the N/O ratio significantly (some effect on N/O may be present if the inflowing gas has low metallicity with low N/O ratio). This scenario would primarily generate a scatter in the horizontal (metallicity) direction of the diagram. A similar model has been proposed by \cite{Koppen2005}.}

\item{\emph{Galactic Fountain}. Yet another explanation is a `galactic fountain' scenario: high metallicity and high N/O gas clouds are expelled from the central region (lying on the `main envelope' traced by the black DR7 contour) and fall back onto the external, low metallicity and low N/O regions of the disc. If, as expected, the gas mass expelled from the central region is small relative to mass of gas in the external disc with which the mixing happens, then the resulting increase of metallicity of the gas in the outer disc would be small, but the mixing would result into a strong increase of the N/O ratio.}

\end{enumerate}

A priori there is no reason to suppose that only one of these mechanisms would be at play. However, for the rest of this section we will be speculating on whether it is possible to use the P-MaNGA data to determine which, if any, of these mechanisms is dominant.

\begin{figure*} 
\includegraphics[width=0.47\textwidth, trim=70 100 0 250, clip]{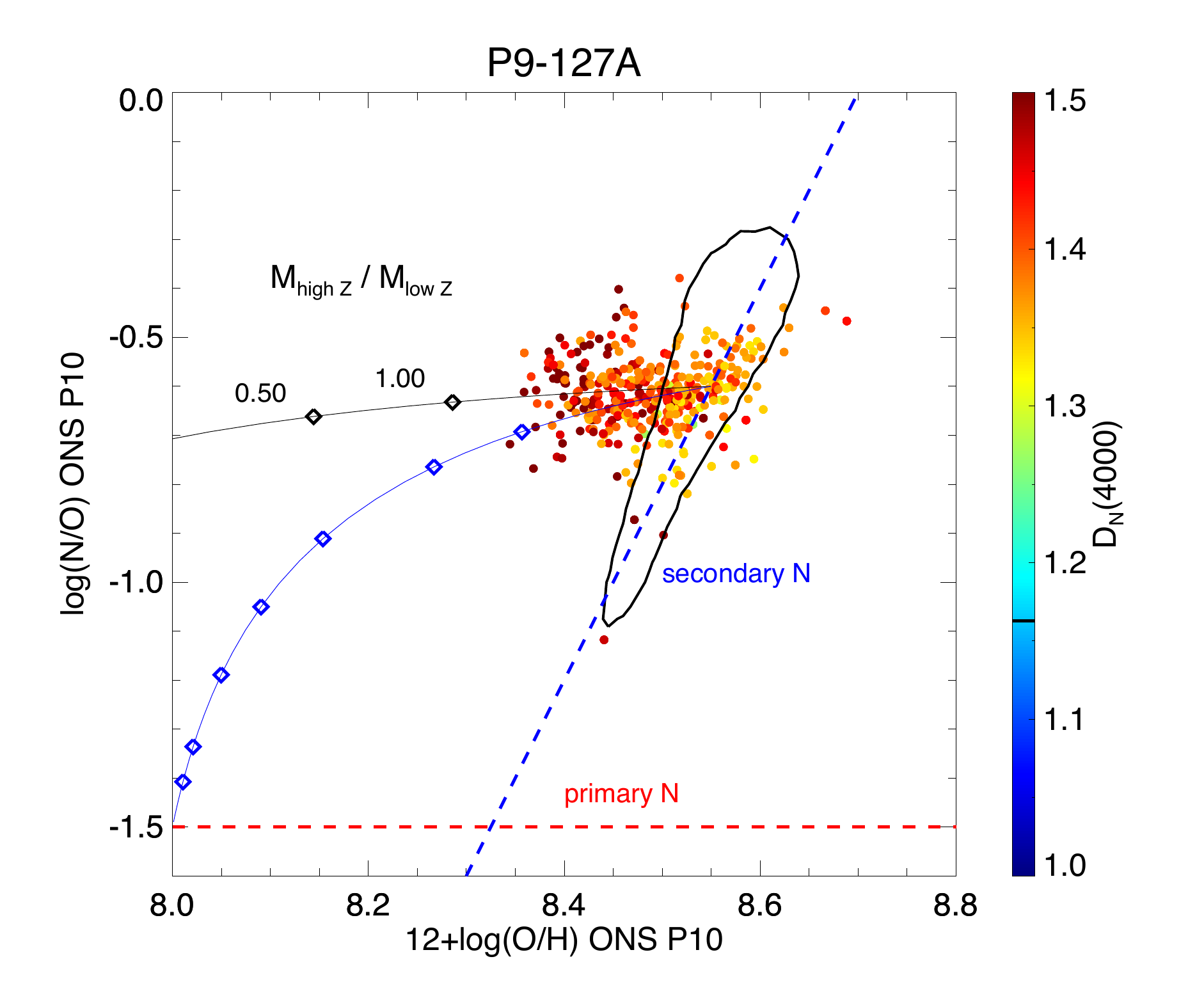}
\includegraphics[width=0.47\textwidth, trim=70 100 0 250, clip]{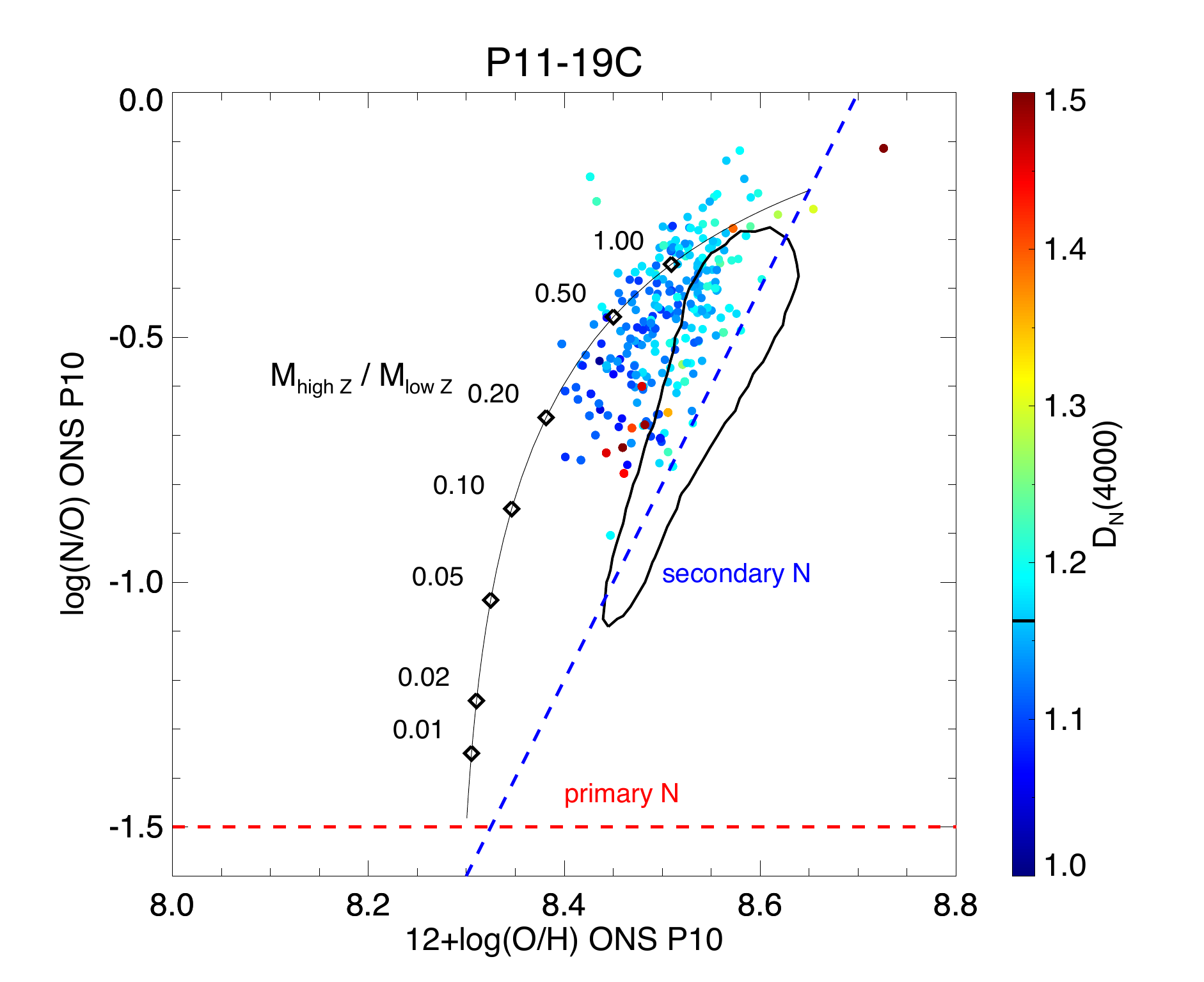}
\includegraphics[width=0.47\textwidth, trim=70 100 0 250, clip]{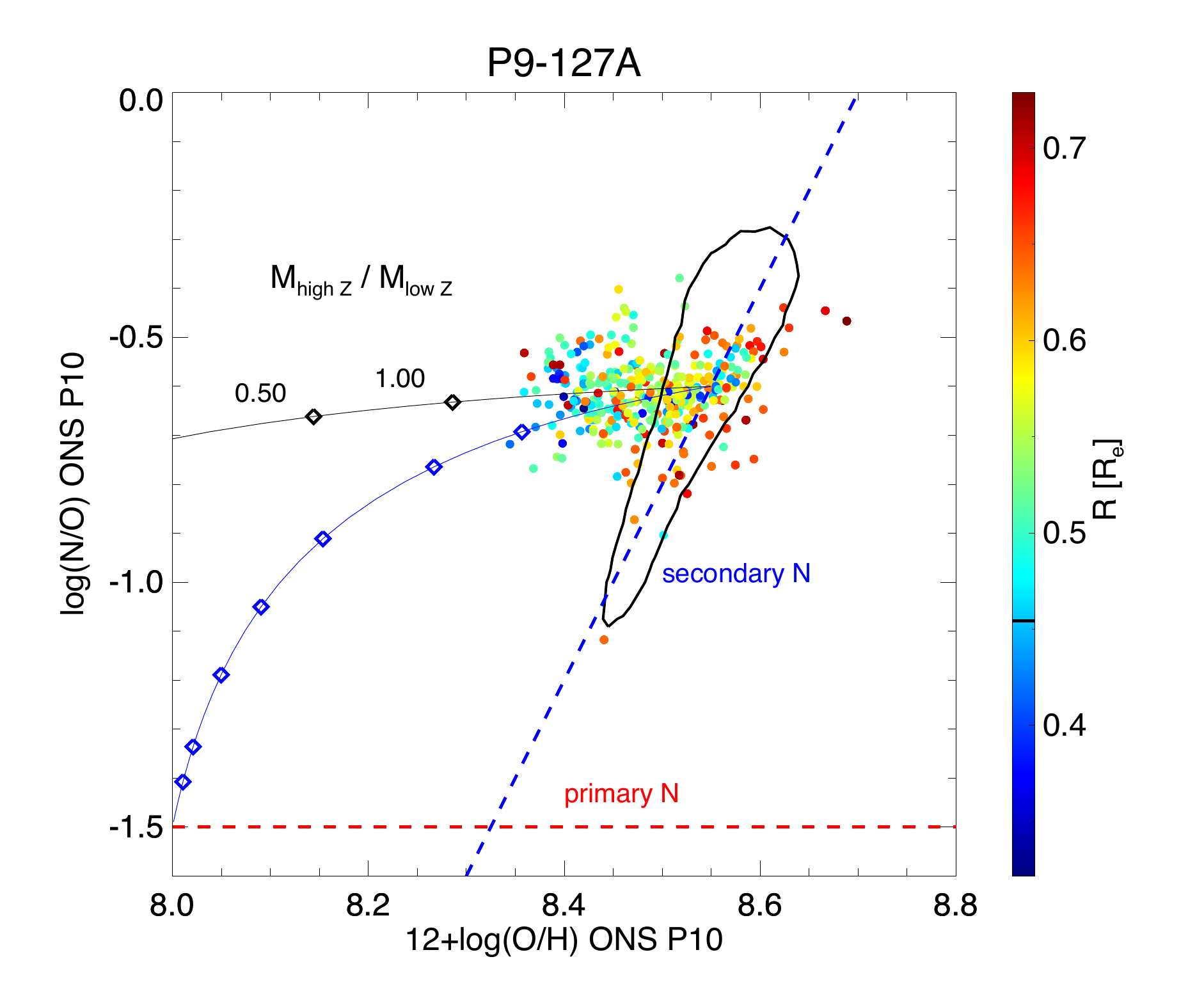}
\includegraphics[width=0.47\textwidth, trim=70 100 0 250, clip]{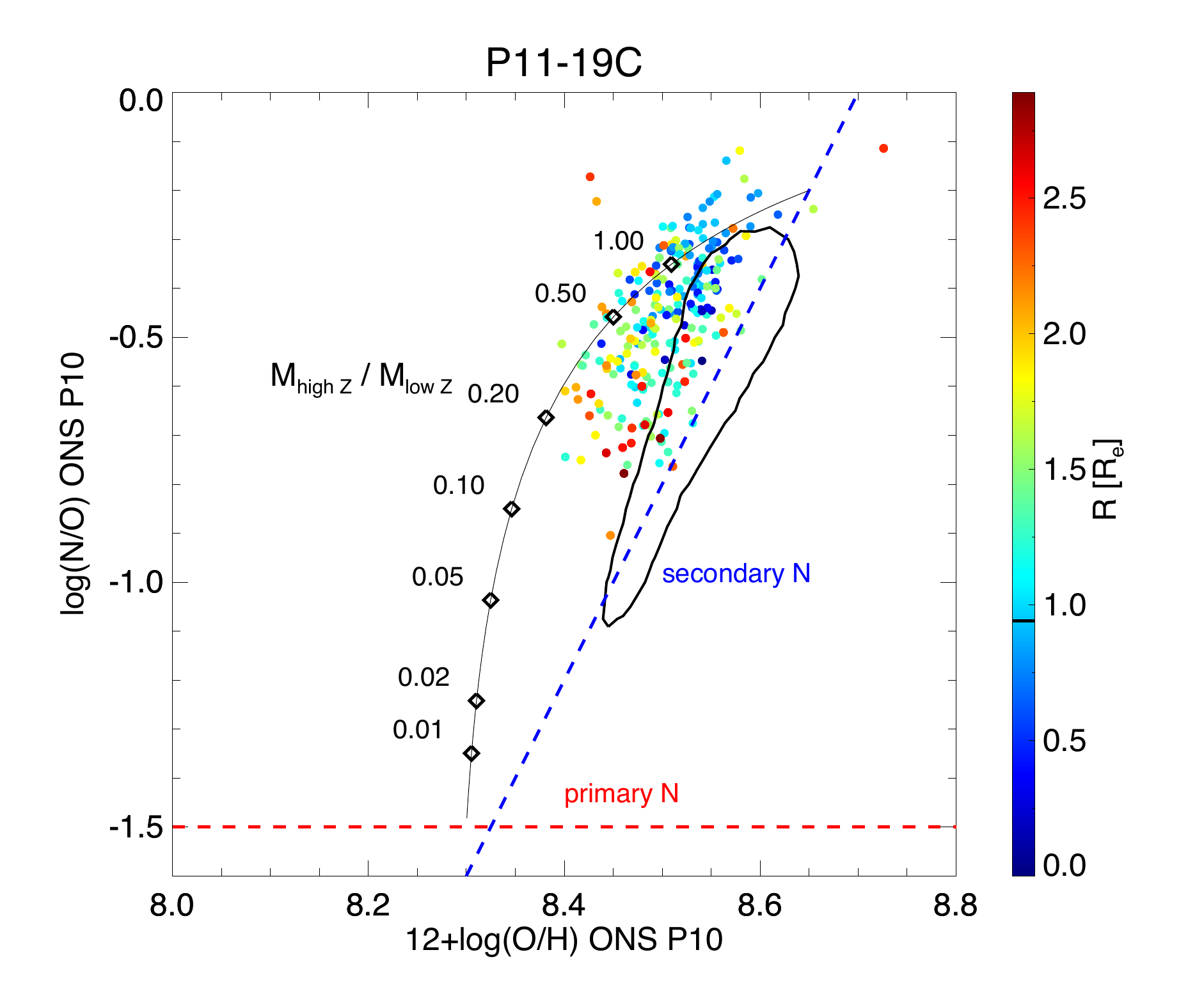}
\caption{N/O vs O/H for P9-127A and P11-19C. Top: Individual points are colour-coded by $\rm D_n(4000)$.  Metallicities and N/O are calculated using the ONS calibration \protect\citep{Pilyugin2010}. The red dashed line represents the N/O expected when nitrogen is only a primary nucleosynthetic product, while the blue dashed line represents a linear fit to the the N/O distribution of galaxies in SDSS DR7 with $\mathrm{12+log(O/H) > 8.3}$. The contour enclosing 50\% of the DR7 galaxies is shown in black. The thin black and blue lines correspond to simple gas mixing models. In these models a gas cloud with high metallicity and high N/O, lying on the envelope traced by the dashed blue line, is mixed with a cloud of low metallicity and low N/O, lying on the primary-N/low-metallicity relation, traced by the horizontal red dashed line. The marks indicate the gas mass ratio between the metal/nitrogen-rich and metal/nitrogen-poor clouds, as indicated in the label next to the marks on the top curve. For P9-127A a dilution scenario, with pristine gas falling into the gas poor central regions, provides a good fit to the data (the black and blue models corresponds respectively to low metallicity cloud with $\rm 12+ log(O/H)=7.5$ and 8.0). The data for P11-19E can be explained by a combination of nitrogen enrichment due to a SF burst and the mixing of some metal rich gas from the nuclear regions onto the disc (a fountain effect, with the black model corresponding to a low-metallicity cloud of $\rm 12+ log(O/H)=8.3$). Bottom: Same as above, but points are colour-coded by the deprojected galactocentric distance (in arcsec).}

\label{nitro_model}
\end{figure*}

In Fig. \ref{all_nitro} we show N/O vs O/H for all the P-MaNGA galaxies individually, with each point colour-coded by $\rm D_n(4000)$. We note that some galaxies fit very well within the SDSS DR7 main contour but there are a number of outliers. In particular, P9-127A, P9-61A, P11-61A and P4-127B show departures from the main DR7 contour in the horizontal (O/H) direction, at roughly constant N/O. These deviations seem to form a sharp sequence. There is some evidence that $\rm D_n(4000)$ increases as we move away from the DR7 contour, implying that the deviating spaxels belong to an older stellar population. These results are at odds with the predictions from the SF burst model, which would expect spaxels to scatter above the DR7 contour (and not horizontally as observed in the galaxies discussed above) and would expect these regions to have been recently experiencing a starburst (which would correspond to a much lower $\rm D_n(4000)$). We therefore conclude that in these galaxies the main driver of the deviation from the DR7 contour is represented by gas flows. 

To quantify this statement we built simple gas mixing toy models. Let us consider a high metallicity gas cloud containing a mass of gas $\rm M_{Z \ high}$ and metallicity $\rm Z_{high}$ being mixed with a cloud of low metallicity gas $\rm Z_{low}$ and gas mass $\rm M_{Z \ low}$. We parametrise the nitrogen abundance sequence as observed from the Sloan DR7 and other observational data by two straight lines in log-log space:
\begin{enumerate}
\item{We assume nitrogen is a primary nucleosynthetic product for $\rm 12+log(O/H) < 8.35$, with regions of lower metallicities having a flat $\rm log(N/O) = -1.5$.}
\item{We assume nitrogen is a secondary nucleosynthetic product for $\rm 12+log(O/H) > 8.35$, and use a linear fit to the Sloan DR7 data to parametrise the relation between N/O and O/H (log(N/O)= 4.0 [12+log(O/H) -8.7]).}
\end{enumerate}

We use this simple model to assign a N/O ratio to the two gas clouds we wish to mix and generate the mixing patterns in N/O vs O/H space. Depending on the choice of metallicities and mass ratio between the two clouds, this model can be used to represent a dilution scenario and/or the effect of a galactic fountain.

In Fig. \ref{nitro_model} we show how dilution models can be used to successfully explain the horizontal patterns observed in P9-127A (or similarly for P9-61A, P11-61A and P4-127B). The models plotted in Fig. \ref{nitro_model} show the mixing pattern generated by mixing a high metallicity cloud ($\rm 12 + log(O/H)=8.55$) with a very low metallicity cloud of metallicity $\rm 12 + log(O/H)=7.5$ (black solid curve) or $\rm 12 + log(O/H)=8.0$ (blue solid curve). The diamonds over plotted on the mixing curves correspond to the final position of the mixed cloud for different gas mass ratios. In the case of P9-127A the points at lower metallicities in the horizontal stripe have higher $\rm D_n(4000)$ (top panels in Fig. \ref{nitro_model}) and are closer to the galactic centre (bottom panel in Fig. \ref{nitro_model}, in which the colour coding gives the deprojected galactocentric distance). If these regions also have lower gas content, as one might reasonably expect for a galaxy whose central regions have older population and no evidence for ongoing SF, we could explain that fact that the older more central regions present the largest deviation from the DR7 contour: it takes only a small mass of pristine gas to dilute their metallicity considerably. In summary, the observed patterns of dispersion in O/H at constant N/O can be explained by the infall of pristine/low-metallicity gas throughout the galaxy, but the effects of metallicity dilution are stronger in those regions characterised by lower gas content, likely associated with older stellar populations.

Galaxies such as P9-19E and P11-19C show a different deviation from the DR7 contour with respect to the galaxies we mentioned above. Most of the spaxels in these two galaxies lie just above the DR7 contour and do not create a horizontal sequence in the N/O vs O/H plane. In both galaxies (but more clearly in P11-19C) we can also see that spaxels with the largest offset from the DR7 contour have the lowest $\rm D_n(4000)$. The most deviating points are also the most distant from the centre (as shown in the bottom right panel of Fig. \ref{nitro_model}). This effect could be explained in the context of the galactic fountain scenario discussed above, i.e. by the mixing of a small amount of metal rich gas from central regions into the outer, younger (more gas rich) and metal poor regions of the disc. This scenario is modelled in Fig. \ref{nitro_model} with a model in which gas clouds expelled from the central region, with initial metallicity of 12+log(O/H)=8.65, are mixed with outer gas with metallicity 8.3. However, other effects, like nitrogen enrichment due to a recent burst of SF, might also explain the observed pattern. More sophisticated chemical evolution models are needed to make further progress and will be presented in future work.

Overall, different gas flow scenarios are likely to apply to different galaxies. It will be interesting, in the context of the full MaNGA sample, to investigate how the resolved N/O vs O/H relation varies with different galactic properties. For the moment, we emphasise that, with spatially resolved spectroscopy, we observe two different behaviours in the distribution of the galactic regions on the N/O vs O/H diagram (horizontal sequence versus steeper sequences, offset from the distribution of the bulk of the galaxy population) and the associated distribution of $\rm D_n(4000)$ and galactocentric distance.

\section{Conclusion}
\label{conc}

We have exploited integral field spectroscopic observations of a sample of 14 local galaxies, obtained with the MaNGA prototype instrument (P-MaNGA), to investigate the spatially resolved properties of the gas excitation and chemical abundances, by tracing the distribution of nebular emission lines.

The main conclusions obtained in our analysis are summarised below:

\begin{enumerate}

\item {We observe extended (up to $\rm 0.5-1 R_{e}$) LINER-like line ratios in the central regions of three galaxies in our sample. These regions present high $\rm D_n(4000)$ and low $\rm EW(H \delta_A)$, consistent with a population of old and metal rich stars and very low levels of ongoing star formation. Although gas is present in these regions (as traced by the presence of nebular emission lines), this gas is not forming stars. The use of $\rm EW( H \alpha)$ to discriminate between AGN photoionisation and ionisation due to evolved stars (following \citealt{CidFernandes2010}) leads to the conclusion that ionisation is dominated by evolved stars. This finding is consistent with galaxies growing `inside-out', and central regions of massive galaxies no longer hosting active star formation.}

\item {P-MaNGA data reveals the presence of star formation in the discs of galaxies whose central regions are dominated by other sources of ionisation (both Seyfert-like and LINER-like). This finding is particularly interesting when compared with previous single-fibre surveys (like SDSS), where such star formation at large radii has been completely missed.}

\item {Edge-on star forming galaxies show evidence for LINER-like emission above and below the galactic plane, which is likely tracing Diffuse Ionised Gas, excited by ionising radiation from the galactic disc, likely hardened by absorption.}

\item {We find a positive correlation between SFR surface density and metallicity on resolved scales, for galaxies with stellar mass $\rm log(M_\star/ M_\odot) < 10.5$. At higher stellar masses the current data does not allow us to draw strong conclusions.}

\item {We investigated the N/O abundance ratio versus metallicity (O/H) relation on spatially resolved scales. We find a broad distribution of metallicities at a given N/O for regions within individual galaxies, relative to the sharp N/O vs O/H sequence traced by the central regions of galaxies, sampled by the bulk of SDSS-DR7 spectra. In several cases such deviations trace a well-defined sequence on the N/O vs O/H plane. Moreover such deviations correlate (in a different way for different galaxies) with the $\rm D_n(4000)$ index and galactocentric distance. We show that these deviations can be explained in terms of infall of low-metallicity/pristine gas, for some galaxies, or in terms of galactic fountains, for other galaxies.}

\end{enumerate}

Overall our findings are consistent with the general picture of galaxies growing `inside out', with nuclear regions of massive galaxies having already processed most of their fuel, and presenting older stellar populations, higher metallicities and very little or no ongoing star formation \citep{Perez2013, GonzalezDelgado2014, SanchezBlazquez2014}. The role of AGN in regulating star formation remains unclear with the statistics currently available. At the moment we are limited by the lack of large observational studies of bona-fide AGN using IFS with a spectral range comparable to that of MaNGA and a large field of view. 

It is also clear that the simple BPT diagnostic, popularised though the study of galaxy spectra from SDSS, might be misleading in its optical classification of Seyfert galaxies. The MaNGA survey, dutifully complemented by observations at other wavelengths and complementary IFS datasets, has the power to shed new light on the AGN census in the nearby Universe.

\section*{Acknowledgements}

We acknowledge Cheng Li and Enci Wang for their contributions to the analysis of the stellar population indices. We thank Michele Cappellari and the referee for their insightful comments.
This work was supported by the STFC, UK and World Premier International Research Center Initiative (WPI Initiative), MEXT, Japan. 

Funding for SDSS-III and SDSS-IV has been provided by the Alfred P.~Sloan Foundation and Participating Institutions.  Additional funding for SDSS-III comes from the National Science Foundation and the U.S.~Department of Energy Office of Science.  Further information about both projects is available at {\tt www.sdss3.org}.

SDSS is managed by the Astrophysical Research Consortium for the Participating Institutions in both collaborations.  In SDSS-III these include the University of Arizona, the Brazilian Participation Group, Brookhaven National Laboratory, Carnegie Mellon University, University of Florida, the French Participation Group, the German Participation Group, Harvard University, the Instituto de Astrofisica de Canarias, the Michigan State/Notre Dame/JINA Participation Group, Johns Hopkins University, Lawrence Berkeley National Laboratory, Max Planck Institute for Astrophysics, Max Planck Institute for Extraterrestrial Physics, New Mexico State University, New York University, Ohio State University, Pennsylvania State University, University of Portsmouth, Princeton University, the Spanish Participation Group, University of Tokyo, University of Utah, Vanderbilt University, University of Virginia, University of Washington, and Yale University.

The Participating Institutions in SDSS-IV are Carnegie Mellon University, University of Colorado Boulder, Harvard-Smithsonian Center for Astrophysics Participation Group, Johns Hopkins University, Kavli Institute for the Physics and Mathematics of the Universe, Max-Planck-Institut fuer Astrophysik (MPA Garching), Max-Planck-Institut fuer Extraterrestrische Physik (MPE), Max-Planck-Institut fuer Astronomie (MPIA Heidelberg), National Astronomical Observatory of China, New Mexico State University, New York University, The Ohio State University, Pennsylvania State University, Shanghai Astronomical Observatory, United Kingdom Participation Group, University of Portsmouth, University of Utah, University of Wisconsin, and Yale University.

\bibliography{bibliography3}
\bibliographystyle{mn2e}

\noindent \hrulefill

\noindent $^1$ University of Cambridge, Cavendish Astrophysics, CB3 0HE, Cambridge, UK.
\\$^2$ University of Cambridge, Kavli Institute for Cosmology, CB3 0HE, Cambridge, UK.
\\$^3$ Kavli Institute for the Physics and Mathematics of the Universe, Todai Institutes for Advanced Study, the University
of Tokyo, Kashiwa, Japan 277-8583 (Kavli IPMU, WPI).
\\$^{4}$ Institute of Cosmology and Gravitation, University of Portsmouth, Dennis Sciama Building, Portsmouth, PO1 3FX, UK.
\\$^{5}$ Instituto de Astronomia, Universidad Nacional Autonoma de Mexico, A.P. 70-264, 04510 Mexico D.F., Mexico.
\\$^6$ University of Winsconsin-Madison, Department of Astronomy, 475 N. Charter Street, Madison, WI 53706-1582, USA.
\\$^{7}$ Observatories of the Carnegie Institution for Science, Pasadena, CA, USA.
\\$^{8}$ Yale University, Physics Department, PO Box 208120, New Haven, CT 06520-8121, USA.
\\$^9$ European Southern Observatory, Karl-Schwarzchild-str., 2, 85748, Garching b. Munchen, Germany.
\\$^{10}$ McDonald Observatory, The University of Texas at Austin, 2515 Speedway, Stop C1402, Austin, TX 78712, USA.
\\$^{11}$Universit\'e Lyon 1, Observatoire de Lyon, Centre de Recherche Astrophysique de Lyon and Ecole Normale
Sup\'erieure de Lyon, 9 avenue Charles Andr\'e, F-69230 Saint-Genis Laval, France.
\\$^{12}$ Department of Physics and Astronomy, University of Iowa, 751 Van Allen Hall, Iowa City, IA 52242, USA.
\\$^{13}$ New York University, Centre for Cosmology and Particle Physics, Department of Physics, New York, NY 10003, USA.
\\$^{14}$ NYU Abu Dhabi, PO Box 129188, Abu Dhabi, UAE.
\\$^{15}$ Space Telescope Science Institute, 3700 San Martin Drive, Baltimore, MD 21218, USA.
\\$^{16}$ South East Physics Network (SEPNet), www.sepnet.ac.uk
\\$^{17}$ Department of Physical Sciences, The Open University, Milton Keynes MK7 6AA, UK.
\\$^{18}$ School of Physics and Astronomy, University of St. Andrews, North Haugh, St. Andrews, KY16 9SS, UK.
\\$^{19}$ Department of Physics and Astronomy, University of Kentucky, 505 Rose Street, Lexington, KY 40506-0055, USA.
\\$^{20}$ Partner Group of Max-Planck Institute for Astrophysics, Shanghai Astronomical Observatory, Nandan Road 80,
Shanghai 200030, China.
\\$^{21}$ Cosmology and Galaxy Center, Shanghai Astronomical Observatory, 1819, Nandan Road 80, Shanghai 200030, China.
\\$^{22}$ Apache Point Observatory, P.O. Box 59, Sunspot, NM, USA.

\appendix
\section{Maps for all Galaxies}
\label{app1}
In this section we present the maps for all the 14 galaxies considered in this paper.
From top to bottom and left to right the maps correspond to:

\begin{enumerate}
\item{A \textit{g-r-i} colour-composite image from SDSS, with the positions of the P-MaNGA IFU (purple hexagon) and the SDSS $3''$-diameter fibre (red box) shown.}
\item{Map of the $\rm H \alpha$ velocity (in units of $ \rm km \ s^{-1}$).}
\item{Map of the $\rm H \alpha$ velocity dispersion ($\sigma$), corrected for instrumental dispersion (in units of $ \rm km \ s^{-1}$).  }
\item{Map of the log of the $\rm H \alpha$ flux, in units of $\rm 10^{-17} \ erg \ s^{-1} \ cm^{-2} \AA^{-1}$.  }
\item{Map of the flux ratio of the [SII] doublet ([SII]$\lambda$6731/[SII]$\lambda$6717). This diagnostic is a proxy for electronic density in the gas.  }
\item{Map of the ionisation parameter (log U) as derived from the extinction-corrected $\rm [OIII] \lambda 5007/ [OII] \lambda 3727$ ratio. Note that for simplicity the [OIII]/[OII] ratio has been converted into ionisation parameter for all the spaxels that meet the S/N cuts, but in regions that are not associated with SF (as detailed below in the resolved BPT diagrams) the conversion is invalid since the line ratio traces also the hardness of the ionisation field. Hence we warn the reader to use particular care in reading these maps and compare with the resolved BPT maps.}
\item{Map of the flux ratio $\mathrm{ log ([NII] \lambda 6584/H \alpha) } $.}
\item{Classical BPT diagram using the $\mathrm{ [NII] \lambda 6584/H \alpha} $ and $\mathrm{ [OIII] \lambda 5007/H \beta} $ line ratio. Each point corresponds to a Voronoi bin within the galaxy and is colour-coded by $\rm D_n (4000)$. The open diamond corresponds to the line ratios obtained from the Sloan $3''$-diameter fibre. The median error is shown in the bottom right corner.}
\item{Map of the excitation mechanism of the gas as traced by the [NII] BPT diagram. The colour-coding reflects the position of each region in the BPT diagram and corresponds to the same colour as the labels (SF, Inter and Sy) in the [NII] BPT diagram.}
\item{Map of the flux ratio $\mathrm{ log ([OIII] \lambda 5007/H \beta)} $.}
\item{Classical BPT diagram using the $\mathrm{ [SII] \lambda \lambda 6717, 31/H \alpha} $ and $\mathrm{ [OIII] \lambda 5007/H \beta} $ line ratio. Each point corresponds to a Voronoi bin within the galaxy and is colour-coded by $\rm D_n (4000)$. The open diamond corresponds to the line ratios obtained from the Sloan $3''$-diameter fibre. The median error is shown in the bottom right corner.}
\item{Map of the excitation mechanism of the gas as traced by the [SII]-BPT diagram. The colour-coding reflects the position of each region in the BPT diagram and corresponds to the same colour as the labels (SF, Sy, LINER) in the [SII] BPT diagram.}

\end{enumerate}
The PSF of the P-MaNGA data is represented in all maps in the bottom right corner.
Maps of the stellar population diagnostics $\rm D_n(4000)$ and $\rm EW( H \delta_A)$ are presented in Li et al. (submitted).

\begin{figure*} 
\includegraphics[width=\textwidth, trim=0 70 0 70, clip]{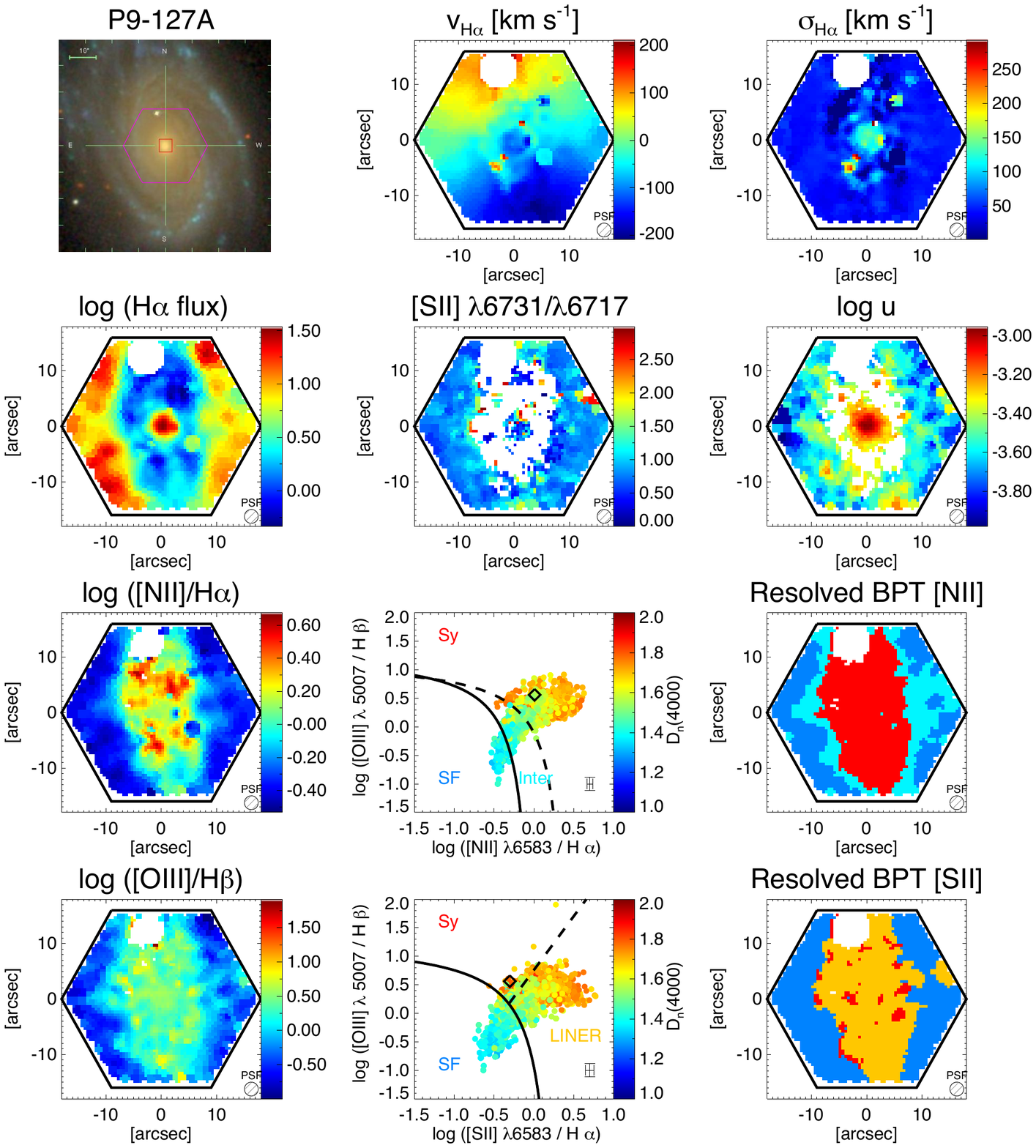} 
\caption{P9-127A, individual maps are described in Appendix \protect\ref{app1} above.}
\end{figure*}

\begin{figure*} 
\includegraphics[width=\textwidth, trim=0 70 0 70, clip]{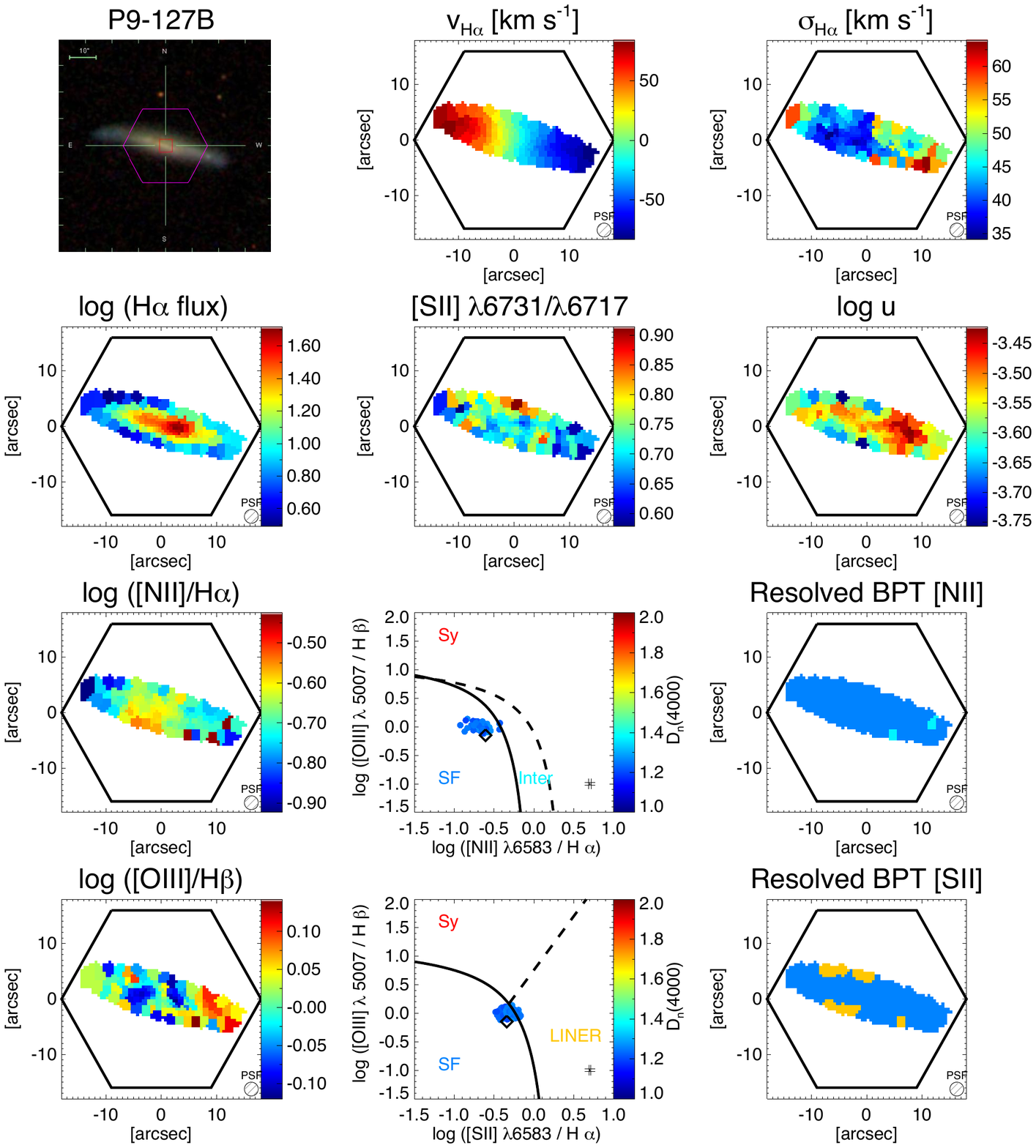} 
\caption{P9-127B}
\end{figure*}

\begin{figure*} 
\includegraphics[width=\textwidth, trim=0 70 0 70, clip]{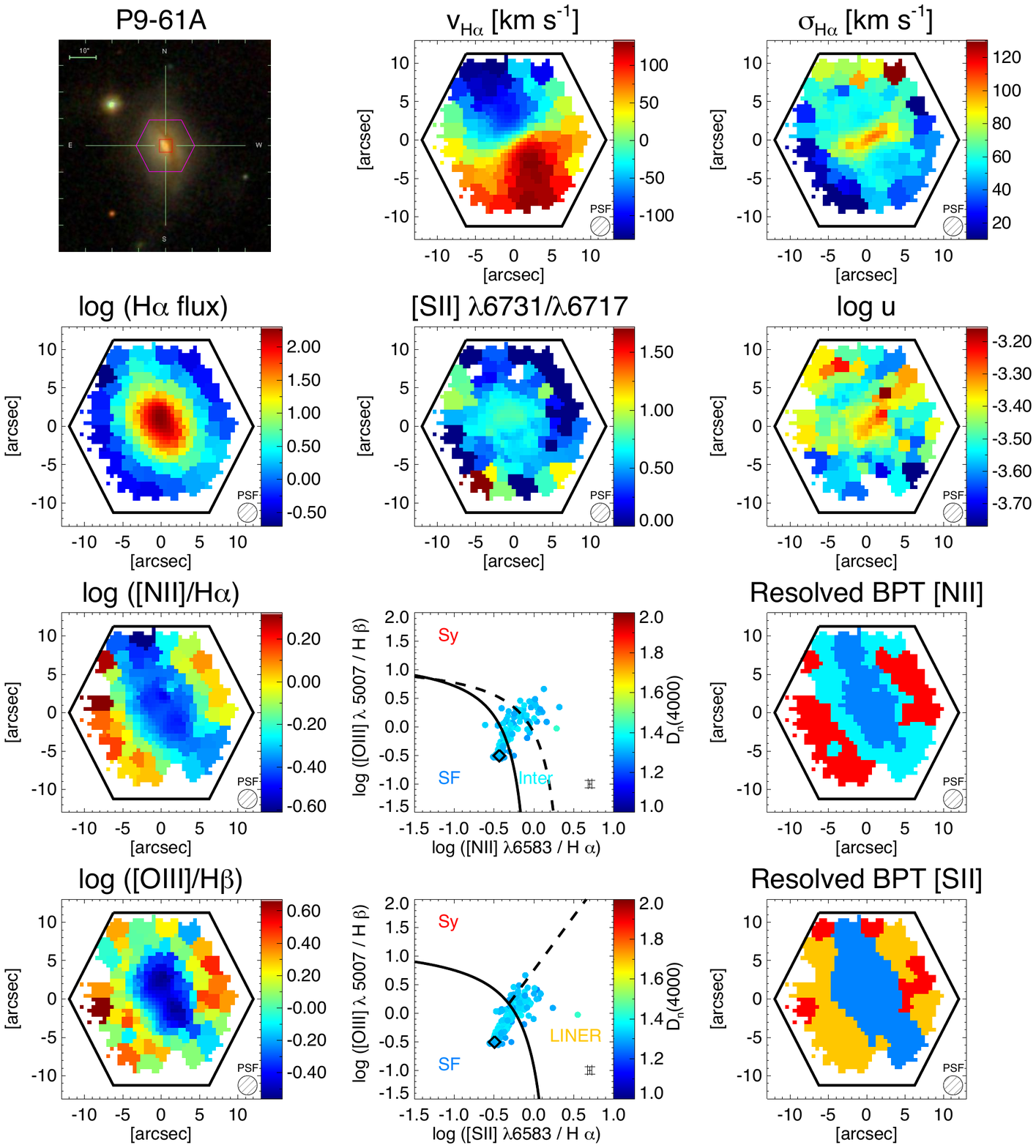} 
\caption{P9-61A}
\end{figure*}
\begin{figure*} 
\includegraphics[width=\textwidth, trim=0 70 0 70, clip]{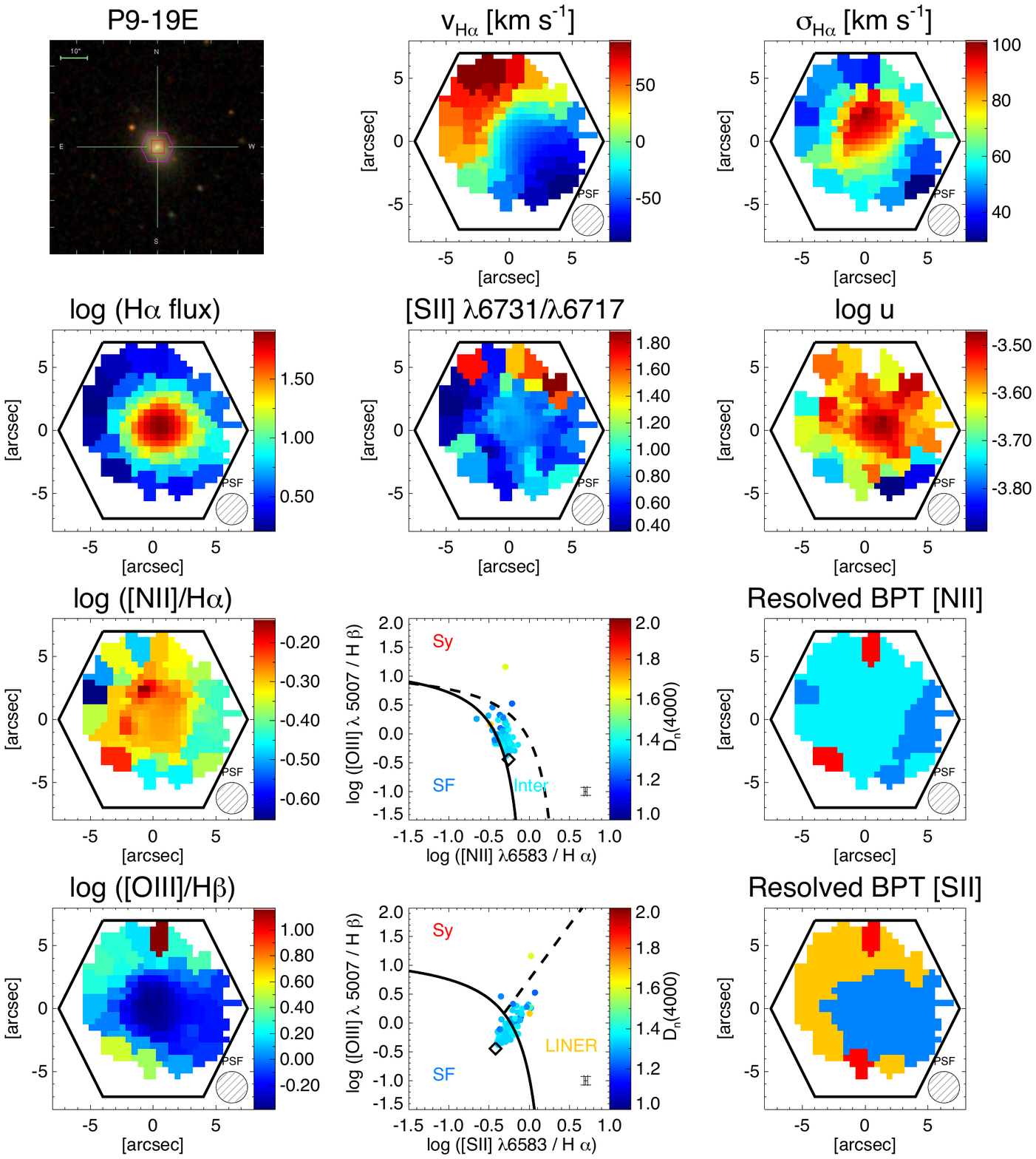} 
\caption{P9-19E}
\end{figure*}

\begin{figure*} 
\includegraphics[width=\textwidth,  trim=0 70 0 70, clip]{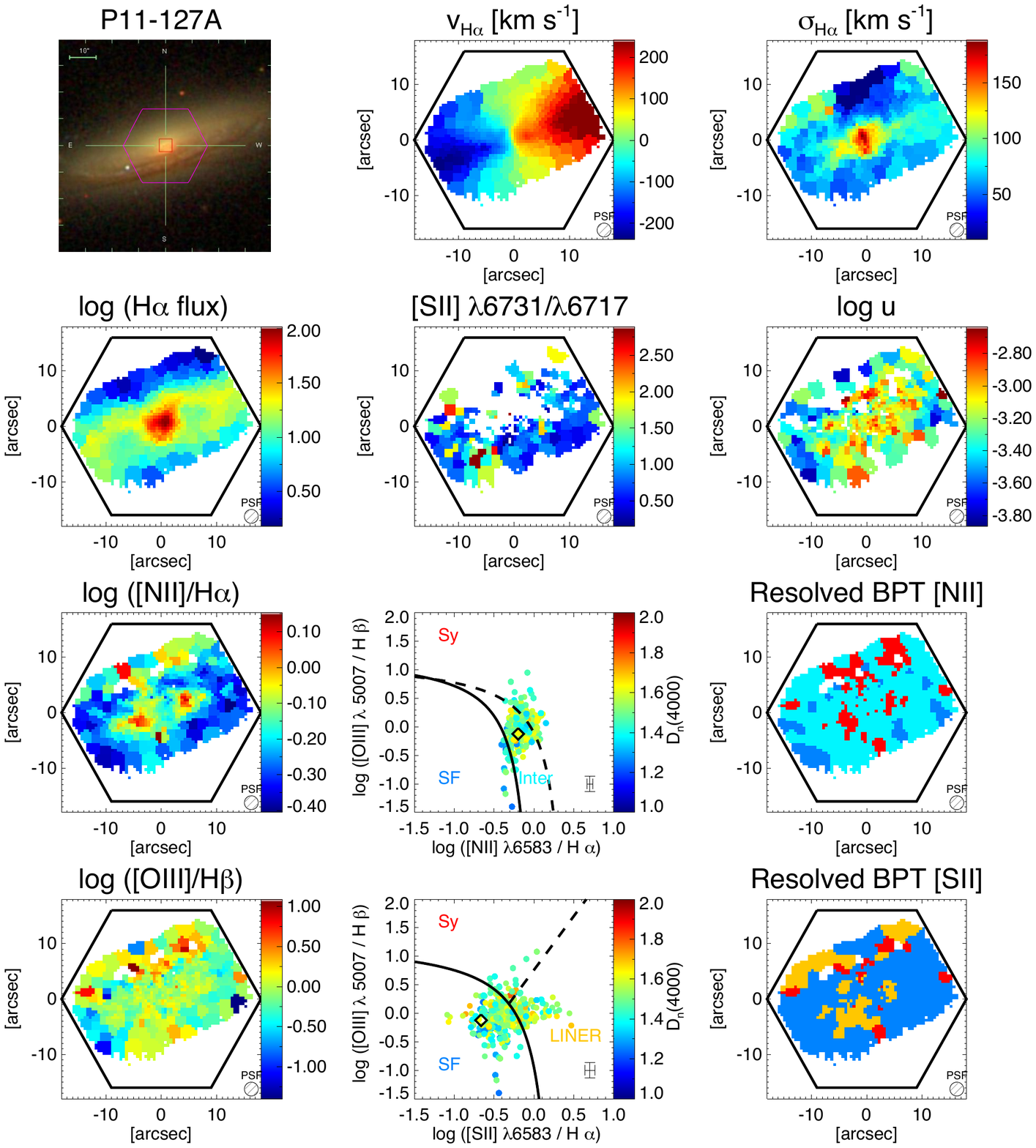} 
\caption{P11-127A}
\end{figure*}

\begin{figure*} 
\includegraphics[width=\textwidth,  trim=0 70 0 70, clip]{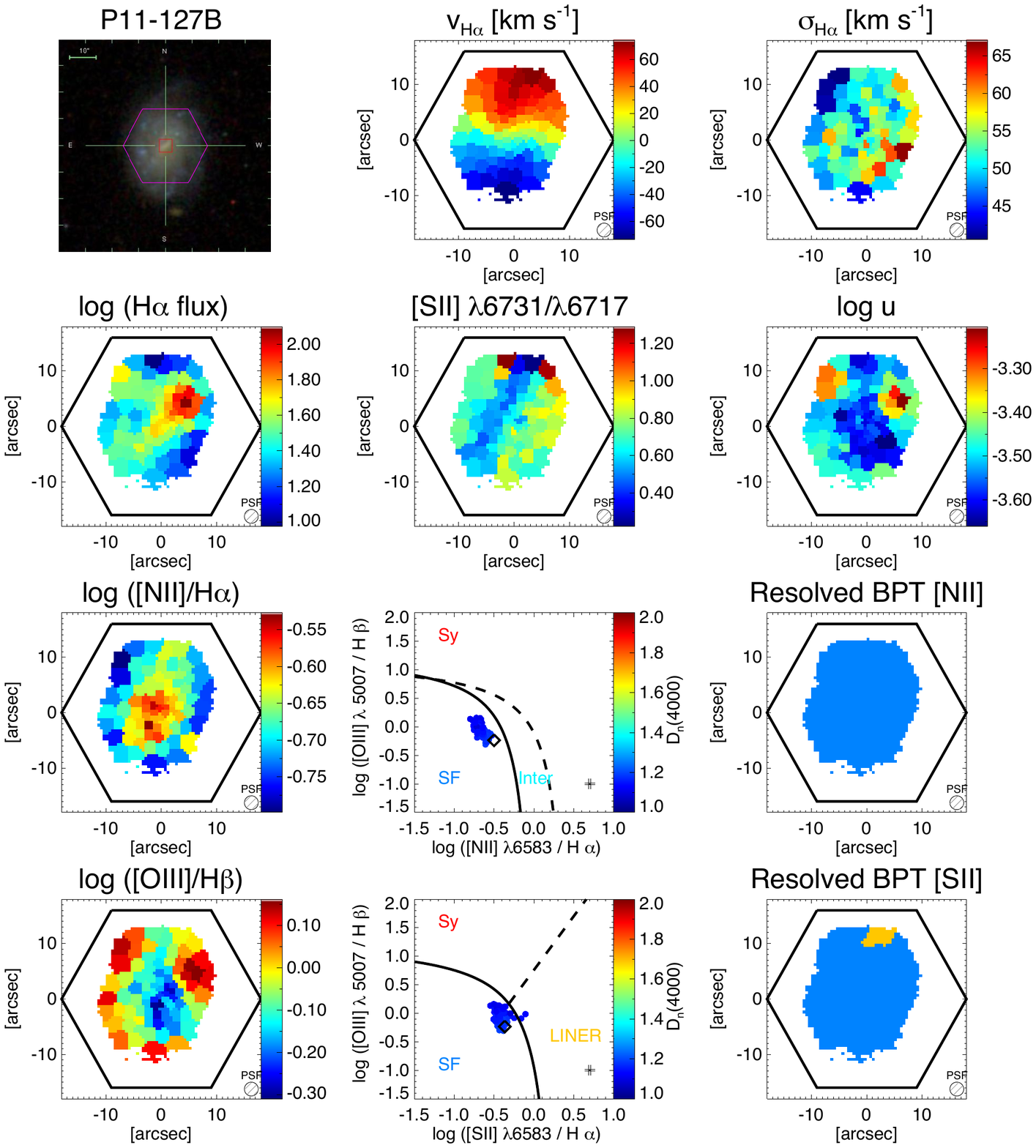} 
\caption{P11-127B}
\end{figure*}

\begin{figure*} 
\includegraphics[width=\textwidth,  trim=0 70 0 70, clip]{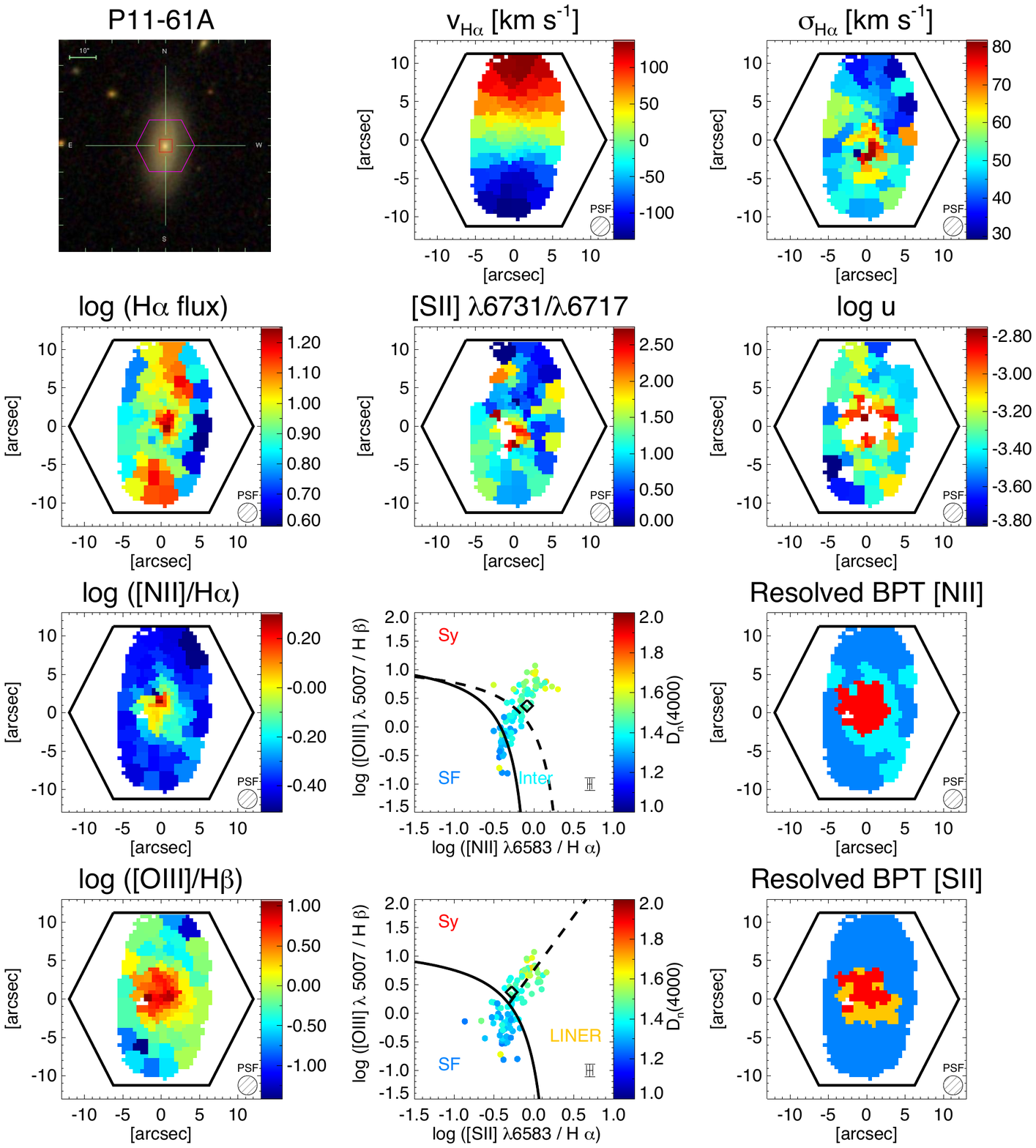} 
\caption{P11-61A}
\end{figure*}

\begin{figure*} 
\includegraphics[width=\textwidth,  trim=0 70 0 70, clip]{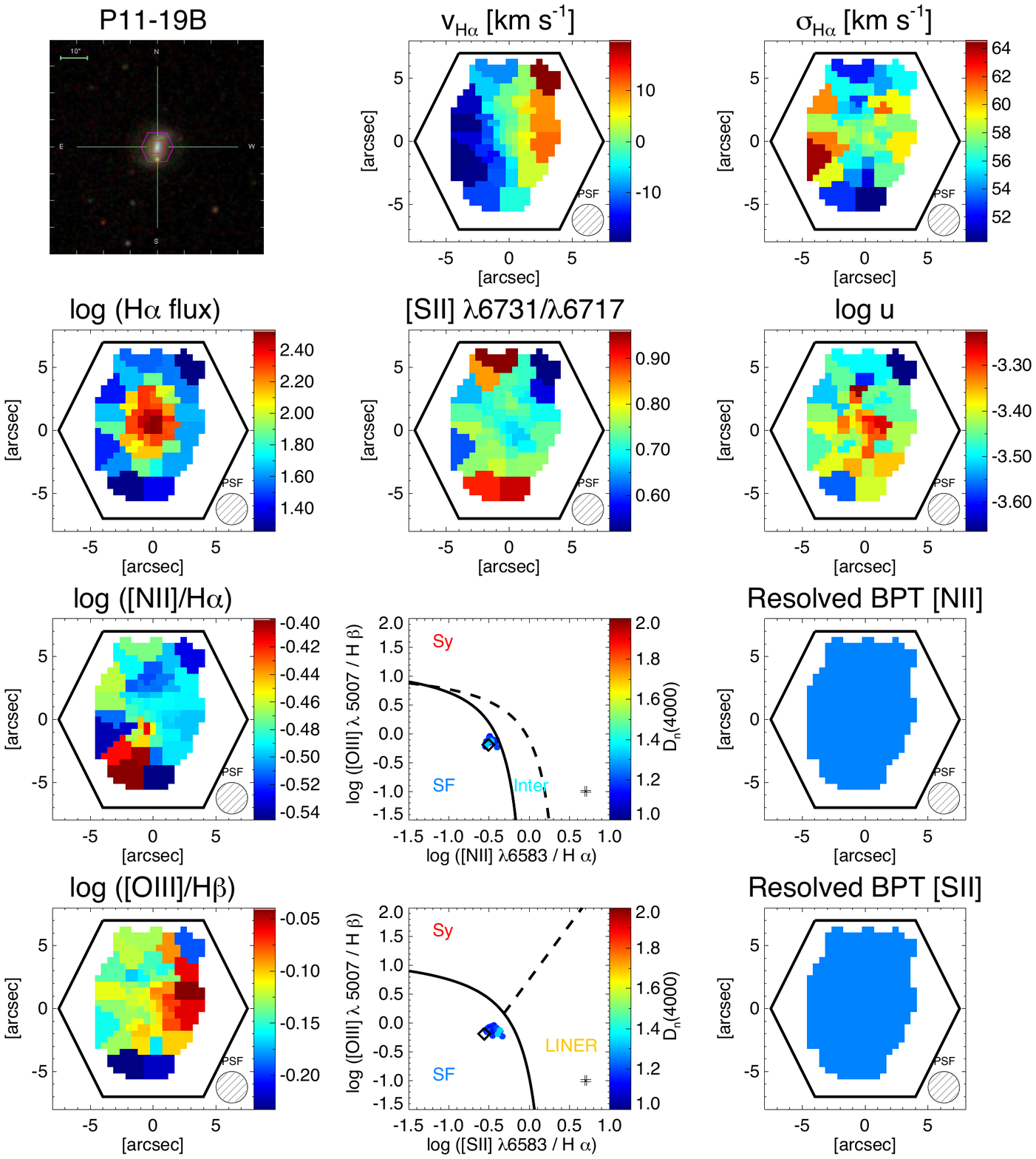} 
\caption{P11-19B}
\end{figure*}

\begin{figure*} 
\includegraphics[width=\textwidth,  trim=0 70 0 70, clip]{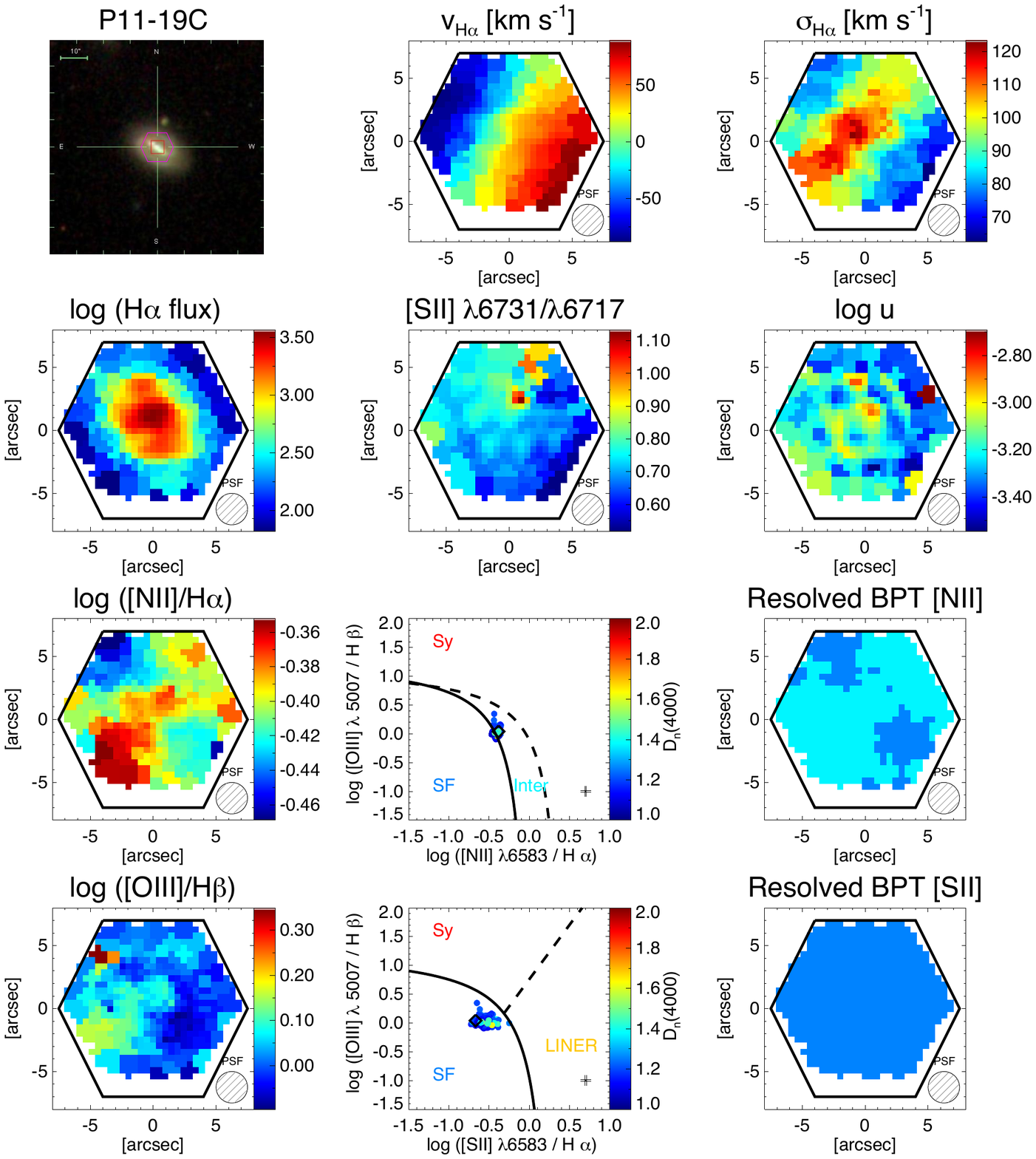} 
\caption{P11-19C}
\end{figure*}

\begin{figure*} 
\includegraphics[width=\textwidth,  trim=0 70 0 70, clip]{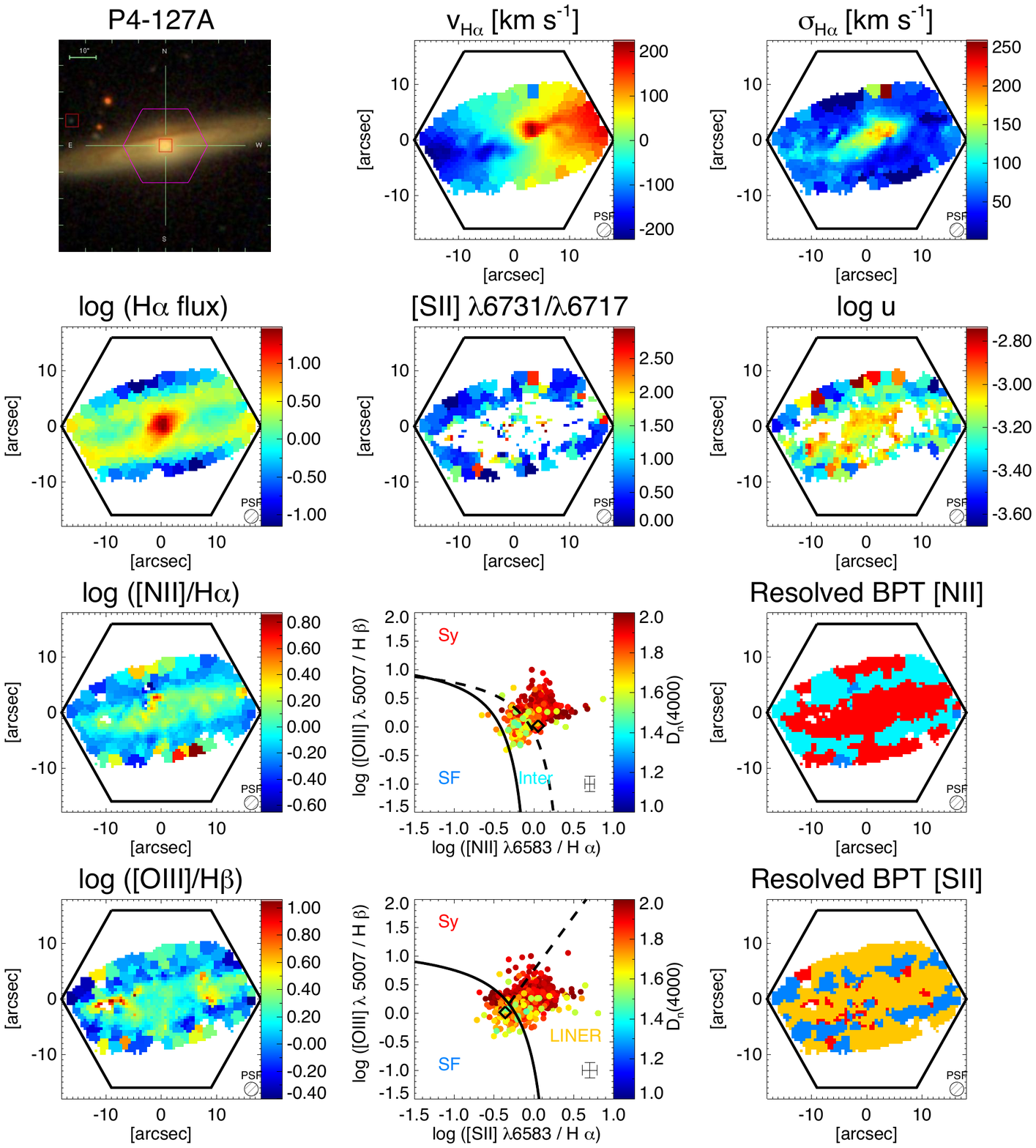} 
\caption{P4-127A}
\end{figure*}

\begin{figure*} 
\includegraphics[width=\textwidth,  trim=0 70 0 70, clip]{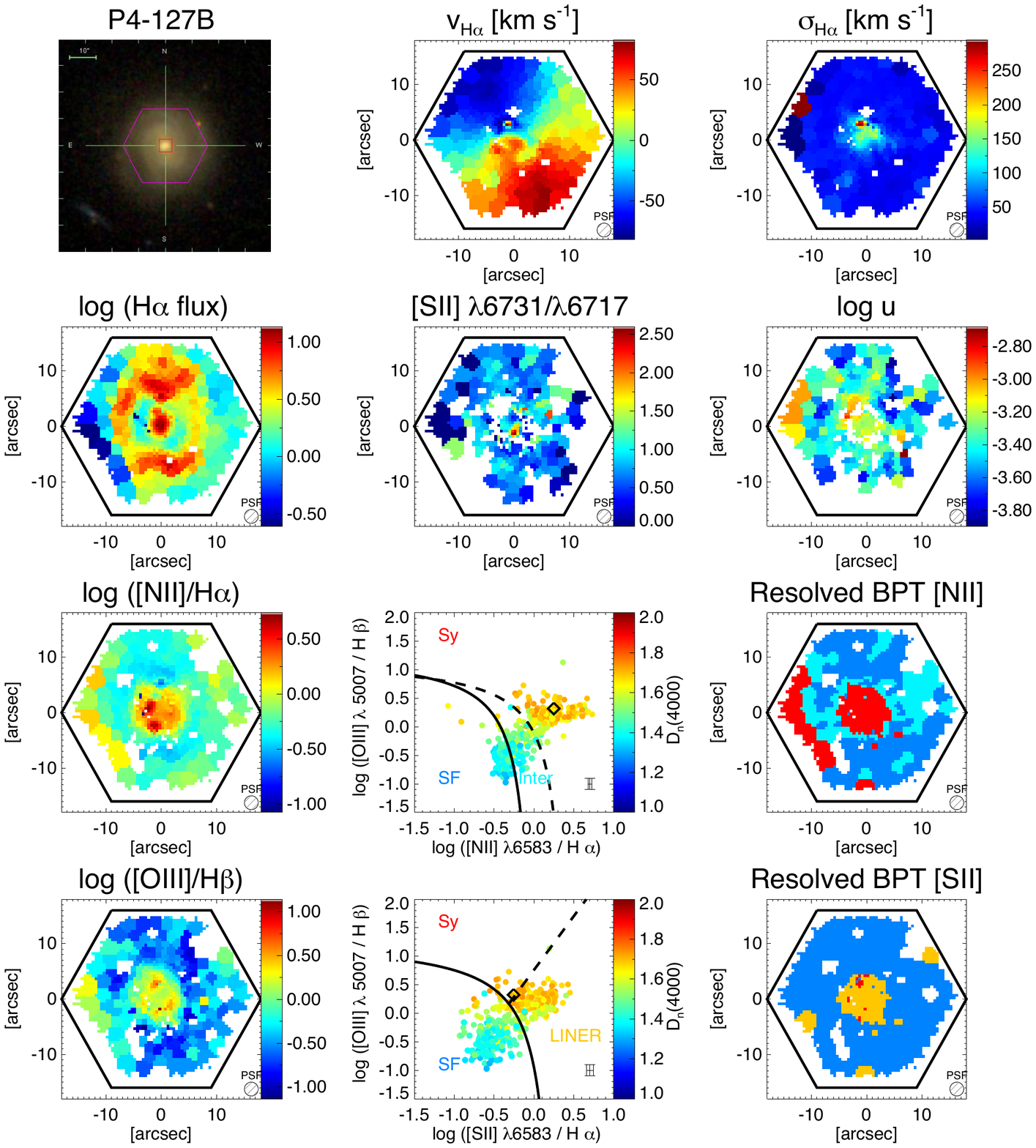} 
\caption{P4-127B}
\end{figure*}

\begin{figure*} 
\includegraphics[width=\textwidth,  trim=0 70 0 70, clip]{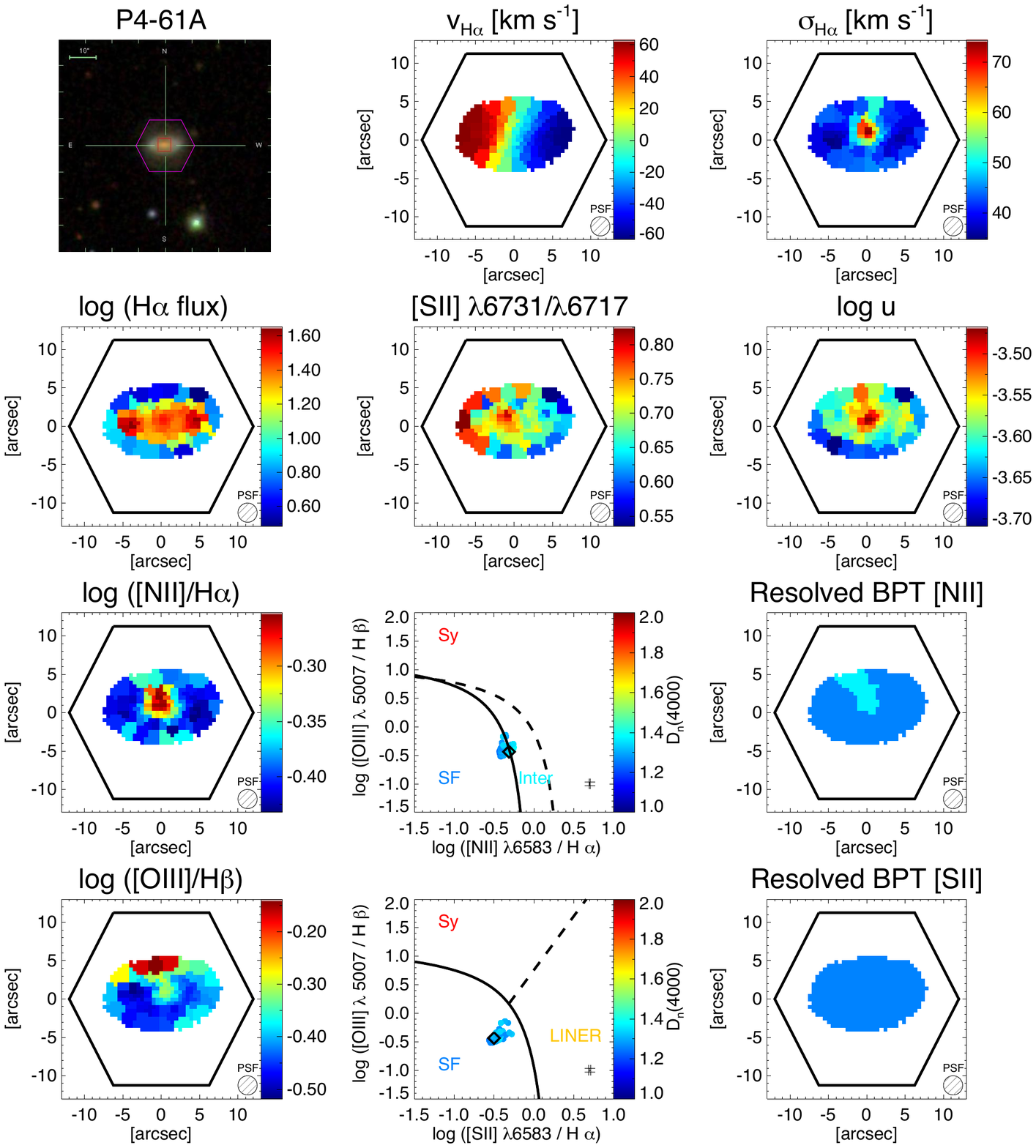} 
\caption{P4-61A}
\end{figure*}

\begin{figure*} 
\includegraphics[width=\textwidth,  trim=0 70 0 70, clip]{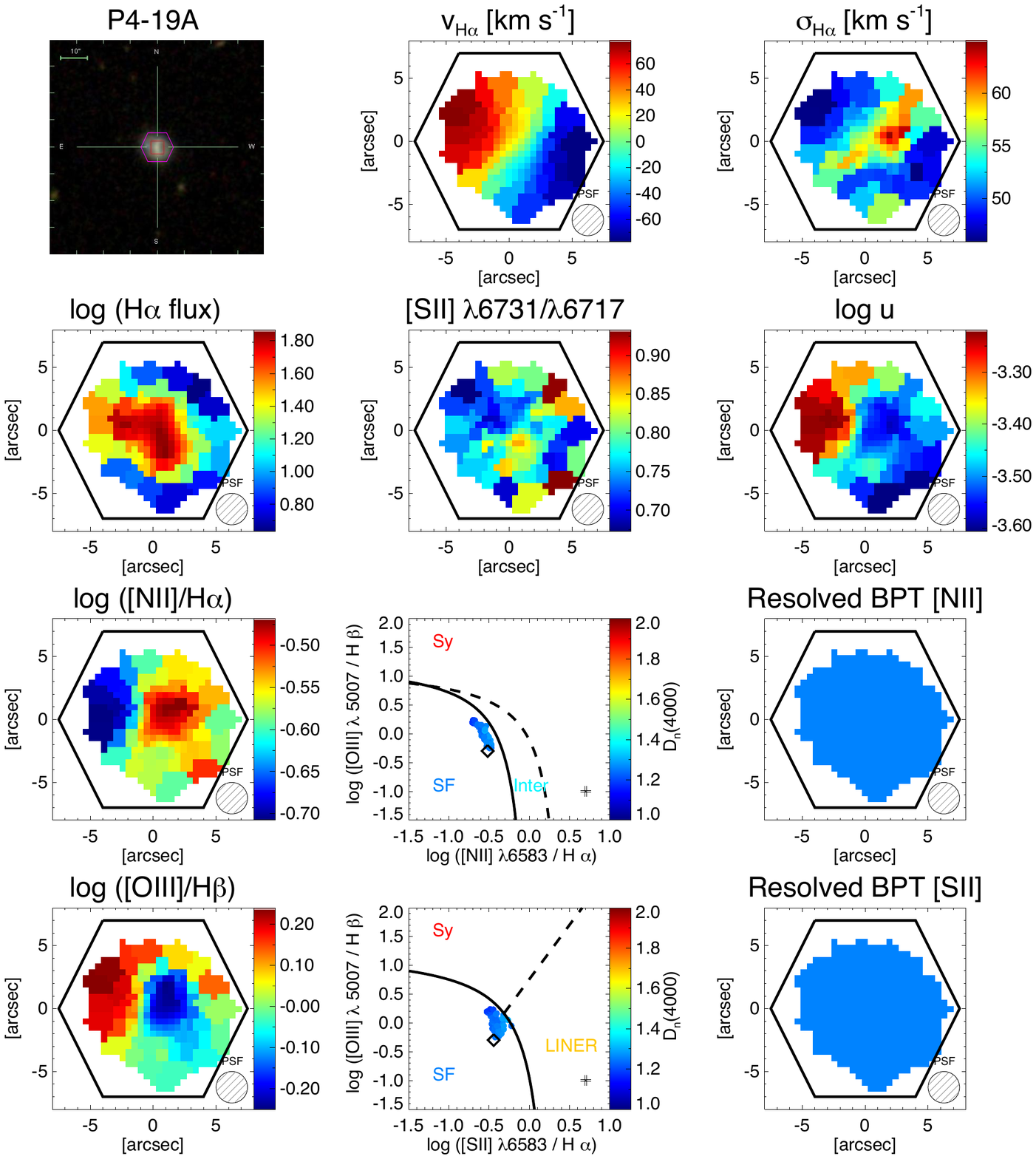} 
\caption{P4-19A}
\end{figure*}

\begin{figure*} 
\includegraphics[width=\textwidth,  trim=0 70 0 70, clip]{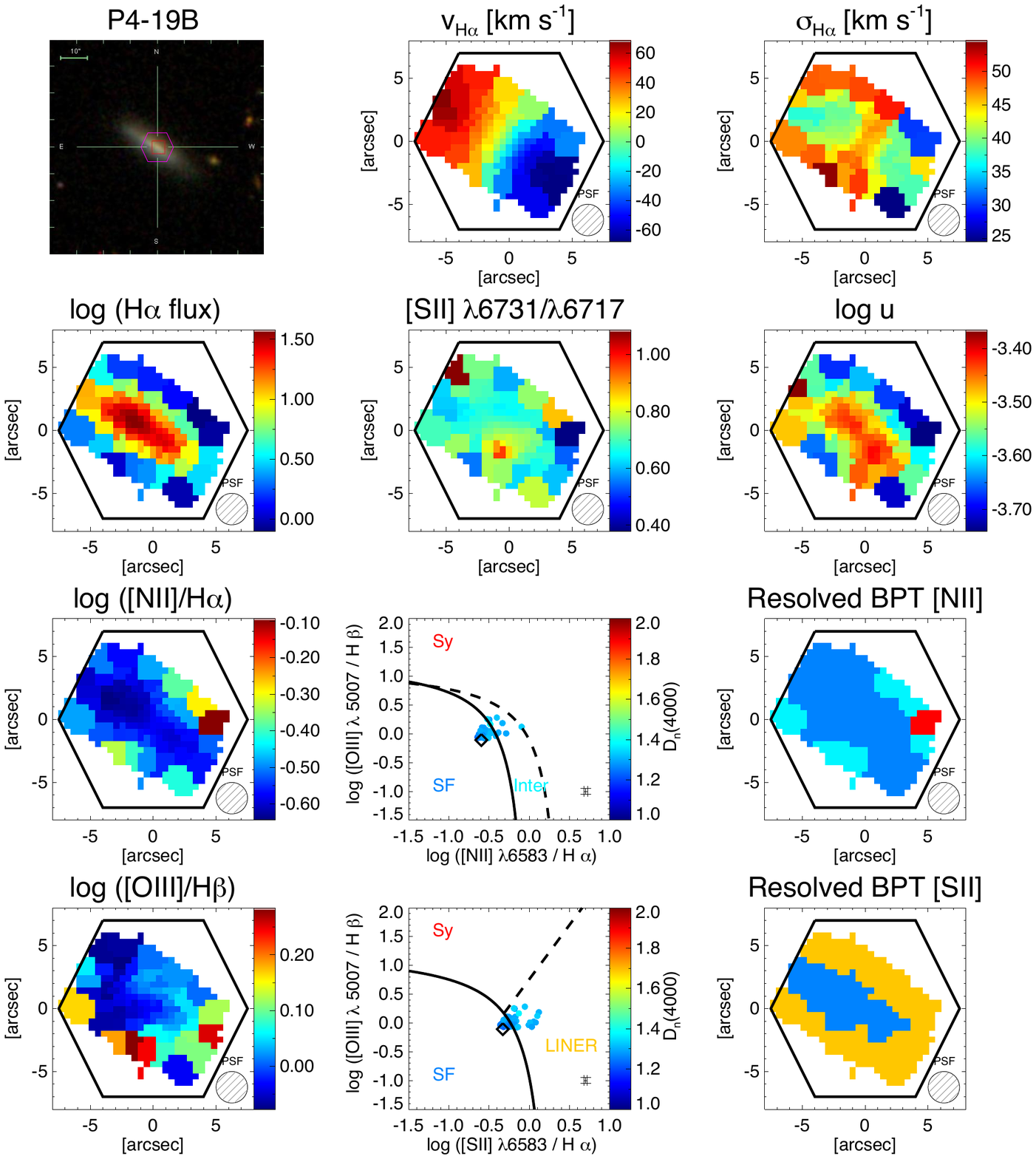} 
\caption{P4-19B}
\end{figure*}

\end{document}